\RequirePackage{fix-cm}
\documentclass[twocolumn, epjc3, comma, sort&compress, natbib]{svjour3}
\bibpunct{[}{]}{,}{n}{}{,} 

\journalname{Eur. Phys. J. C}

\usepackage{latexsym}
\usepackage{amsmath}
\usepackage{amssymb}
\usepackage{amsfonts}
\usepackage{CJKutf8}

\usepackage[mathscr,scaled=1.15]{urwchancal}
\DeclareFontFamily{OT1}{pzc}{}
\DeclareFontShape{OT1}{pzc}{m}{it}%
{<-> s * [1.15] pzcmi7t}{}
\DeclareMathAlphabet{\mathpzc}{OT1}{pzc}{m}{it}

\usepackage{color}

\usepackage{supertabular}
\usepackage{placeins}
\usepackage{epsfig}
\usepackage{graphicx}

\definecolor{purple}{rgb}{0.5,0,0.5}
\definecolor{blue}{rgb}{0.0,0,0.9}
\definecolor{prdblue}{rgb}{0.133,0.118,0.498}
\usepackage[colorlinks=true, pdfstartview=FitV, linkcolor=prdblue, citecolor= prdblue, urlcolor=prdblue]{hyperref}

\begin{document}

\begin{CJK}{UTF8}{song}

\title{$\,$\\[-6ex]\hspace*{\fill}{\normalsize{\sf\emph{Preprint no}.\ NJU-INP 083/24}}\\[1ex]%
Contact interaction study of proton parton distributions}

\author{Yang Yu\thanksref{NJU,INP}
        $\,^{\href{https://orcid.org/0009-0008-8011-3430}{\textcolor[rgb]{0.00,1.00,0.00}{\sf ID}}}$
\and
    Peng Cheng\thanksref{NJU,INP}
        $\,^{\href{https://orcid.org/0000-0002-6410-9465}{\textcolor[rgb]{0.00,1.00,0.00}{\sf ID}}}$
\and
    Hui-Yu\ Xing\thanksref{NJU,INP}
    $\,^{\href{https://orcid.org/0000-0002-0719-7526}{\textcolor[rgb]{0.00,1.00,0.00}{\sf ID}}}$
\and
    \\ Fei Gao\thanksref{BIT}
    $\,^ {\href{https://orcid.org/0000-0001-5925-5110}{\textcolor[rgb]{0.00,1.00,0.00}{\sf ID}}}$
\and
        Craig D.\ Roberts\thanksref{NJU,INP}%
       $\,^{\href{https://orcid.org/0000-0002-2937-1361}{\textcolor[rgb]{0.00,1.00,0.00}{\sf ID}}}$
}

\authorrunning{Yang Yu \emph{et al}.} 

\institute{School of Physics, Nanjing University, Nanjing, Jiangsu 210093, China \label{NJU}
           \and
           Institute for Nonperturbative Physics, Nanjing University, Nanjing, Jiangsu 210093, China \label{INP}
           \and
           School of Physics, Beijing Institute of Technology, Beijing, 100811, P.R. China \label{BIT}
\\[1ex]
Email:
\href{mailto:fei.gao@bit.edu.cn}{fei.gao@bit.edu.cn} (F. Gao);
\href{mailto:cdroberts@nju.edu.cn}{cdroberts@nju.edu.cn} (C. D. Roberts)
            }

\date{2024 July 09}

\maketitle

\end{CJK}

\begin{abstract}
Using a symmetry-preserving formulation of a vector$\,\times\,$vector contact interaction (SCI) and treating the proton as a quark + interacting-diquark bound state,  whose structure is obtained by solving a Poincar\'e-covariant Faddeev equation, we provide a comprehensive, coherent set of predictions for unpolarised and polarised proton parton distribution functions (DFs): valence, glue, and four-flavour separated sea.
The results enable many themes to be addressed, including:
the asymmetry of antimatter in the proton;
the neutron:proton structure function ratio;
helicity retention in hard scattering processes;
the charm quark momentum fraction;
the sign and size of the polarised gluon DF;
and the origin of the proton spin.
In all cases where sound analyses of data are available, SCI predictions are semiquantitatively in agreement with the results.  Those mismatches which exist are typically attributable to the momentum-independence of the underlying interaction.
Judiciously interpreted, the SCI delivers a sound and insightful explanation of proton structure as expressed in DFs.
\end{abstract}


%
\section{Introduction}
Viewed from a Standard Model perspective, the proton is a composite system, with unit positive electric charge, built from three valence quarks ($u+u+d$), whose associated state vector is labelled by three Poincar\'e invariant quantities: mass squared, $m_p^2$; total angular momentum squared (spin), $J^2=J(J+1)=3/4$; and parity $P=+1$.  At this point, all seems straightforward.  However, it has long been known that much of the proton's spin is not carried by its valence quarks \cite{EuropeanMuon:1987isl, Aidala:2012mv, Deur:2018roz} and science is awakening to the fact that the Higgs boson alone is responsible for only $\approx 1$\% of the proton mass \cite{Bashir:2012fs, Roberts:2016vyn}.  Thus, today, the proton is recognised to be an object which is so complex that modern and next generation accelerators accommodate dedicated programmes focused on collecting data that will deliver answers to the following questions \cite{Aguilar:2019teb, Chen:2020ijn, Anderle:2021wcy, Arrington:2021biu, AbdulKhalek:2021gbh, Quintans:2022utc, Carman:2023zke}: How do the mass and spin of the proton arise?

One approach to developing answers to these questions is found in charting the distribution of valence quark, sea quark, and gluon partons within the proton.  Efforts centred on relevant measurements have long been a focus of experiments \cite[Vol.\,1, Sec.\,18]{Workman:2022ynf}.  Theoretically, in order to correlate and understand the data obtained, one would ideally employ a single framework that simultaneously delivers predictions for both unpolarised and polarised parton distributions and connects them with the strong interaction dynamics that generates a proton mass scale on the order of 1\,GeV in the absence of Higgs boson couplings to the quarks \cite{Binosi:2022djx, Papavassiliou:2022wrb, Ding:2022ows, Ferreira:2023fva}.  Lattice Schwinger function methods have come to provide one such approach \cite{Lin:2017snn}.  Progress with that framework is complemented by advances made using continuum Schwinger function methods (CSMs), which have delivered parameter-free, unifying predictions for all proton (unpolarised and polarised), pion and kaon parton distribution functions (DFs) -- valence, sea, and glue \cite{Cui:2020tdf, Chang:2022jri, Lu:2022cjx, Cheng:2023kmt}, and pion fragmentation functions \cite{Xing:2023pms}.  \linebreak Where valid comparisons are possible, the continuum and lattice DF predictions agree -- see, \emph{e.g}., Refs.\,\cite{Chang:2021utv, Lu:2022cjx, Cui:2021mom, Cui:2022bxn, Lu:2023yna}.

A merit of continuum analyses is the ability to test the sensitivity of a given body of results to statements about dynamics.  This has proved useful in connection with hadron elastic and transition form factors; for instance, highlighting those features of observables which are most sensitive to the character of the quark-quark interaction \cite{GutierrezGuerrero:2010md, Roberts:2010rn, Roberts:2011wy, Wilson:2011aa, Chen:2012txa, Segovia:2013uga, Xu:2015kta, Bedolla:2016yxq, Raya:2021pyr, Xing:2022sor, Cheng:2022jxe, Xing:2022jtt}.  Analogous studies have been completed for the distribution amplitudes (DAs) of some two-body systems \cite{Lu:2021sgg} and the light-front momentum fraction $x\simeq 1$ ratios of nucleon DFs \cite{Roberts:2013mja}.  We consider it valuable to extend the latter analyses to calculations of the pointwise behaviour of nucleon DFs because this will enable comparisons with reliable phenomenological analyses of data to be used as a means of exposing connections between DFs and the interaction that generates them.

Herein, therefore, we calculate all proton DFs, unpolarised and polarised, using the symmetry-preserving formulation of a vector$\,\times\,$vector contact interaction \linebreak (SCI) introduced in Refs.\,\cite{GutierrezGuerrero:2010md, Roberts:2010rn, Roberts:2011wy, Wilson:2011aa}.  Direct comparison will be made with what may be called the QCD-kindred predictions in Refs.\,\cite{Lu:2022cjx, Cheng:2023kmt}, obtained using Schwinger functions that realistically express basic features of quantum chromodynamics (QCD), and relevant, available inferences from data.

An additional merit of this effort is the fact that SCI analyses are largely algebraic.  Hence, all calculations possess a degree of transparency that enables robust insights to be drawn; not just about the SCI studies themselves, but also concerning results obtained using more sophisticated frameworks.  This is because SCI results typically deliver both a useful first estimate of a given observable and a means of checking the validity of algorithms employed in calculations that rely upon high performance computing.

Section~\ref{SecUDF} outlines the derivation of all formulae used in the SCI calculation of hadron-scale, $\zeta_{\cal H}$, proton fla\-vour-non\-sing\-let (valence) DFs in the quark + interac\-ting-diquark picture of baryon structure \cite{Barabanov:2020jvn}, which is a widely used, efficacious approximation to the full three-body problem.  The sketches are augmented by background and reference material in \ref{AppendixSM} and \ref{AppendixDFs}.
Algebraic and numerical results pertaining to the $\zeta_{\cal H}$  valence DFs are reported in Sec.\,\ref{SeczH}.

Comparisons with data require scale evolution to $\zeta > \zeta_{\cal H}$.  We employ the all-orders (AO) scheme detailed elsewhere \cite{Yin:2023dbw}, which has proved efficacious in numerous applications, \emph{e.g}., delivering unified predictions for all pion, kaon, and proton (unpolarised and polarised) DFs \cite{Cui:2020tdf, Chang:2022jri, Lu:2022cjx, Cheng:2023kmt}, and pion fragmentation functions \cite{Xing:2023pms}, that agree with much available data.  In this approach, all flavour singlet DFs are zero at $\zeta_{\cal H}$ -- this feature is definitive of the hadron scale.  On $\zeta>\zeta_{\cal H}$, they are generated via evolution from the nonsinglet hadron-scale distributions.  In Sec.\,\ref{EvolvedDFs}, thus obtained, SCI results for all DFs are compared with existing QCD-kindred predictions and available data.
Thereafter, having calculated the gluon helicity DF, the spin of the proton is discussed in Sec.\,\ref{SecSpin}.

Section~\ref{epilogue} provides a summary and perspective.

\section{Distribution Functions: Algebraic Formulae}
\label{SecUDF}
\subsection{Helicity independent}
In formulating the calculation of helicity-independent proton DFs, we follow Ref.\,\cite[Sec.\,2]{Chang:2022jri}.  The first step is to identify the hadron scale, $\zeta_{\cal H}$, as that probe resolution at which valence degrees-of-freedom carry all properties of the proton, including its light-front momentum.  This is a basic tenet of the AO DF evolution scheme \cite{Yin:2023dbw}.

One then supposes that a hadron scale proton, with total momentum $K$, is effectively a two-body system, \emph{viz}.\ in some practical respects, built from dressed valence-quarks and nonpointlike, interacting isoscalar-scalar and isovector-axial\-vector quark + quark (diquark) correlations \cite{Barabanov:2020jvn}.
Such a quark + diquark picture of the proton represents the bound-state via the following canonically-normalised Faddeev amplitude \cite{Chen:2012qr}:
\begin{equation}
\label{FAproton}
\psi(\ell;K) = \sum_{J^P=0^+,1_{\{uu\}}^+,1_{\{ud\}}^+} a_{J^P} \psi^{J^P}(\ell;K)\,.
\end{equation}
Here: $a_{0^+},a_{1_{\{uu\}}^+},a_{1_{\{ud\}}^+}$ measure the relative strength of the scalar and axialvector diquark terms [$a_{1^+} := a_{1_{\{ud\}}^+} = -a_{1_{\{uu\}}^+}/\surd 2$ owing to isospin Clebsch-Gordon couplings]; $\psi^{0^+,1^+}(\ell;K)$ describe the dependence of the amplitude on the quark momentum, $\ell$; and the Lorentz index associated with the $1^+$ correlations has been suppressed.

In such a proton, the hadron-scale valence $u$ quark DF, ${\mathpzc u}_V^p(x;\zeta_{\cal H})$, receives four distinct contributions:
\begin{equation}
{\mathpzc u}_V^p(x;\zeta_{\cal H}) = \sum_{t=Q_0,Q_1,D_0,D_1} {\mathpzc u}_{V_t}^p(x;\zeta_{\cal H})\,.
\label{Equproton}
\end{equation}
Term $1$ is provided by the probe striking a $u$-quark that is accompanied by a bystander scalar diquark
($\hat\delta_n^{xK} = n\cdot K \delta(n\cdot \ell - x n\cdot K)$, $n^2=0$, $n\cdot P=-m_p$ in the proton rest frame):
\begin{align}
\Lambda_+ & \gamma\cdot n  {\mathpzc u}_{V_{Q_0}}^p(x;\zeta_{\cal H}) \Lambda_+   =
\int\! \tfrac{d^4\ell}{(2\pi)^4} \,
\hat\delta_n^{xK} \;
\Lambda_+ Q_0  \Lambda_+\,, \label{uproton1}
\end{align}
with
$\Lambda_+ =(m_p-i\gamma \cdot K)/(2m_p)$, $K^2=-m_p^2$, $m_p$ being the proton mass; and
\begin{align}
Q_0 & = a_{0^+}^2 \,
\bar\psi^{0^+}(\ell-\tfrac{1}{2}K;-K) S(\ell)  \nonumber \\
& \quad \times \gamma\cdot n S(\ell) \psi^{0^+}(\ell-\tfrac{1}{2}K;K) \Delta^{0^+}(\ell-K)\,,
\label{EqQ0}
\end{align}
where $S(\ell)$ is the propagator of the dressed valence-quark and $\Delta^{0^+}(\ell-K)$ is the scalar diquark propagator.

Term 2 is generated by the probe striking a $u$-quark that is accompanied by an axialvector $\{ud\}$ diquark:
\begin{align}
\Lambda_+ & \gamma\cdot n  {\mathpzc u}_{V_{Q_1}}^p(x;\zeta_{\cal H}) \Lambda_+   =
\int\! \tfrac{d^4\ell}{(2\pi)^4} \,
\hat\delta_n^{xK} \;
\Lambda_+ Q_1  \Lambda_+\,, \label{uproton11}
\end{align}
with
\begin{align}
Q_1 & = a_{1_{\{ud\}}^+}^2 \,
\bar\psi_\rho^{1_{\{ud\}}^+}(\ell-\tfrac{1}{2}K;-K) S(\ell)  \nonumber \\
& \quad \times  \gamma\cdot n S(\ell) \psi_\sigma^{1_{\{ud\}}^+}(\ell-\tfrac{1}{2}K;K) \Delta_{\rho\sigma}^{1^+}(\ell-K)\,,
\label{EqQ1}
\end{align}
where $\Delta_{\rho\sigma}^{1^+}(\ell-K)$ is the axialvector propagator.

Term 3 exposes the $u$-quark within the scalar diquark and is expressed as a convolution:
\begin{align}
{\mathpzc u}_{V_{D_0}}^p(x;\zeta_{\cal H}) = \int_x^1\,\frac{dy}{y}\,{\mathpzc s}_{0^+}^p(y;\zeta_{\cal H})
{\mathpzc u}_V^{0^+}(x/y;\zeta_{\cal H})\,. \label{convolution0}
\end{align}
(The $1/y$ factor was inadvertently omitted in the written formulae of Ref.\,\cite[Sec.\,2]{Chang:2022jri}, although it was used in calculations.)
Here, ${\mathpzc u}_V^{0^+}(x;\zeta_{\cal H})$ is the valence $u$-quark DF in a scalar diquark; and the probability density for finding a scalar diquark carrying a light-front fraction $x$ of the proton's momentum is
\begin{subequations}
\begin{align}
\Lambda_+ & \gamma\cdot n  {\mathpzc s}_{0^+}^p(x;\zeta_{\cal H}) \Lambda_+   =
\int\! \tfrac{d^4\ell}{(2\pi)^4} \,
\hat\delta_n^{xK} \;
\bar\Lambda_+ D_{0}  \bar\Lambda_+\,,\\
D_{0} & =  a_{0^+}^2  i n\cdot \partial^\ell
\left[\bar\psi^{0^+}(\ell-\tfrac{1}{2}K;K)
 \right. \nonumber \\
& \qquad \left. \underline{S(\ell-K)} \psi^{0^+}(\ell-\tfrac{1}{2}K;-K) \Delta^{0^+}(\ell)\right]\,;
\end{align}
\end{subequations}
the underlined propagator is \underline{not} differentiated, so $D_{0}$ has three distinct terms;
and $\bar\Lambda_+ = (m_p+i\gamma \cdot K)/(2m_p)$.

{\allowdisplaybreaks
Term 4 reveals the $u$-quarks in the $1^+$ diquarks.  Its form in the isospin-symmetry limit follows:
\begin{subequations}
\label{convolution1}
\begin{align}
{\mathpzc u}_{V_{D_1}}^p(x;\zeta_{\cal H}) &= 5 {\mathpzc q}_{V_{D_1}}^p(x;\zeta_{\cal H})\,, \\
{\mathpzc q}_{V_{D_1}}^p(x;\zeta_{\cal H})& =  \int_x^1\,\frac{dy}{y}\,{\mathpzc s}_{1^+}^p(y;\zeta_{\cal H})
{\mathpzc u}_V^{1^+}(x/y;\zeta_{\cal H})\,,
\end{align}
\end{subequations}
where the ``5'' owes to isospin Clebsch-Gordon couplings
and the presence of two active $u$-quarks in $\{uu\}$, and
\begin{subequations}
\label{uproton3}
\begin{align}
 \Lambda_+ & \gamma\cdot n  {\mathpzc s}_{1^+}^p(x;\zeta_{\cal H}) \Lambda_+
= \int\! \tfrac{d^4\ell}{(2\pi)^4} \, \hat\delta_n^{xK}\;
\bar \Lambda_+ D_{1}  \bar \Lambda_+\,,\\
D_{1} & =   a_{1_{\{ud\}}^+}^2 i n\cdot \partial^\ell
\left[\bar\psi_\rho^{1_{\{ud\}}^+}(\ell-\tfrac{1}{2}K; K) \right. \nonumber \\
& \qquad \left.
\underline{S(\ell-K)} \psi_\sigma^{1_{\{ud\}}^+}(\ell-\tfrac{1}{2}K;-K) \Delta_{\rho\sigma}^{1^+}(\ell)\right]\,.
\end{align}
\end{subequations}
Again, the underlined propagator is \underline{not} differentiated.
}

In each case, the integrals above are evaluated using the relevant expressions and results from \ref{AppendixSM} and the standard SCI regularisation procedures.  One thereby arrives at the formulae in \ref{AppendixDFs}.

The helicity-independent proton valence $d$-quark DF is similarly obtained:
\begin{equation}
{\mathpzc d}_V^p(x;\zeta_{\cal H}) = \sum_{t=Q_1,D_0,D_1} {\mathpzc d}_{V_t}^p(x;\zeta_{\cal H})\,, \label{dquarkDF}
\end{equation}
where, owing to isospin symmetry,
\begin{subequations}
\label{dpDFs}
\begin{align}
{\mathpzc d}_{V_{Q_1}}^p(x;\zeta_{\cal H}) & = 2 {\mathpzc u}_{V_{Q_1}}^p(x;\zeta_{\cal H})\,, \\
{\mathpzc d}_{V_{D_0}}^p(x;\zeta_{\cal H}) & =  {\mathpzc u}_{V_{D_0}}^p(x;\zeta_{\cal H})\,, \\
{\mathpzc d}_{V_{D_1}}^p(x;\zeta_{\cal H}) & = {\mathpzc q}_{V_{D_1}}^p(x;\zeta_{\cal H})\,.
\end{align}
\end{subequations}

The following identities are readily established:
\begin{equation}
\label{Equsexchange}
{\mathpzc u}_{V_{Q_j}}^p(x;\zeta_{\cal H})  = {\mathpzc s}_{j^+}^p(1-x;\zeta_{\cal H})\,,
\; j=0,1\,.
%
\end{equation}
As shown elsewhere \cite{Chang:2022jri}, these identities and canonical normalisation of the proton Faddeev amplitude guarantee conservation of baryon number and momentum:
\begin{subequations}
\label{SumRules}
\begin{align}
\int_0^1 dx\, {\mathpzc u}_V^p(x;\zeta_{\cal H})  & = 2\,,\;
\int_0^1 dx\, {\mathpzc d}_V^p(x;\zeta_{\cal H})   = 1\,, \label{baryonnumber} \\
\langle x \rangle_{{\mathpzc u}_p}^{\zeta_{\cal H}}+
\langle x \rangle_{{\mathpzc d}_p}^{\zeta_{\cal H}} & :=
\int_0^1 dx\, x [{\mathpzc u}_V^p(x;\zeta_{\cal H}) +{\mathpzc d}_V^p(x;\zeta_{\cal H})] \nonumber \\
& = 1\,. \label{momentumsumrule}
\end{align}
\end{subequations}
Whereas Eqs.\,\eqref{baryonnumber} are valid for any scale $\zeta$, Eq.\,\eqref{momentumsumrule} is definitive of the hadron scale: on $\zeta >\zeta_{\cal H}$, evolution shifts momentum into sea and glue DFs.

\subsection{Helicity dependent}
\label{SecPDF}
Generalisation of the formulae in Sec.\,\ref{SecUDF} to the case of hadron-scale helicity-dependent DFs is straightforward.   Since a $0^+$ diquark cannot be polarised, there is no analogue of Term 3.  On the other hand, there is a new contribution associated with axial current induced $0^+ \leftrightarrow 1^+$ transitions.  Thus,
\begin{equation}
\Delta{\mathpzc u}_V^p(x;\zeta_{\cal H}) = \sum_{t=Q_0,Q_1,D_1,D_{01}} \Delta{\mathpzc u}_{V_t}^p(x;\zeta_{\cal H})\,,
\label{EqDuproton}
\end{equation}
where Term~$1_\Delta$ is
{\allowdisplaybreaks
\begin{align}
\Lambda_+  \gamma_5\gamma\cdot n  \Delta{\mathpzc u}_{V_{Q_0}}^p&(x;\zeta_{\cal H}) \Lambda_+   \nonumber\\
& = \int\! \tfrac{d^4\ell}{(2\pi)^4} \,
\hat\delta_n^{xK} \;
\Lambda_+ \Delta Q_0  \Lambda_+\,, \label{Duproton1}
\end{align}
with
\begin{align}
\Delta & Q_0 = a_{0^+}^2 \,
\bar\psi^{0^+}(\ell-\tfrac{1}{2}K;-K) S(\ell)  \nonumber \\
& \quad \times {\mathpzc A}_0 \gamma_5 \gamma\cdot n S(\ell) \psi^{0^+}(\ell-\tfrac{1}{2}K;K) \Delta^{0^+}(\ell-K)\,.
\label{EqDQ0}
\end{align}
Here, ${\mathpzc A}_0$ is the dressed-quark axial charge, obtained from the forward limit of the quark-axialvector vertex \cite{Chang:2012cc}.
}

Term $2_\Delta$, \emph{i.e}., probing a $u$ quark in the presence of an axialvector $\{ud\}$ spectator:
\begin{align}
\Lambda_+ \gamma_5 \gamma\cdot n  \Delta {\mathpzc u}_{V_{Q_1}}^p& (x;\zeta_{\cal H}) \Lambda_+   \nonumber  \\
& = \int\! \tfrac{d^4\ell}{(2\pi)^4} \,
\hat\delta_n^{xK} \;
\Lambda_+ \Delta Q_1  \Lambda_+\,, \label{uproton11D}
\end{align}
with
\begin{align}
\Delta & Q_1 = a_{1_{\{ud\}}^+}^2 \,
\bar\psi_\rho^{1_{\{ud\}}^+}(\ell-\tfrac{1}{2}K;-K) S(\ell)  \nonumber \\
& \times  {\mathpzc A}_0 \gamma_5 \gamma\cdot n S(\ell) \psi_\sigma^{1_{\{ud\}}^+}(\ell-\tfrac{1}{2}K;K) \Delta_{\rho\sigma}^{1^+}(\ell-K)\,.
\label{EqDQ1}
\end{align}

Term $3_\Delta$ reveals the $u$ quarks in the axialvector diquarks, $\{uu\}$, $\{ud\}$:
\begin{subequations}
\label{convolutionD1}
\begin{align}
\Delta{\mathpzc u}_{V_{D_1}}^p(x;\zeta_{\cal H}) &= 5 \Delta{\mathpzc q}_{V_{D_1}}^p(x;\zeta_{\cal H})\,, \\
\Delta{\mathpzc q}_{V_{D_1}}^p(x;\zeta_{\cal H})& =  \int_x^1\,\frac{dy}{y}\,\Delta{\mathpzc s}_{1^+}^p(y;\zeta_{\cal H})
{\mathpzc u}_V^{1^+}(x/y;\zeta_{\cal H})\,,
\end{align}
\end{subequations}
where, following Ref.\,\cite{Cheng:2022jxe}, the proton light-front axialvector diquark helicity fraction number density is
\begin{subequations}
\label{uprotonD3}
\begin{align}
 \Lambda_+ & \gamma_5\gamma\cdot n  \Delta{\mathpzc s}_{1^+}^p(x;\zeta_{\cal H}) \Lambda_+
= \int\! \tfrac{d^4\ell}{(2\pi)^4} \, \hat\delta_n^{xK}\;
\bar \Lambda_+ \Delta D_{1}  \bar \Lambda_+\,,\\
\Delta & D_{1}  =   a_{1_{\{ud\}}^+}^2
\bar\psi_\rho^{1_{\{ud\}}^+}(\ell-\tfrac{1}{2}K;K)S(\ell-K) \Delta_{\rho\alpha}^{1^+}(\ell) \nonumber \\
&
 \times n_\mu \Gamma_{5\mu;\alpha\beta}^{AA}(\ell,\ell)\Delta_{\beta\sigma }^{1^+}(\ell)
\psi_\sigma^{1_{\{ud\}}^+}(\ell-\tfrac{1}{2}K;-K)\,,
\end{align}
\end{subequations}
with $\Gamma_{5\mu;\alpha\beta}^{AA}$ given elsewhere \cite[Eq.\,(A37b)]{Cheng:2022jxe}.

The final contribution, Term $4_\Delta$, is associated with the $u$ quark exposed in the probe-induced scalar-axial\-vector diquark transition:
\begin{align}
\Delta{\mathpzc u}^p_{V_{D_{01}}}(x;\zeta_{\cal H}) = \int_x^1 \frac{dy}{y}
\Delta{\mathpzc s}_{0^+1^+}^p(y;\zeta_{\cal H})
{\mathpzc u}_V^{0^+1^+}(x/y;\zeta_{\cal H})\,, \label{Eq01TDF}
\end{align}
where ${\mathpzc u}_V^{0^+1^+}(x;\zeta_{\cal H})$ is the valence $u$-quark DF in the scalar-axialvector diquark transition and
\begin{subequations}
\label{uprotonD4}
\begin{align}
 \Lambda_+ & \gamma_5\gamma\cdot n  \Delta{\mathpzc s}_{0^+1^+}^p(x;\zeta_{\cal H}) \Lambda_+ \nonumber \\
& = \int\! \tfrac{d^4\ell}{(2\pi)^4} \, \hat\delta_n^{xK}\;
\bar \Lambda_+ \Delta D_{01}  \bar \Lambda_+\,,\\
\Delta D_{01} & =   a_{0^+} a_{1_{\{ud\}}^+}
\left[
\bar\psi_\rho^{1_{\{ud\}}^+}(\ell-\tfrac{1}{2}K;K) S(\ell-K) \Delta_{\rho\alpha}^{1^+}(\ell)
\right . \nonumber \\
& \qquad
 \times n_\mu \Gamma_{5\mu;\alpha}^{SA}(\ell,\ell)\Delta^{0^+}(\ell)
\psi^{0^+}(\ell-\tfrac{1}{2}K;-K)  \nonumber \\
& \quad + \bar\psi^{0^+}(\ell-\tfrac{1}{2}K;K) S(\ell-K) \Delta^{0^+}(\ell) n_\mu \Gamma_{5\mu;\alpha}^{SA}(\ell,\ell) \nonumber \\
& \qquad \times \left.  \Delta_{\alpha\rho}^{1^+}(\ell)
\bar\psi_\rho^{1_{\{ud\}}^+}(\ell-\tfrac{1}{2}K;-K) \right]\,,
\end{align}
\end{subequations}
with $\Gamma_{5\mu;\alpha}^{SA}$ given elsewhere \cite[Eqs.\,(A35), (A39b)]{Cheng:2022jxe}.

Similarly, the helicity-dependent proton valence $d$-quark DF is:
\begin{equation}
\Delta{\mathpzc d}_V^p(x;\zeta_{\cal H}) = \sum_{t=Q_1,D_1,D_{01}} \Delta{\mathpzc d}_{V_t}^p(x;\zeta_{\cal H})\,, \label{dquarkPDF}
\end{equation}
where, owing to isospin symmetry,
\begin{subequations}
\label{dpPDFs}
\begin{align}
\Delta{\mathpzc d}_{V_{Q_1}}^p(x;\zeta_{\cal H}) & = 2 \Delta{\mathpzc u}_{V_{Q_1}}^p(x;\zeta_{\cal H}) \,, \\
\Delta{\mathpzc d}_{V_{D_1}}^p(x;\zeta_{\cal H}) & =  \Delta{\mathpzc q}_{V_{D_1}}^p(x;\zeta_{\cal H}) \,, \\
\Delta{\mathpzc d}_{V_{D_{01}}}^p(x;\zeta_{\cal H}) & = -\Delta{\mathpzc u}_{V_{D_{01}}}^p(x;\zeta_{\cal H})\,.
\end{align}
\end{subequations}

\section{Distribution Functions: Results at the Hadron Scale}
\label{SeczH}
Using the framework described in Sec.\,\ref{SecUDF}, with the DF formulae evaluated as summarised in \ref{AppendixSM} and \ref{AppendixDFs}, one arrives at numerical results for the hadron-scale proton unpolarised and polarised DFs that can reliably be interpolated using the following function:
\begin{align}
\label{InterpolateDFs}
{\mathpzc f}(x;\zeta_{\cal H}) & = {\mathpzc n}_{\mathpzc f} \, 6 (1-x) \sum_{n=0}^{10}a_n^{\mathpzc f} C_n^{3/2}(1-2x)\,,
\end{align}
where $a_0^{\mathpzc f}=1$ and the remaining coefficients are listed in Table~\ref{Icoeffs}.  One readily finds that Eqs.\,\eqref{SumRules} are satisfied.
Moreover, upon inspection, it is apparent that all SCI proton valence-quark DFs have the same large-$x$ behaviour:
\begin{equation}
\label{SCIpower}
\mbox{SCI:} \quad {\mathpzc f}(x;\zeta_{\cal H}) \stackrel{x\simeq 1}{\propto} (1-x)^1 .
\end{equation}
As in QCD, this power is one greater than that on the SCI valence-quark DFs for pseudoscalar mesons \cite{Brodsky:1994kg, Yuan:2003fs, Cui:2021mom, Cui:2022bxn, Lu:2022cjx}:
\begin{subequations}
\label{LargeX}
\begin{align}
\mbox{QCD:} \quad {\mathpzc f}^p(x;\zeta_{\cal H}) & \stackrel{x\simeq 1}{\propto} (1-x)^3\,,\\
\mbox{QCD:} \quad {\mathpzc f}^\pi(x;\zeta_{\cal H}) & \stackrel{x\simeq 1}{\propto} (1-x)^2\,.
\end{align}
\end{subequations}

Turning to the other endpoint,
\begin{equation}
\mbox{SCI:} \quad {\mathpzc f}(x;\zeta_{\cal H}) \stackrel{x\simeq 0}{\propto} x^0\,, {\rm \emph{i.e}.,~nonzero~constant.}
\end{equation}
That the $\zeta_{\cal H}$ unpolarised and polarised valence-quark DFs each separately exhibit the same power law behaviour is consistent with QCD-based expectations. \linebreak
However, in large-$\zeta$ DF phenomenology, it is typically argued that there to is no correlation between the helicity of the struck quark and that of the parent proton, in which case the polarised:unpolarised ratio of DFs should vanish as $x \to 0$.  Using the SCI, this is not the case at $\zeta_{\cal H}$: $\Delta {\mathpzc u}/{\mathpzc u}|_{x\simeq 0} = 0.34$; $\Delta {\mathpzc d}/{\mathpzc d}|_{x\simeq 0} = -0.31$.  Moreover, as described in connection with Eqs.\,\eqref{DFratios} below, this ratio is invariant under evolution; so, it pointwise maintains the same nonzero values $\forall \zeta$.


\subsection{Helicity independent at $\zeta_{\cal H}$}
\begin{figure}[t]
\centerline{\includegraphics[width=0.44\textwidth]{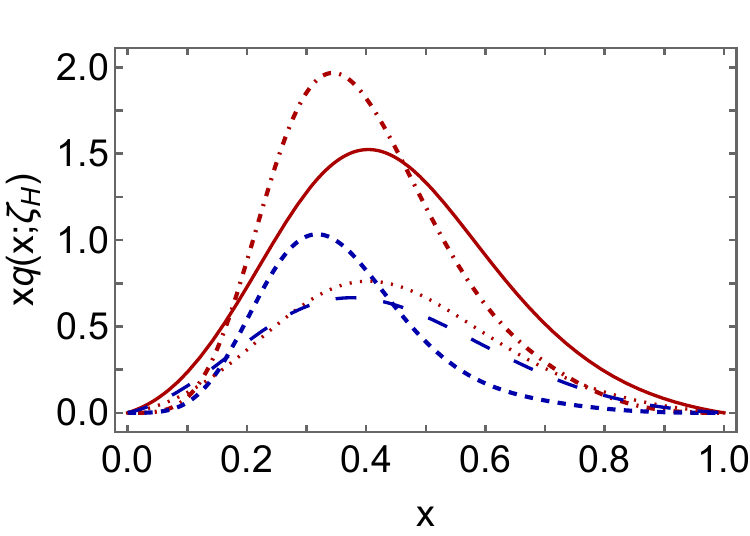}}
\caption{\label{FigUnPolarisedzH}
Hadron scale helicity-independent valence quark distributions:
solid red curve -- $\mathpzc u$ quark (SCI);
dotted red curve -- $0.5{\mathpzc u}^p(x;\zeta_{\cal H}) $ (SCI);
and long-dashed blue curve -- $\mathpzc d$ quark (SCI).
For comparison, the QCD-kindred from Ref.\,\cite[Fig.\,1A]{Chang:2022jri} are also displayed: dot-dashed red curve -- $\mathpzc u$ quark; and short-dashed blue curve -- $\mathpzc d$ quark.
}
\end{figure}

The SCI predictions for the proton hadron-scale unpolarised valence-quark DFs are drawn in Fig.\,\ref{FigUnPolarisedzH}.  Compared with the QCD-kindred predictions \cite{Chang:2022jri}, they are dilated and possess markedly different endpoint beha\-viour.  Notwithstanding that, the momentum fraction lodged with each valence quark species is the same:
\begin{subequations}
\label{MomFracCompare}
\begin{align}
\langle x \rangle_{{\mathpzc u}_p}^{\zeta_{\cal H}} = 0.686_{\rm SCI} & \quad {\it cf.}\quad
 0.687_{\mbox{\rm\footnotesize \cite{Chang:2022jri}}} \,,\\
\langle x \rangle_{{\mathpzc d}_p}^{\zeta_{\cal H}} = 0.314_{\rm SCI} & \quad {\it cf.}\quad
 0.313_{\mbox{\rm\footnotesize \cite{Chang:2022jri}}}\,.
\end{align}
\end{subequations}

Figure\,\ref{FigUnPolarisedzH} also displays $0.5{\mathpzc u}^p(x;\zeta_{\cal H}) $.  Evidently, revealing an impact of diquark correlations in the proton wave function, this curve is different from ${\mathpzc d}^p(x;\zeta_{\cal H}) $.  Similar observations pertain to the QCD-kindred results.  We subsequently return to this point in connection with the data in Ref.\,\cite[MARATHON]{Abrams:2021xum}.

\subsection{Helicity dependent at $\zeta_{\cal H}$}
\begin{figure}[t]
\centerline{\includegraphics[width=0.44\textwidth]{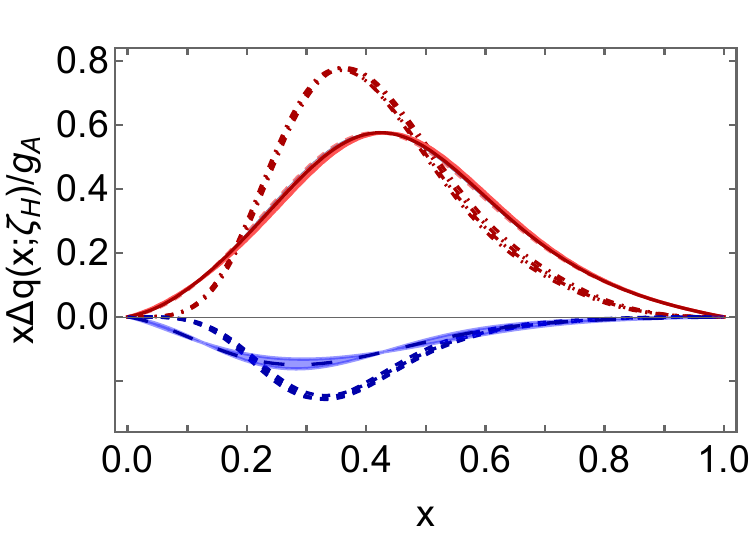}}
\caption{\label{FigPolarisedzH}
Hadron-scale helicity-dependent valence quark DFs, each normalised by the respective, computed value of the proton axial charge: solid red curve -- $\mathpzc u$ quark (SCI); and long-dashed blue curve -- $\mathpzc d$ quark (SCI).  The band surrounding each curve displays the sensitivity of the result to the choice of $0^+\leftrightarrow 1^+$ transition DF, Eqs.\,\eqref{TransitionDFModels}.
For comparison, the bands of QCD-kindred predictions from Ref.\,\cite[Fig.\,1]{Cheng:2023kmt} are also drawn:
dot-dashed red curve -- $\mathpzc u$ quark; and dashed blue curve -- $\mathpzc d$ quark.
}
\end{figure}

Figure \ref{FigPolarisedzH} displays the SCI predictions for the proton hadron-scale polarised valence-quark DFs.  Evidently, answering a question that might be asked in connection with Term $4_\Delta$, Eq.\,\eqref{Eq01TDF}, there is negligible sensitivity to the choice of $0^+\leftrightarrow 1^+$ transition DF: Eqs.\,\eqref{TransitionDFCentral}, \eqref{TransitionDFModels} yield practically equivalent results.
Once again, compared with QCD-kindred predictions \cite{Cheng:2023kmt}, the SCI results are dilated and possess markedly different endpoint behaviour.  Notwithstanding that, the displayed results yield
\begin{subequations}
\label{gACompare}
\begin{align}
\langle  \Delta {\mathpzc u}_p \rangle^{\zeta_{\cal H}}/g_A = \phantom{-}0.747_{\rm SCI} & \quad {\emph cf.}\quad
 \;\;\phantom{-}0.762_{\mbox{\rm\footnotesize \cite{Cheng:2023kmt}}} \,, \\
\langle  \Delta {\mathpzc d}_p \rangle^{\zeta_{\cal H}}/g_A  =
-0.253_{\rm SCI} & \quad {\emph cf.}\quad
 -0.238(6)_{\mbox{\rm\footnotesize \cite{Cheng:2023kmt}}}\,,
\end{align}
\end{subequations}
with $g_A = 0.92$  (SCI) \emph{cf}.\ $g_A = 1.25$ \cite{Cheng:2023kmt}.
We employ a static approximation, Eq.\,\eqref{static}, in treating the proton Faddeev equation and axialvector current.
Consequently, contributions associated with Ref.\,\cite[Fig.\,2-Diagram~4]{Cheng:2022jxe} are omitted; so, $g_A$ is underestimated.
Complete insensitivity of the SCI zeroth moments in Eq.\,\eqref{gACompare} to the form of the $0^+\leftrightarrow 1^+$ transition DF is guaranteed by Eq.\,\eqref{Eq01TDF}.
Equations\,\eqref{gACompare} mean that the $u$ quark contributes a positive fraction of the proton spin whereas the $d$ quark fraction is negative and one-third the size.

\subsection{Behaviour at very high $x$}
\label{SubSecLargeX}
Here, given that the $x=1$ value of any ratio of valence quark DFs is scale-independent \cite{Holt:2010vj, Cui:2020dlm}, we provide an update of Ref.\,\cite[Table~1]{Roberts:2013mja} -- see Table~\ref{x1UpdateN} herein.
Evidently, the SCI predictions are broadly consistent with results obtained using realistic DFs \cite{Chang:2022jri, Cheng:2023kmt} and the ``Faddeev'' column estimates.  The latter are obtained by using the simple formulae introduced in Ref.\,\cite{Roberts:2013mja} for use in analysing nucleon Faddeev wave functions to obtain $x\to 1$ values of DF ratios without the need for calculating the $x$-dependence of any DF.
On the other hand, the SCI predictions argue against both the scalar diquark only and SU$(4)$-symmetric picture of the nucleon wave function.

\begin{table}[t]
\begin{center}
\begin{tabular*}
{\hsize}
{
l@{\extracolsep{0ptplus1fil}}
c@{\extracolsep{0ptplus1fil}}
c@{\extracolsep{0ptplus1fil}}
c@{\extracolsep{0ptplus1fil}}
c@{\extracolsep{0ptplus1fil}}
c@{\extracolsep{0ptplus1fil}}
c@{\extracolsep{0ptplus1fil}}}\hline
 & herein & realistic & Faddeev & SC only & SU$(4)$ & pQCD \\\hline
$\rule{0ex}{3ex}\frac{F_2^n}{F_2^p}$ & $\phantom{-}0.56$ & $\phantom{-}0.45(5)$ & $\phantom{-}0.49$ & $\frac{1}{4}$ & $\phantom{-}\frac{2}{3}$ &$\frac{3}{7}$   \\[1ex]
$\frac{d}{u}$ & $\phantom{-}0.36$ & $\phantom{-}0.23(6)$ & $\phantom{-}0.28$ & 0 & $\phantom{-}\frac{1}{2}$ & $\frac{1}{5}$ \\[1ex]
$\frac{\Delta d}{\Delta u}$ & $-0.12$ & $-0.14(3)$ & $-0.11$ & 0 & $-\frac{1}{4}$ & $\frac{1}{5}$ \\[1ex]
$\frac{\Delta u}{u}$ & $\phantom{-}0.50$  &  $\phantom{-}0.63(8)$ & $\phantom{-}0.65$ & 1 & $\phantom{-}\frac{2}{3}$ & 1 \\[1ex]
$\frac{\Delta d}{d}$ & $-0.16$  & $-0.38(7)$ & $-0.26$ & 0 & $-\frac{1}{3}$ & 1 \\[1ex]
$A_1^n$ & $\phantom{-}0.11$ & $\phantom{-}0.15(5)$ & $\phantom{-}0.17$ & 1 & $\phantom{-}0$ &  1 \\[1ex]
$A_1^p$ & $\phantom{-}0.45$  & $\phantom{-}0.58(8)$ & $\phantom{-}0.59$ & 1 & $\phantom{-}\frac{5}{9}$ & 1 \\\hline
\end{tabular*}
\caption{\label{x1UpdateN}
Predictions for $x=1$ value of the indicated quantities.
Column ``realistic'' collects predictions from Refs.\,\cite{Chang:2022jri, Cheng:2023kmt}.
``Faddeev'' reproduces the DSE-1 results in Ref.\,\cite{Roberts:2013mja}, obtained using simple formulae, expressed in terms of diquark appearance and mixing probabilities.
The next two columns are, respectively, results drawn from Ref.\,\protect\cite{Close:1988br} -- proton modelled as being built using an elementary scalar diquark (no axialvector); and Ref.\,\cite{Hughes:1999wr} -- proton described by a SU$(4)$ spin-flavour wave function.
The last column, ``pQCD,'' lists predictions made in Refs.\,\protect\cite{Farrar:1975yb, Brodsky:1994kg}, which assume an SU$(4)$ spin-flavour wave function for the proton's valence-quarks and that a hard photon may interact only with a quark that possesses the same helicity as the target.  ($3/7 \approx 0.43$.)
}
\end{center}
\end{table}

Regarding the ``pQCD'' column, the conclusions are mixed.  SCI predictions for the ratio of unpolarised DFs are loosely consistent with the listed pQCD results \cite{Farrar:1975yb, Brodsky:1994kg}.  However, there are significant disagreements in the ratios involving polarised DFs.  The discrepancies are tied to the question of helicity retention in hard scattering processes \cite{Farrar:1975yb, Brodsky:1994kg}.
If that notion is correct, then $\Delta {\mathpzc d}/{\mathpzc d} = 1=\Delta {\mathpzc u}/{\mathpzc u}$ on $x\simeq 1$, as evident in Table~\ref{x1UpdateN}-column~6.
Such an outcome requires a zero in $\Delta {\mathpzc d}^p(x;\zeta_{\cal H})$, but this is not seen in the SCI prediction for $\Delta{\mathpzc d}^p(x;\zeta_{\cal H})$, Fig.\,\ref{FigPolarisedzH}.
Moreover, extant precision data indicate that if such a zero exists, then it must lie on $x\gtrsim 0.6$ \cite[HERMES]{HERMES:2004zsh}, \cite[COMPASS]{COMPASS:2010hwr},
\cite[CLAS EG1]{CLAS:2006ozz, CLAS:2008xos, CLAS:2015otq, CLAS:2017qga},
\cite[E06-014]{JeffersonLabHallA:2014mam},
\cite[E99-117]{JeffersonLabHallA:2003joy, JeffersonLabHallA:2004tea}.
Plainly, in deciding this issue, data relating to the polarised valence quark distributions on $x \gtrsim 0.6$ are desirable.  They exist \cite[CLAS RGC]{E1206109}, \cite[E12-06-110]{Zheng:2006} and completed analyses can reasonably be expected within a few years.

It is also worth including additional remarks on the $d/u$ (equivalently $F_2^n/F_2^p$) ratio in Table~\ref{x1UpdateN}.
Within mutual uncertainties, columns ``realistic'', ``Faddeev'' and ``pQCD'' agree.
The first two derive from similarly based calculations of the proton's Poincar\'e-covariant wave \linebreak function that include scalar and axialvector diquarks with dynamically determined relative strengths.
These wave functions correspond to a structured leading-twist had\-ron-scale proton DA \cite{Bali:2015ykx, Mezrag:2017znp}.
As the resolving scale, $\zeta$, is increased, however, it is anticipated that all such structure is eliminated as the DA approaches its asymptotic form \cite{Lepage:1980fj}; so, the wave function exhibits SU$(4)$ spin-flavour symmetry on $m_p/\zeta \simeq 0$ \cite{Lepage:1980fj}.
Supposing that outcome, then since $d/u(x=1)$ is invariant under evolution, the ``pQCD'' prediction may be interpreted as a constraint on the relative strength of scalar and axialvector diquark contributions to the canonical normalisation within a given computational framework.
The wave functions behind the ``realistic'' and ``Faddeev'' column results possess a value for this ratio ($\approx 0.6$) which provides an explanation \cite{Chang:2022jri, Lu:2022cjx} of modern data on $F_2^n(x)/F_2^p(x)$ \cite[MARATHON]{Abrams:2021xum} and its extrapolation \cite{Cui:2021gzg}: on $x\simeq 1$, $F_2^n/F_2^p = 0.437(85)$.

In this context, the SCI prediction for $F_2^n/F_2^p$ appears somewhat high, despite the fact that the axialvector:scalar diquark ratio in the wave function normalisation is low, \emph{viz}.\ $\approx 0.3$.  The enhanced $x=1$ value of $F_2^n/F_2^p$  must therefore be attributable to other features of the SCI framework, which distinguish it from approaches that use realistic momentum dependent interactions.  In fact, the SCI prediction is materially influenced by the somewhat unequal treatment of $0^+$, $1^+$ diquark correlation amplitudes: the SCI supports an axialvector component of the scalar diquark, $F_{0^+}\neq 0$, but not a pseudotensor part of the axialvector diquark \cite{Roberts:2011wy}.  For instance, if one sets $F_{0^+}=0$, then $F_2^n/F_2^p = 0.51$, in line with results in the ``realistic'' and ``Faddeev'' columns of Table~\ref{x1UpdateN} and the extrapolation of MARATHON data \cite{Cui:2021gzg}.

\section{Distribution Functions: Evolved to Experiment Scales}
\label{EvolvedDFs}
Whilst each $x=1$ result in Table~\ref{x1UpdateN} is a fixed point under QCD evolution \cite{Holt:2010vj, Cui:2020dlm}, at any other $x\in (0,1)$, the DFs evolve as $\zeta$ is increased above $\zeta_{\cal H}$.  Thus, for comparisons with inferences made from data, the SCI hadron-scale DFs must be evolved.  Herein, that is implemented using the AO scheme explained in Ref.\,\cite{Yin:2023dbw}, which has proved efficacious in many applications -- see, \emph{e.g}., Refs.\,\cite{Cui:2020tdf, Chang:2022jri, Lu:2022cjx, Cheng:2023kmt, Xing:2023pms}.

\begin{table*}[!t]
\caption{
\label{TabMoments}
Low-order Mellin moments, $\langle x^m\rangle_{{\mathpzc p}_p}^{\zeta_3}$, of the DFs drawn in Figs.\,\ref{FigUnPolarisedz3}\,--\,\ref{ImageGlue}, measured in \%.
The upper block lists SCI predictions.
The lower block lists QCD-kindred predictions \cite{Lu:2022cjx}.
The $m=1,2,3$ moments of the proton isovector distribution, $[u-d]$, in Fig.\,\ref{FigUnPolarisedz3}B are
$16.4$\%, $4.7$\%, $1.8$\% \emph{cf}.\ Ref.\,\cite{Lu:2022cjx}:
$16.6(7)$\%, $4.5(3)$\%, $1.6(1)$\%.
}
\begin{tabular*}
{\hsize}
{
l@{\extracolsep{0ptplus1fil}}
c@{\extracolsep{0ptplus1fil}}
c@{\extracolsep{0ptplus1fil}}
c@{\extracolsep{0ptplus1fil}}
c@{\extracolsep{0ptplus1fil}}
c@{\extracolsep{0ptplus1fil}}
c@{\extracolsep{0ptplus1fil}}
c@{\extracolsep{0ptplus1fil}}}\hline
%
%
SCI & ${\mathpzc u}^p \ $ & ${\mathpzc d}^p \ $ & ${\mathpzc g}^p\ $ &
${\mathpzc S}_p^u\ $ & ${\mathpzc S}_p^{d}\ $ & ${\mathpzc S}_p^s\ $ & ${\mathpzc S}_p^c\ $ \\ \hline
$\langle x \rangle^{\zeta_3}\ $ & $30.2\phantom{(10)}\ $ & $13.9\phantom{(10)}\ $ & $43.8\phantom{(10)}\ $& $3.7\phantom{(1)}\ $ & $4.5\phantom{(1)}\ $ & $2.9\phantom{(11)}\ $& $1.4\phantom{(1)}\ $ \\
$\langle x^2\rangle^{\zeta_3}\ $ & $8.4\phantom{(1)}\ $ & $3.7\phantom{(1)}\ $ & $2.5\phantom{(1)}\ $& $\phantom{1}0.18\phantom{(1)}\ $& $\phantom{1}0.25\phantom{(1)}\ $ & $0.15\phantom{(1)}\ $ & $\phantom{1}0.06\phantom{(1)}\ $\\
$\langle x^3\rangle^{\zeta_3}\ $ & $3.1\phantom{(1)}\ $ & $1.3\phantom{(1)}\ $ & $0.44\phantom{1)}\ $& $0.03\phantom{1,}\ $ & $0.04\phantom{1,}\ $ & $0.02\phantom{(1)}\ $ & $0.010\phantom{1}\ $ \\
\hline
realistic \mbox{\protect\cite{Lu:2022cjx}}
& ${\mathpzc u}^p \ $ & ${\mathpzc d}^p \ $ & ${\mathpzc g}^p\ $ &
${\mathpzc S}_p^u\ $ & ${\mathpzc S}_p^{d}\ $ & ${\mathpzc S}_p^s\ $ & ${\mathpzc S}_p^c\ $ \\ \hline
$\langle x \rangle^{\zeta_3}\ $ & $30.4(1.3) $ & $13.8(0.6)\ $ & $42.8(1.0)\ $& $3.3(3)\ $ & $4.1(3)\ $ & $3.0(2)\phantom{11}\ $& $1.82(6)\ $ \\
$\langle x^2\rangle^{\zeta_3}\ $ & $7.7(5)\ $ & $3.2(2)\ $ & $2.2(1)\ $& $\phantom{1}0.15(1)\ $& $\phantom{1}0.21(1)\ $ & $\phantom{1}0.14(0)\phantom{11}\ $ & $\phantom{1}0.075(2)\ $\\
$\langle x^3\rangle^{\zeta_3}\ $ & $2.5(2)\ $ & $0.9(1)\ $ & $\phantom{1}0.35(2)\ $& $\phantom{11}0.019(0)\ $ & $\phantom{11}0.028(0)\ $ & $\phantom{11}0.019(0)\phantom{11}\ $ & $\phantom{1}0.010(1)\ $ \\
\hline
\end{tabular*}
\end{table*}

It is, perhaps, worth recapitulating some basic features of the AO approach.  There are two principal elements.
The first is the concept of an effective charge \cite{Grunberg:1980ja, Grunberg:1982fw, Deur:2023dzc}, \emph{viz}.\ a QCD running coupling defined by any single observable via the formula which expresses that observable to first-order in the perturbative coupling.  Such a coupling implicitly incorporates terms of arbitrarily high order in the perturbative coupling and is: consistent with the QCD renormalisation group; renormalisation scheme independent; everywhere analytic and finite; and supplies an infrared completion of any standard running coupling.

Regarding DFs, the AO scheme posits that there is a charge, $\alpha_{1\ell}(k^2)$, that, when used to integrate the leading-order perturbative DGLAP equations, defines an evolution scheme for \emph{every} DF that is all-orders exact.  This definition is uncommonly broad because it refers to an entire class of observables.  It is worth stressing that the pointwise form of $\alpha_{1\ell}(k^2)$ is largely irrelevant.  On the other hand, the process-independent strong running coupling defined and computed in Refs.\ \cite{Binosi:2016nme, Cui:2019dwv} has all the required properties.

The second key to AO evolution is a specification of the hadron scale, $\zeta_{\cal H}<m_p$, \emph{i.e}., the starting scale for evolution.  So as to eliminate all ambiguity, it is natural to associate $\zeta_{\cal H}$ with that scale at which all properties of a given hadron are carried by its valence degrees-of-freedom.  This means, \emph{e.g}., that all of a hadron's light-front momentum is carried by valence degrees-of-freedom at  $\zeta_{\cal H}$.  Furthermore, that DFs associated with gluons and sea quarks are identically zero at $\zeta_{\cal H}$.  Working with the charge discussed in Refs.\,\cite{Binosi:2016nme, Cui:2019dwv, Deur:2023dzc, Brodsky:2024zev}, the value of the hadron scale is a prediction \cite{Cui:2021mom}:
\begin{equation}
  \zeta_{\cal H} = 0.331(2)\,{\rm GeV}\,.
\end{equation}
This value is confirmed by numerical simulations of lattice-regularised QCD (lQCD) \cite{Lu:2023yna}.

When evolving singlet and glue DFs, a Pauli blocking factor is included in the gluon splitting function, as discussed elsewhere \cite[Sec.\,6]{Chang:2022jri}:
\begin{equation}
\label{gluonsplit}
P_{f \leftarrow g}(x;\zeta) \to P_{f \leftarrow g}(x) +
 \sqrt{3}  (1 - 2 x) \frac{ {\mathpzc g}_{f} }{1+(\zeta/\zeta_H-1)^2}\,,
\end{equation}
where $P_{f \leftarrow g}(x) $ is the usual one-loop gluon splitting function, ${\mathpzc g}_{s,\bar s}=0={\mathpzc g}_{c,\bar c}$, and ${\mathpzc g}_{d,\bar d}= 0.34 =  -{\mathpzc g}_{u,\bar u}=:{\mathpzc g}$ is a strength parameter.
This term conserves baryon number; but shifts momentum into $d+\bar d$ from $u+\bar u$, otherwise leaving the total sea momentum fraction unchanged.  It vanishes with increasing $\zeta$, reflecting the waning effect of valence quarks as the proton's glue and sea content increases.

In addition, we implement a flavour threshold effect, multiplying the glue$\,\to\,$quark splitting function by the following factor \cite{Lu:2022cjx}:
\begin{equation}
\label{MassDependent}
 {\cal P}_{qg}^\zeta = \tfrac{1}{2} \left(1+\tanh[ (\zeta^2 - \delta_q^2)/\zeta_{\cal H}^2] \right)\,,
\end{equation}
$\delta_{u,d}\approx 0$, $\delta_s \approx 0.1\,$GeV, $\delta_c \approx 0.9\,$GeV.
This threshold function ensures that a given quark flavour only participates in DF evolution when the resolving energy scale exceeds a value determined by the quark's mass.

\begin{figure}[t]
\vspace*{0.5ex}

\leftline{\hspace*{0.5em}{\large{\textsf{A}}}}
\vspace*{-3ex}
\includegraphics[width=0.41\textwidth]{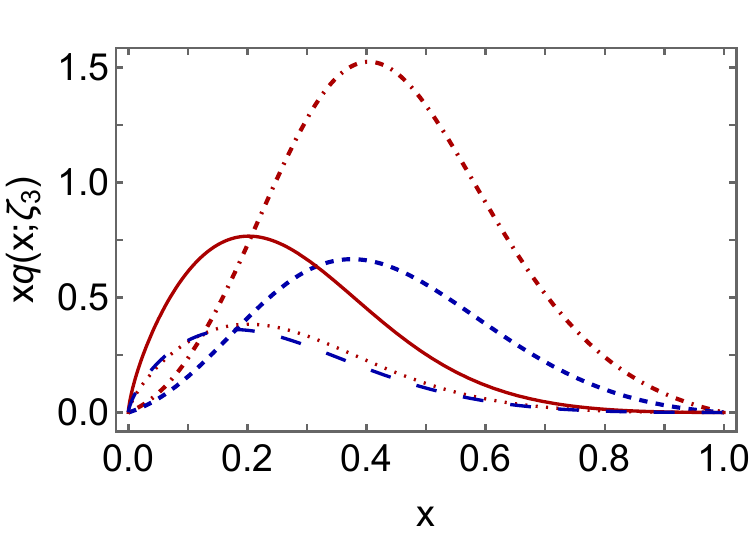}
\vspace*{-1ex}

\leftline{\hspace*{0.5em}{\large{\textsf{B}}}}
\vspace*{-1ex}
\includegraphics[width=0.41\textwidth]{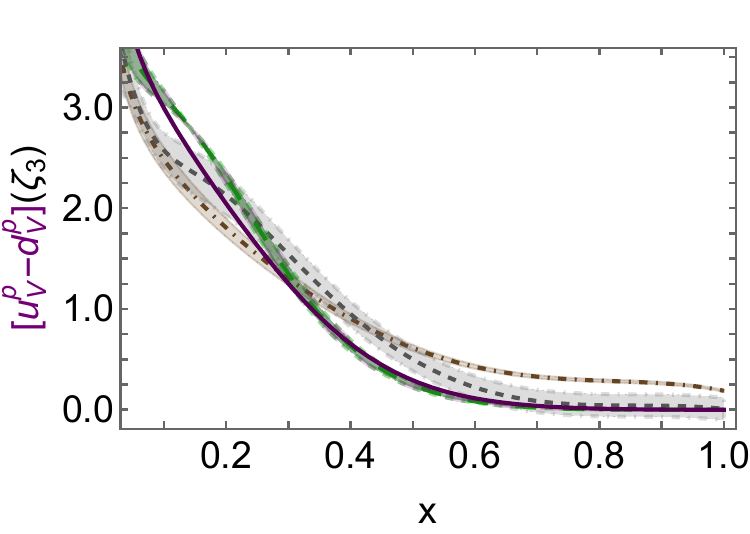}


%
\caption{\label{FigUnPolarisedz3}
{\sf Panel A}.
SCI helicity-independent valence quark DFs $\zeta_{\cal H} \to \zeta_3$:
solid red curve -- $\mathpzc u$ quark;
dotted red curve -- $0.5{\mathpzc u}^p(x;\zeta_3) $;
and long-dashed blue curve -- $\mathpzc d$ quark.
The initial $\zeta_{\cal H}$ DFs are also displayed: dot-dashed red curve -- $\mathpzc u$ quark; and dashed blue curve -- $\mathpzc d$ quark.
{\sf Panel B}.
Isovector distribution $[{\mathpzc u}^p(x;\zeta_{3}) - {\mathpzc d}^p(x;\zeta_{3})]$ (solid purple curve).
Comparisons:
realistic interaction \cite{Lu:2022cjx} (long-dashed green curve within like-coloured uncertainty band);
lQCD result from Ref.\,\cite{Lin:2020fsj} (dashed grey curve and band);
and lQCD result from \cite{Alexandrou:2021oih} (dot-dashed brown curve and band).
%
%
}
\end{figure}

\subsection{Helicity independent: $\zeta = \zeta_3$ predictions}
%
%
Low-order Mellin moments of the SCI proton helicity-independent valence quark DFs evolved $\zeta_{\cal H} \to \zeta_3=3.097\,$GeV are listed in Table~\ref{TabMoments}.  The DFs themselves are drawn in Fig.\,\ref{FigUnPolarisedz3}.
Both the table and figure show that the influence of diquark correlations in the proton wave function is still strong at this scale.  Furthermore, Table~\ref{TabMoments} reveals that good precision is required if low-order moments are to be used effectively in distinguishing between distinct DF profiles.

The calculation of individual valence quark DFs using lQCD is problematic \cite{Alexandrou:2013cda} because of difficulties in handling so-called disconnected contributions.  Consequently, lQCD results are typically only available for isovector distributions, to which disconnected diagrams do not contribute in the continuum limit.  Therefore, Fig.\,\ref{FigUnPolarisedz3}B displays the isovector distribution $[{\mathpzc u}^p(x;\zeta_{3}) - {\mathpzc d}^p(x;\zeta_{3})]$, calculated from the curves in Fig.\,\ref{FigUnPolarisedz3}A, along with lQCD results from Ref.\,\cite{Lin:2020fsj, Alexandrou:2021oih}.
The level of agree\-ment is encouraging.

On the other hand, given the evident similarity between the SCI and QCD-kindred predictions, it seems this isovector difference is not an especially discriminating probe of the underlying theory.  This is because any realistic approach to proton structure must always produce ${\mathpzc u}(x;\zeta) \propto 2 {\mathpzc d}(x;\zeta)$ on $x\gtrsim 0.4$, with a proportionality constant that is on the order of unity -- see Fig.\,\ref{ImageSeaQuest} below; and on the complementary domain, evolution-induced population of sea quark DFs eliminates any dependence on hadron-scale differences between valence-quark DFs.

\begin{figure}[!t]
\vspace*{0.5ex}

\leftline{\hspace*{0.5em}{\large{\textsf{A}}}}
\vspace*{-3ex}
\includegraphics[width=0.41\textwidth]{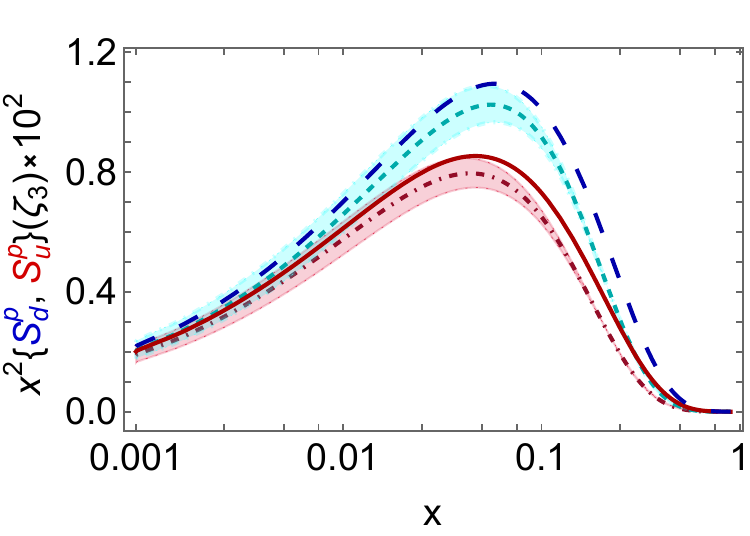}
\vspace*{-1ex}

\leftline{\hspace*{0.5em}{\large{\textsf{B}}}}
\vspace*{-3ex}
\includegraphics[width=0.41\textwidth]{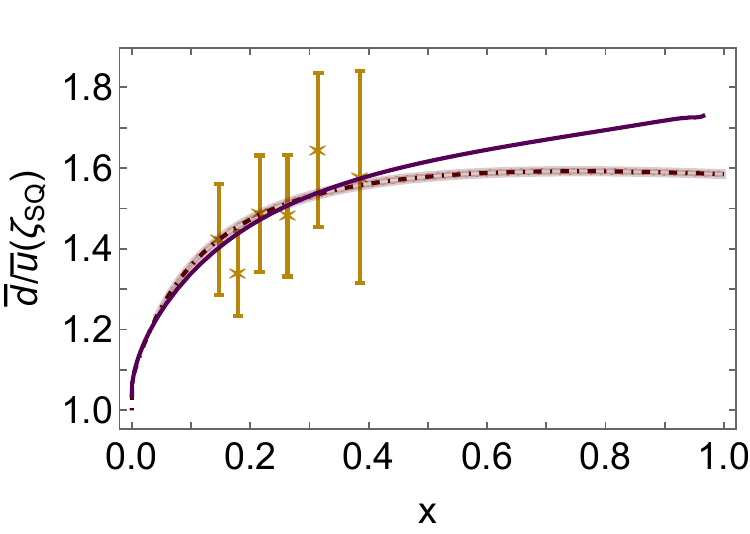}
\caption{\label{ImageSeaUD}
%
%
{\sf Panel A}.
Light quark sea DFs --
$x^2 {\mathpzc S}_u^p(x;\zeta_{3})$: SCI, solid red curve; and QCD-kindred \cite{Lu:2022cjx}, dot-dashed crimson.
$x^2 {\mathpzc S}_d^p(x;\zeta_{3})$: SCI long-dashed blue curve; and realistic interaction \cite{Lu:2022cjx}, dashed cyan.
{\sf Panel B}.
Ratio of light antiquark DFs at $\zeta_{\rm SQ}^2=30\,$GeV$^2$.
Data from Ref.\,\cite[E906]{SeaQuest:2021zxb}.
Solid purple curve: SCI result, obtained from the valence-quark DFs in Fig.\,\ref{FigUnPolarisedz3}A after evolution \cite{Yin:2023dbw} to $\zeta^2=\zeta_{\rm SQ}^2 = 30\,$GeV$^2$.
Dot-dashed maroon curve: QCD-kindred prediction \cite{Lu:2022cjx}.
}
\end{figure}

\begin{figure}[!t]
\vspace*{0.5ex}

\leftline{\hspace*{0.5em}{\large{\textsf{A}}}}
\vspace*{-3ex}
\includegraphics[width=0.41\textwidth]{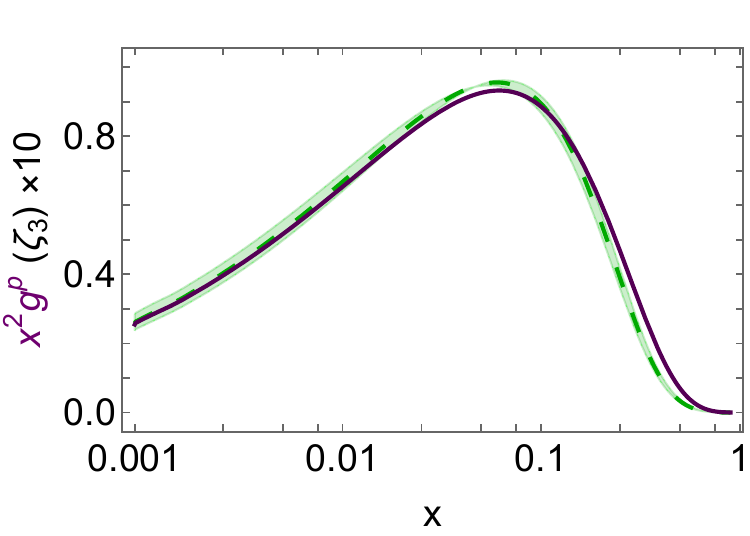}
\vspace*{-1ex}


\leftline{\hspace*{0.5em}{\large{\textsf{B}}}}
\vspace*{-3ex}
\includegraphics[width=0.41\textwidth]{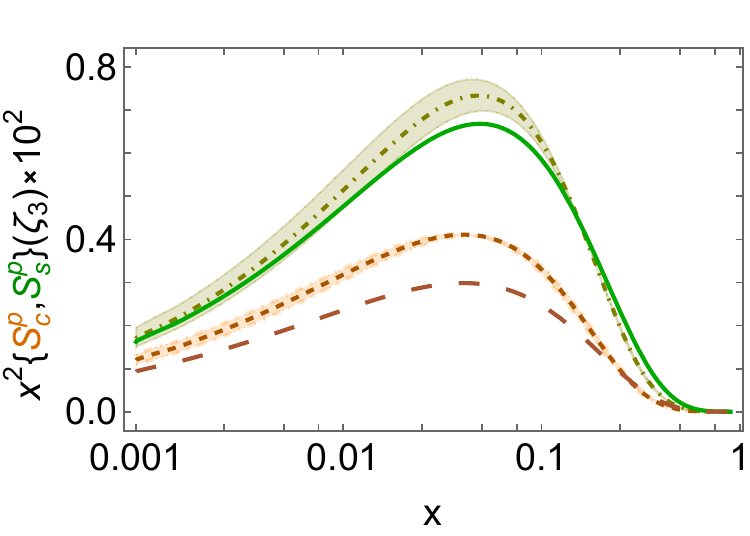}
\caption{\label{ImageGlue}
Glue and heavier sea DFs at $\zeta=\zeta_3$.
{\sf Panel A}.
Glue DFs -- $x^2 {\mathpzc g}$: SCI, solid purple curve; and realistic interaction \cite{Lu:2022cjx}, long-dashed green.  
%
%
{\sf Panel B}.
Heavier quark sea DFs --
$x^2 {\mathpzc S}_s^p(x;\zeta_{3})$: SCI solid green curve; and realistic interaction \cite{Lu:2022cjx}, dot-dashed olive curve.
$x^2 {\mathpzc S}_c^p(x;\zeta_{3})$: SCI, long-dashed burnt-orange curve; and realistic interaction \cite{Lu:2022cjx}, dashed orange.
}
\end{figure}

As in Ref.\,\cite{Lu:2022cjx}, the Pauli blocking factor, Eq.\,\eqref{gluonsplit}, generates an in-proton separation between $\bar d$ and $\bar u$ -- apparent in Fig.\,\ref{ImageSeaUD}.  Consequently, consistent with experiment \cite{NewMuon:1991hlj, NewMuon:1993oys, NA51:1994xrz, NuSea:2001idv, SeaQuest:2021zxb}, the Gottfried sum rule \cite{Gottfried:1967kk, Brock:1993sz} is violated.
Using the DFs in Fig.\,\ref{ImageSeaUD}A, one obtains
\begin{equation}
\label{gottfried}
\int_{0.004}^{0.8} dx\,[\bar {\mathpzc d}(x;\zeta_3) - \bar {\mathpzc u}(x;\zeta_3)]
=0.123
\end{equation}
for the Gottfried sum rule discrepancy on the domain covered by the measurements in Refs.\,\cite{NewMuon:1991hlj, NewMuon:1993oys}.
This value may be compared with that obtained using the QCD-kindred DFs \cite{Lu:2022cjx}, \emph{viz}.\ $0.116(12)$, and inferred from recent fits to a variety of high-precision data ($\zeta = 2\,$GeV) \cite[CT18]{Hou:2019efy}: 0.110(80).
%
After evolution to $\zeta_{\rm SQ}^2=30\,$GeV$^2$, the value in Eq.\,\eqref{gottfried} drops by 5\%, becoming $0.117$.

As shown by Fig.\,\ref{ImageSeaUD}B, the SCI result for the ratio $\bar {\mathpzc d}(x;\zeta_{\rm SQ})/\bar {\mathpzc u}(x;\zeta_{\rm SQ})$ matches modern data \cite[E906]{SeaQuest:2021zxb}.  Furthermore, it is practically indistinguishable from the QCD-kindred result \cite{Lu:2022cjx} on the domain of existing measurements.  Even beyond that domain, the difference is modest.  (On this valence-quark domain, the contrasting large-$x$ power-law behaviours has an influence: Eq.\,\eqref{SCIpower} \emph{cf}.\ Eq.\,\eqref{LargeX}.)
Such insensitivity to details of the pointwise behaviour of the hadron-scale valence quark DFs suggests that Pauli blocking, as expressed in Eq.\,\eqref{gluonsplit}, provides a viable explanation for the asymmetry of antimatter in the proton.  That Pauli blocking might be sufficient was first discussed in Ref.\,\cite{Field:1976ve}.

The SCI prediction for the $\zeta=\zeta_3$ glue DF in the proton is drawn in Fig.\,\ref{ImageGlue}A.  Compared with the QCD-kindred prediction \cite{Lu:2022cjx}, it is harder: writing the DF as $\propto (1-x)^\beta$ on $x\geq 0.9$, then $\beta \approx 3.7$ (SCI) \emph{cf}.\ $5.5$ (realistic).  As one expects from QCD evolution, each of these powers is roughly one unit larger than that of the associated valence quark DF at this scale.

SCI predictions for the light and heavier quark sea DFs are drawn in Figs.\,\ref{ImageSeaUD}A, \ref{ImageGlue}B, respectively.  Each of these DFs is harder than the kindred realistic interaction result \cite{Lu:2022cjx} and, again in accordance with evolution equation expectations, the large-$x$ behaviour of these DFs is characterised by an exponent that is one unit larger than that of the related glue DF: $\beta \approx 4.7$ (SCI) \emph{cf}.\ $6.4$ (realistic).

\begin{figure}[t]


\includegraphics[width=0.41\textwidth]{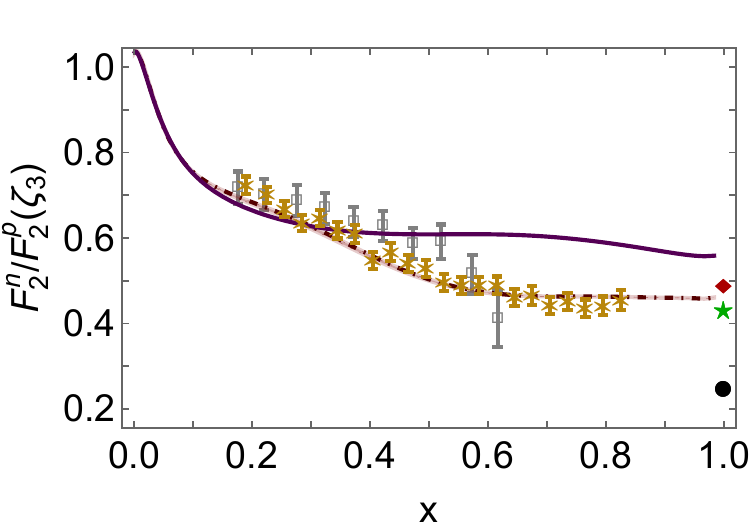}
\caption{\label{ImageSeaQuest}
%
%
%
Neutron-to-proton structure function ratio.
Data: gold asterisks \cite[MARATHON]{Abrams:2021xum}; and
open grey squares \cite[BoNuS]{CLAS:2014jvt}.
Solid purple curve, SCI prediction; and dot-dashed maroon curve, QCD-kindred result \cite{Lu:2022cjx}.
Other predictions:
green star -- helicity conservation in the QCD parton model \cite{Farrar:1975yb, Brodsky:1994kg};
red diamond -- ``Faddeev'' result from Table~\ref{x1UpdateN}\,--\,column 3;
and retaining only scalar diquarks in the proton wave function, which produces a large-$x$ value for this ratio that lies in the neighbourhood of the filled circle \cite{Close:1988br, Xu:2015kta}.
%
%
}
\end{figure}

Regarding Fig.\,\ref{ImageGlue}B, it is worth noting that the heavier sea quark DFs are commensurate in size with those of the light-quark sea.  Furthermore, in comparison with the QCD-kindred predictions, the SCI results are larger and harder on the valence quark domain and, like the glue DF, smaller in magnitude on the complementary lower-$x$ domain.

Whilst discussing heavier quark sea, it notable that both the SCI and QCD-kindred analyses predict a measurable charm quark momentum fraction in the proton at $\zeta=\zeta_3$ -- see Table~\ref{TabMoments} -- without recourse to ``intrinsic charm'' \cite{Brodsky:1980pb}.  This outcome, which is here seen to be independent of details concerning the pointwise behaviour of the proton valence quark DFs, is a challenge to the suggestions presented in Ref.\,\cite{Ball:2022qks}.  (It is also largely insensitive to the explicit form of the effective charge used to complete AO evolution.)  Notwithstanding the size of these calculated fractions, we stress that ${\mathpzc S}^{p}_c(x)$ has a sea-quark profile -- see Figs.\,\ref{ImageSeaUD}A, \ref{ImageGlue}B.

Using the SCI results for valence and sea DFs, one can calculate the neutron-proton structure function ratio:
\begin{align}
\label{F2nF2p}
\frac{F_2^n(x;\zeta)}{F_2^p(x;\zeta)} =
\frac{
{\mathpzc U}(x;\zeta) + 4 {\mathpzc D}(x;\zeta) + \Sigma(x;\zeta)}
{4{\mathpzc U}(x;\zeta) + {\mathpzc D}(x;\zeta) + \Sigma(x;\zeta)}\,,
\end{align}
where, in terms of quark and antiquark DFs,
${\mathpzc U}(x;\zeta) = {\mathpzc u}(x;\zeta)+\bar {\mathpzc u}(x;\zeta)$,
${\mathpzc D}(x;\zeta) = {\mathpzc d}(x;\zeta)+\bar {\mathpzc d}(x;\zeta)$,
$\Sigma(x;\zeta) = {\mathpzc s}(x;\zeta)+\bar {\mathpzc s}(x;\zeta)
  +{\mathpzc c}(x;\zeta)+\bar {\mathpzc c}(x;\zeta)$.
The $\zeta=\zeta_3$ SCI prediction is drawn in Fig.\,\ref{ImageSeaQuest}.  In comparison with modern data \cite[MARATHON]{Abrams:2021xum} and the QCD-kindred result \cite{Lu:2022cjx}, it matches both on the sea-dominated domain, $x\lesssim 0.25$.
On the complementary valence-quark domain, however, there are quantitative differences.  This confirms that the behaviour of this ratio on the valence-quark domain is a useful discriminator between pictures of proton structure.

\begin{figure}[!t]
\vspace*{0.5ex}

\leftline{\hspace*{0.5em}{\large{\textsf{A}}}}
\vspace*{-3ex}
\includegraphics[width=0.41\textwidth]{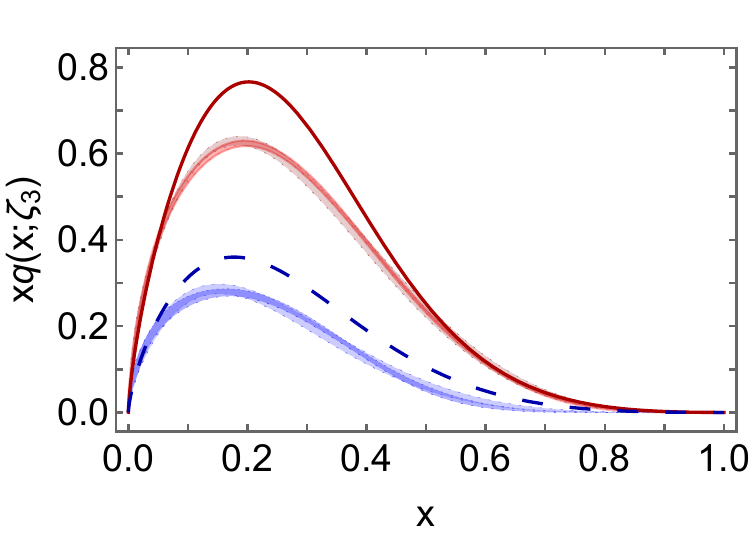}
\vspace*{-1ex}


\leftline{\hspace*{0.5em}{\large{\textsf{B}}}}
\vspace*{-3ex}
\includegraphics[width=0.41\textwidth]{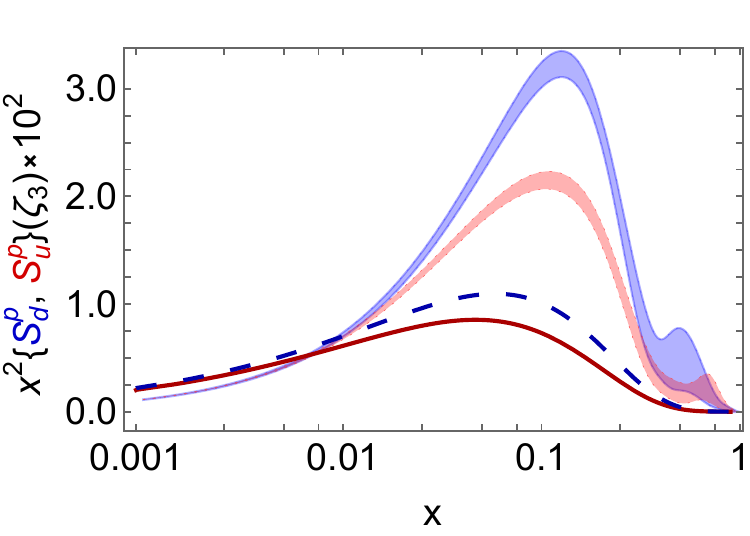}
\caption{\label{ImageFits1}
{\sf Panel A}.
SCI predictions for valence quark DFs ($u$ -- solid red curve, $d$ -- dashed blue curve) compared with DFs inferred from data as described in Ref.\,\cite[NNPDF4.0]{NNPDF:2021njg} -- like-coloured shaded bands.
{\sf Panel B}.
Similar comparison involving $u$, $d$ sea DFs.
(All DFs at $\zeta=\zeta_3$.)
}
\end{figure}

\begin{figure}[!t]
\vspace*{0.5ex}

\leftline{\hspace*{0.5em}{\large{\textsf{A}}}}
\vspace*{-3ex}
\includegraphics[width=0.41\textwidth]{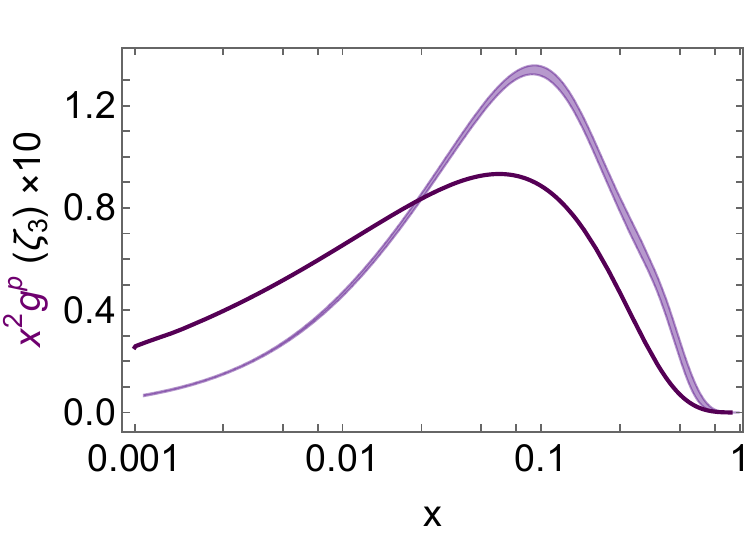}
\vspace*{-1ex}


\leftline{\hspace*{0.5em}{\large{\textsf{B}}}}
\vspace*{-3ex}
\includegraphics[width=0.41\textwidth]{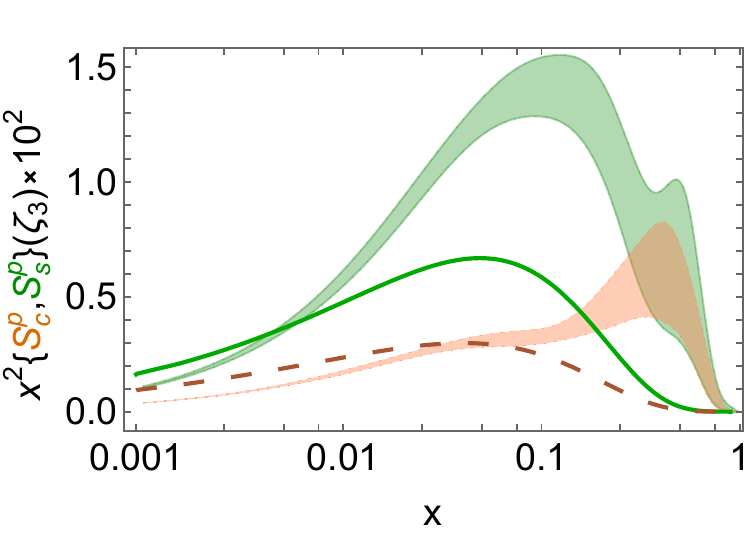}
\caption{\label{ImageFits2}
{\sf Panel A}.
SCI prediction for $x^2 {\mathpzc g}(x)$ (solid purple curve) compared with analogous DF inferred from data as described in Ref.\,\cite[NNPDF4.0]{NNPDF:2021njg} -- like-coloured shaded band.
{\sf Panel B}.
Similar comparison for heavier quark sea DFs.
SCI. $x^2 {\mathpzc S}_s^p(x;\zeta_{3})$ -- solid green curve; and
$x^2 {\mathpzc S}_c^p(x;\zeta_{3})$ -- dashed burnt-orange curve.
Like-coloured bands -- results inferred from data as described in Ref.\,\cite[NNPDF4.0]{NNPDF:2021njg}.
(All DFs at $\zeta=\zeta_3$.)
}
\end{figure}

Following these comparisons with data and other calculations, it is interesting to contrast the SCI predictions for proton DFs with those inferred via fits to available data.  As explained elsewhere \cite{Cui:2021mom}, the results of such fits are sensitive to the form used for the hard scattering kernel in the convolution integrals that provide the bridge between data and DFs.  Comparisons with inferences described in  Ref.\,\cite[NNPDF4.0]{NNPDF:2021njg} are presented in Figs.\,\ref{ImageFits1}, \ref{ImageFits2}.  Semiquantitatively similar curves are typically obtained by other collaborations.

Regarding the comparisons, there are some similarities.
For instance, in both cases, ${\mathpzc u}(x) \neq 2 {\mathpzc d}(x)$; and relative to the maximum in $x {\mathpzc d}(x)$, that in $x {\mathpzc u}(x)$ is shifted to larger $x$ by roughly 15\%.  Existing CSM analyses attribute this shift to the existence of diquark correlations within the proton.
Further, again in both cases, the $d$ quark sea DF is uniformly larger than that associated with $u$ quarks,
the $s$ quark sea is next greatest in magnitude, and the $c$ quark sea is smallest.

However, there are also marked differences.
Although the glue momentum fraction is practically identical to that predicted by the SCI -- see Table~\ref{TabMoments}, the DFs inferred from fits to data lodge a noticeably smaller light-front momentum fraction with valence $u$ and $d$ quarks: respectively, $\approx 15$\%, $25$\% less than in the SCI predictions.
This is compensated in the fits by locating additional momentum with the sea: each lighter quark flavour carries roughly 80\% more momentum and the $c$ quark sea carries up to 26\% more.
Additionally, the singlet DFs inferred from data typically possess more support than the SCI predictions on the valence quark domain $x\gtrsim 0.25$.
On this domain, too, the singlet DFs inferred from fits exhibit remarkable structure.  This is especially true of the sea DFs, which are not guaranteed to be monotonically decreasing.

In these outcomes, the inferred proton DFs exhibit similar curious features to those seen earlier in comparisons between continuum predictions for pion DFs and those inferred via fits to data \cite[Fig.\,9]{Cui:2021mom}.

Regarding the implications of these comparisons, we note that the SCI results are parameter-free predictions obtained via AO evolution from the hadron scale valence quark DFs detailed in \ref{AppendixDFs}.  As evident from Figs.\,\ref{FigUnPolarisedz3}\,--\,\ref{ImageGlue}, these evolved DFs are, in many respects, semiquantitatively compatible with predictions obtained using a more realistic starting point.  All CSM DFs are monotonically decreasing on the valence quark domain, whereupon, too, they satisfy endpoint \linebreak con\-straints appropriate to the quark+quark interaction employed and subsequent DGLAP evolution.

In contrast, the fits are affected by the scarcity of relevant data at small $x$ and on the valence quark domain.  The impact of the latter is apparent, \emph{e.g}., in both the large uncertainties and ``wiggles'' revealed by Figs.\,\ref{ImageFits1}B, \ref{ImageFits2}B.
Further, as noted above, the fitting outcomes are particularly sensitive to the form of hard scattering kernel used.  Soft gluon (next-to-leading logarithm threshold) resummation is ignored in analyses of proton DFs.  This should be included before a detailed comparison of predictions and fits is possible, especially the pointwise behaviour.  (Leading order or next-to-leading order in the coupling is far less important.)  At that point, one could also begin to ask about the relevance and/or impact on inferred DFs of QCD constraints on the endpoint behaviour of DFs -- see, \emph{e.g}., Ref.\,\cite[Table~1]{Lu:2022cjx}.

In our view, with continuum and lattice Schwinger function methods now beginning to deliver predictions for the pointwise behaviour of hadron DFs, one may anticipate a significant expansion of constructive feedback between theory and phenomenology, enabling more rapid progress in both areas.

\begin{figure}[t]
\includegraphics[width=0.41\textwidth]{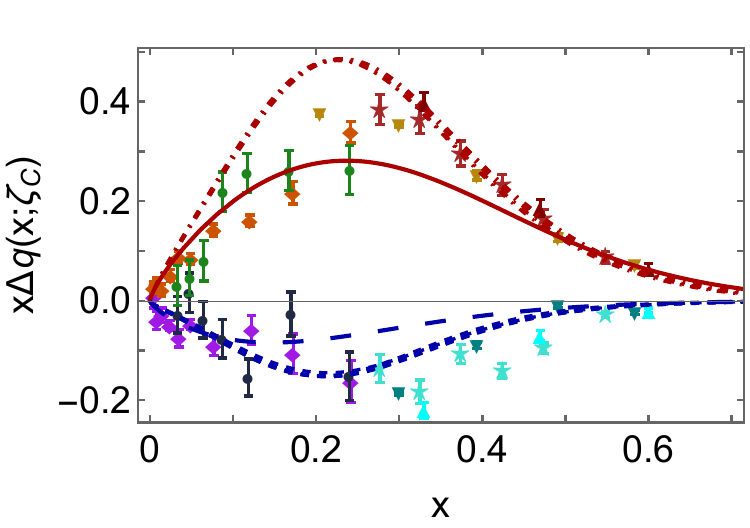}
\caption{\label{polzetaCa}
%
Polarised quark DFs.
SCI: $\Delta {\mathpzc u}(x;\zeta_{\rm C})$ -- solid red curves; and $\Delta {\mathpzc d}(x;\zeta_{\rm C})$ -- long-dashed blue curves.
Like-coloured dot-dashed and dashed curves -- QCD-kindred results for the same quantities.
Data: \cite[HERMES]{HERMES:2004zsh} -- circles;
\cite[COMPASS]{COMPASS:2010hwr} -- diamonds;
\cite[CLAS EG1]{CLAS:2006ozz, CLAS:2008xos, CLAS:2015otq, CLAS:2017qga} -- filled down-triangles;
\cite[E06-014]{JeffersonLabHallA:2014mam} -- five-pointed stars;
\cite[E99-117]{JeffersonLabHallA:2003joy, JeffersonLabHallA:2004tea} -- filled up-triangles.
}
\end{figure}

\subsection{Helicity dependent: $\zeta = \zeta_{\rm C}$ predictions}
\label{SubSecH}
When discussing proton polarised DFs, it is useful to present results at $\zeta=\zeta_{\rm C} = \surd 3\,$GeV \cite[COMPASS]{COMPASS:2010hwr}.  So, beginning with the SCI hadron-scale proton polarised valence-quark DFs in Fig.\,\ref{FigPolarisedzH} and employing the AO evolution scheme \cite{Yin:2023dbw}, one arrives at the $\zeta_{\rm C}=\surd 3\,$GeV polarised quark DFs drawn in Fig.\,\ref{polzetaCa}.

\begin{figure}[t]
\vspace*{0.5ex}

\leftline{\hspace*{0.5em}{\large{\textsf{A}}}}
\vspace*{-3ex}
\includegraphics[width=0.41\textwidth]{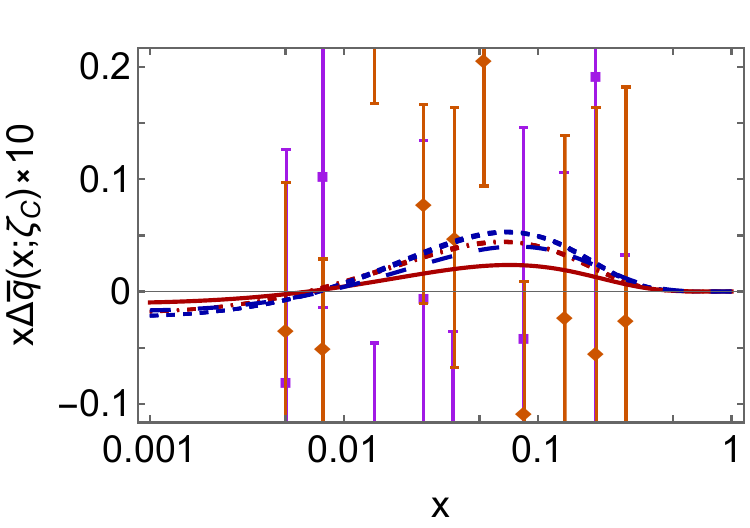}
\vspace*{-1ex}

\leftline{\hspace*{0.5em}{\large{\textsf{B}}}}
\vspace*{-3ex}
\includegraphics[width=0.41\textwidth]{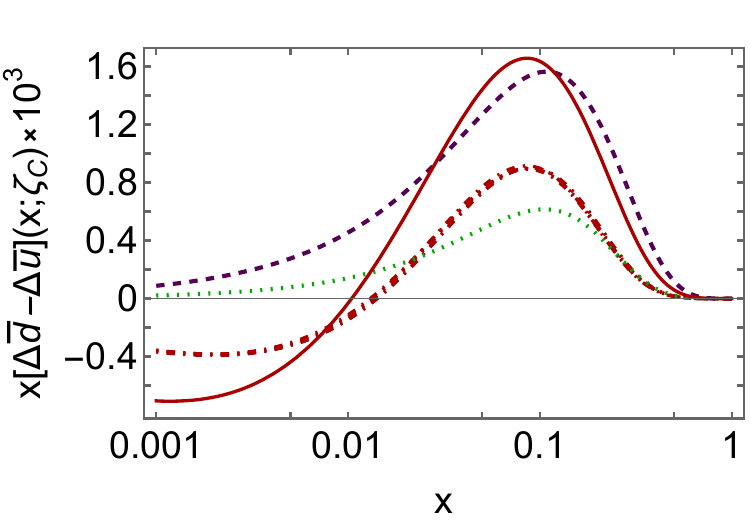}
\caption{\label{polzetaCB}
{\sf Panel A}.
Polarised antiquark DFs.  SCI: $x\Delta \bar{\mathpzc u}(x;\zeta_{\rm C})$ -- solid red curves; and $x\Delta \bar{\mathpzc d}(x;\zeta_{\rm C})$ -- long-dashed blue curves.
Like-coloured dot-dashed and dashed curves -- QCD-kindred results for the same quantities \cite{Cheng:2023kmt}.
Data \cite[COMPASS]{COMPASS:2010hwr}: $\bar{\mathpzc u}$ -- red diamonds; and $\bar{\mathpzc d}$ -- purple squares.
{\sf Panel B}.
$x[\Delta \bar{\mathpzc d}(x;\zeta_{\rm C})-\Delta \bar{\mathpzc u}(x;\zeta_{\rm C})]$: SCI -- solid red curves; and QCD-kindred -- dot-dashed red.
For comparison: $x^2[\bar{\mathpzc d}(x;\zeta_{\rm C})- \bar{\mathpzc u}(x;\zeta_{\rm C})]$: SCI -- dashed purple curved; QCD-kindred -- dotted green curve.
(This is also the comparison presented in Ref.\,\cite[Fig.\,3B]{Cheng:2023kmt} -- only the caption therein is incorrect.)
%
%
}
\end{figure}

\begin{figure}[t]
\vspace*{0.5ex}

\leftline{\hspace*{0.5em}{\large{\textsf{A}}}}
\vspace*{-3ex}
\includegraphics[width=0.41\textwidth]{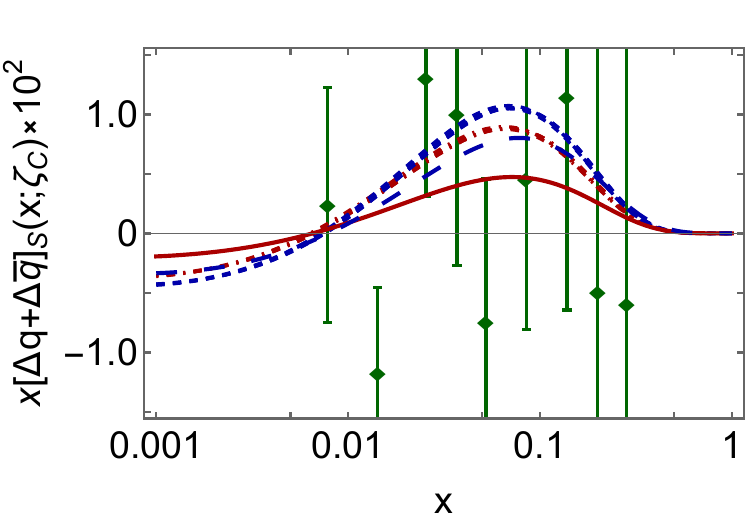}
\vspace*{-1ex}

\leftline{\hspace*{0.5em}{\large{\textsf{B}}}}
\vspace*{-3ex}
\includegraphics[width=0.41\textwidth]{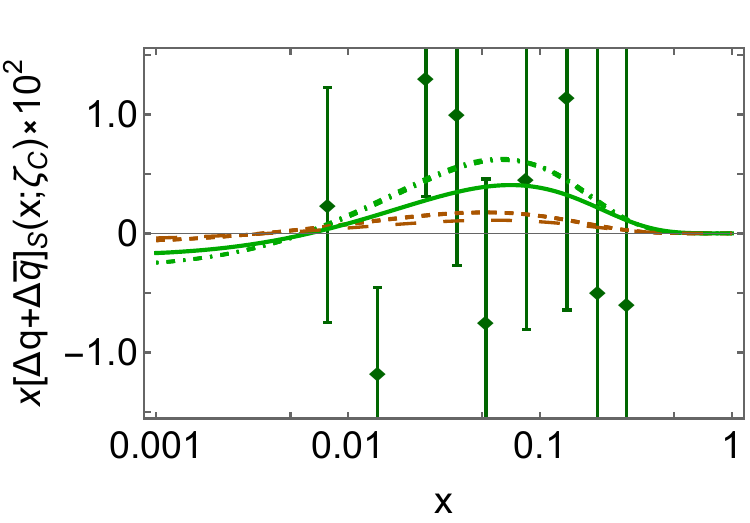}
\caption{\label{FigPolarisedSea}
Polarised sea quark distributions at $\zeta_{\rm C}$.
{\sf Panel A}.  $x[\Delta{\mathpzc u}+\Delta\bar{\mathpzc u}]_S$: SCI -- solid red curves; QCD-kindred \cite{Cheng:2023kmt} -- dot-dashed red curves.
$x[\Delta{\mathpzc d}+\Delta\bar{\mathpzc d}]_S$: SCI -- long-dashed blue curves; QCD-kindred -- dashed blue curves.
{\sf Panel B}.
$x[\Delta{\mathpzc s}+\Delta\bar{\mathpzc s}]_S$: SCI -- solid green curves; QCD-kindred -- dot-dashed green curves.
$x[\Delta{\mathpzc c}+\Delta\bar{\mathpzc c}]_S$: SCI -- long-dashed orange curves; QCD-kindred -- dashed orange curves.
Data: values of $2x\Delta {\mathpzc s}(x;\zeta_{\rm C})$ from Ref.\cite[COMPASS]{COMPASS:2010hwr}, wherein $x\Delta {\mathpzc s}(x;\zeta_{\rm C}) \approx x\Delta \bar {\mathpzc s}(x;\zeta_{\rm C})$.
}
\end{figure}

The SCI results in Fig.\,\ref{polzetaCa} are semiquantitatively in agreement with available inferences from data \cite[HERMES]{HERMES:2004zsh}, \cite[COMPASS]{COMPASS:2010hwr},
\cite[CLAS EG1]{CLAS:2006ozz, CLAS:2008xos, CLAS:2015otq, CLAS:2017qga},
\cite[E06-014]{JeffersonLabHallA:2014mam},
\cite[E99-117]{JeffersonLabHallA:2003joy, JeffersonLabHallA:2004tea}.
Regarding the COMPASS results, lying on $x\lesssim 0.2$, the collaboration's extrapolations yield $g_A^d=-0.34(5)$, $g_A^u=0.71(4)$, $g_A=g_A^u-g_A^d=1.05(6)$.
The SCI values are $g_A^d=-0.23$, $g_A^u=0.69$ and the QCD-kindred results are \cite{Cheng:2023kmt} $g_A^d=-0.30(1)$, $g_A^u=0.95(2)$.
These comparisons explain why the SCI curves match the COMPASS results for $\Delta {\mathpzc u}(x;\zeta_{\rm C})$ and are smaller in magnitude than those for $\Delta {\mathpzc d}(x;\zeta_{\rm C})$, whereas the reverse is true for the QCD-kindred predictions.
Notably, like the SCI value, the COMPASS result for $g_A$ is markedly smaller than ($0.83(5)$-times) the empirical value from neutron $\beta$-decay.
(\emph{N.B}.\, Polarised antiquark DFs are negligible at this scale -- see Figs.\,\ref{polzetaCB}, \ref{FigPolarisedSea}.)

It is worth remarking that the AO evolution kernels for unpolarised and polarised nonsinglet DFs are identical.  Hence, for $q=u,d$:
\begin{subequations}
\label{DFratios}
\begin{align}
\Delta{\mathpzc q}(x;\zeta)/\Delta{\mathpzc q}(x;\zeta_{\cal H}) & = {\mathpzc q}(x;\zeta)/{\mathpzc q}(x;\zeta_{\cal H})\,, \\
\Rightarrow \quad \Delta{\mathpzc q}(x;\zeta)/{\mathpzc q}(x;\zeta)
& = \Delta{\mathpzc q}(x;\zeta_{\cal H})/{\mathpzc q}(x;\zeta_{\cal H}) \,.
\end{align}
\end{subequations}
Consequently, the pointwise behaviour of the DF ratio polarised:unpolarised is invariant under evolution and so, therefore, are the endpoint ($x\simeq 0,1$) values.

Polarised antiquark DFs are drawn in Fig.\,\ref{polzetaCB} alongside values from Ref.\,\cite[COMPASS]{COMPASS:2010hwr}.  Evidently, the data have large uncertainties; hence, can only be used to set plausible bounds on the size of these distributions.

Figure\,\ref{polzetaCB}B depicts $x[\Delta \bar{\mathpzc d}(x;\zeta_{\rm C})-\Delta \bar{\mathpzc u}(x;\zeta_{\rm C})]$ compared with results for $x^2[\bar{\mathpzc d}(x;\zeta_{\rm C})- \bar{\mathpzc u}(x;\zeta_{\rm C})]$.  Regarding each difference, the SCI result is larger than that obtained using the QCD-kindred DFs.  Moreover, independent of the interaction, each compared difference pair has the same magnitude and their trend is similar on $x\gtrsim 0.01$.
Here, no comparison with data is reported because the uncertainties on available results \cite[HERMES]{HERMES:2004zsh}, \cite[COMPASS]{COMPASS:2010hwr} are too great for such to be meaningful.

SCI predictions for all polarised sea quark DFs are collected in Fig.\,\ref{FigPolarisedSea}.  As noted in connection with Eq.\,\eqref{MassDependent}, we have thresholds at which heavier quarks begin to play a role in evolution.  Hence, there is a separation between flavours.  Compared with QCD-kindred predictions, the SCI results are almost everywhere smaller in magnitude.
Results on $2x\Delta {\mathpzc s}(x;\zeta_{\rm C})$ inferred from data in Ref.\cite[COMPASS]{COMPASS:2010hwr} are also included in Fig.\,\ref{FigPolarisedSea}.  Owing to the large uncertainties, one can at best conclude that the magnitudes of the empirical inferences and the theory predictions are similar.

Using SCI DFs, one finds
\begin{equation}
\mbox{SCI:} \quad \int_{0.004}^{0.3} dx \,\Delta{\mathpzc s}_S(x;\zeta_{\rm C}) = 0.0048\,.
\end{equation}
This value is smaller than the QCD-kindred result: \linebreak
$\int_{0.004}^{0.3} dx \,\Delta{\mathpzc s}_S(x;\zeta_{\rm C}) = 0.0072(1)$.
Yet both are consistent with the inferred empirical value \cite[COMPASS]{COMPASS:2010hwr}: $-0.01\pm 0.01\pm 0.01$, owing to its large uncertainty.

\begin{figure}[t]
\centerline{\includegraphics[width=0.44\textwidth]{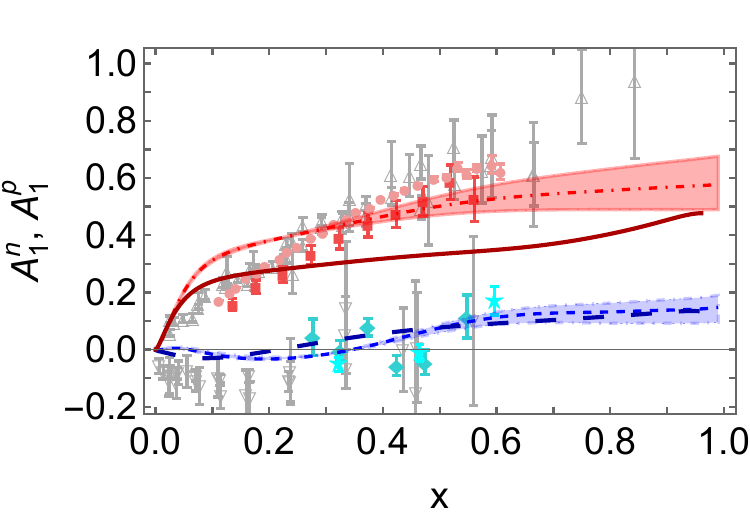}}
\caption{\label{FigA1}
SCI predictions for proton (solid red) and neutron (long-dashed blue) longitudinal spin asymmetries at $\zeta_{\rm C}$.
Like-coloured dot-dashed and short-dashed curves are QCD-kindred results for the same observables \cite{Cheng:2023kmt}.
Data.
$A_1^p$:
red squares -- Refs.\,\cite[CLAS EG1]{CLAS:2006ozz, CLAS:2008xos, CLAS:2015otq, CLAS:2017qga};
pink circles -- Ref.\,\cite{CLAS:2014qtg};
grey up-triangles -- Refs.\,\cite{E155:1999pwm, E155:2000qdr, HERMES:1998cbu, E143:1995clm, E143:1996vck, E143:1998hbs, SpinMuonSMC:1994met, SpinMuonSMC:1997mkb}.
$A_1^n$: turquoise diamonds -- Ref.\,\cite[E06-014]{JeffersonLabHallA:2014mam};
aqua five-pointed stars -- Refs.\,\cite[E99-117]{JeffersonLabHallA:2003joy, JeffersonLabHallA:2004tea};
grey down-triangles -- Refs.\,\cite{HERMES:1997hjr, E154:1997xfa, E154:1997ysl, E142:1996thl}.
}
\end{figure}

SCI predictions for nucleon longitudinal spin asymmetries -- defined, \emph{e.g}., as in Ref.\,\cite[Ch.\,4.7]{Ellis:1991qj}, are drawn in Fig.\,\ref{FigA1}.
(The charm quark contribution is practically negligible at $\zeta=\zeta_{\rm C}$.)
For context, the image also includes results inferred from data collected within the past vicennium \cite{CLAS:2006ozz, CLAS:2008xos, CLAS:2015otq, CLAS:2017qga, CLAS:2014qtg, JeffersonLabHallA:2014mam, JeffersonLabHallA:2003joy, JeffersonLabHallA:2004tea} and selected earlier results \cite{E155:1999pwm, E155:2000qdr, HERMES:1998cbu, E143:1995clm, E143:1996vck, E143:1998hbs, SpinMuonSMC:1994met, SpinMuonSMC:1997mkb, HERMES:1997hjr, E154:1997xfa, E154:1997ysl, E142:1996thl}.
Comparison with the QCD-kindred predictions \cite{Cheng:2023kmt} highlights that much improved precision is required in the extraction of these observables from data before the empirical results can unambiguously distinguish between competing pictures of proton structure.

In this connection, it is worth noting that the mismatch between the predictions and inferences at low-$x$ may reflect known discrepancies between continuum predictions for sea quark DFs and those produced by phenomenological fits \cite{Cui:2020tdf, Lu:2022cjx}.
%

As remarked when discussing helicity retention in Sec.\,\ref{SubSecLargeX}, new experiments able to return DF information on $x\gtrsim 0.6$ are desirable.  Thus, analyses of data collected recently \cite[CLAS RGC]{E1206109}, \cite[E12-06-110]{Zheng:2006} are eagerly awaited; especially because the SCI and QCD-kindred predictions are qualitatively and semiquantitatively consistent, with neither supporting helicity retention in hard scattering processes.  (A qualitatively similar outcome is obtained in a QCD-informed light-front Hamiltonian approach \cite{Xu:2023nqv}.)

\subsection{Polarised gluon distribution at $\zeta = \zeta_{\rm C}$}
\begin{figure}[t]
\vspace*{0.5ex}

\leftline{\hspace*{0.5em}{\large{\textsf{A}}}}
\vspace*{-3ex}
\includegraphics[width=0.41\textwidth]{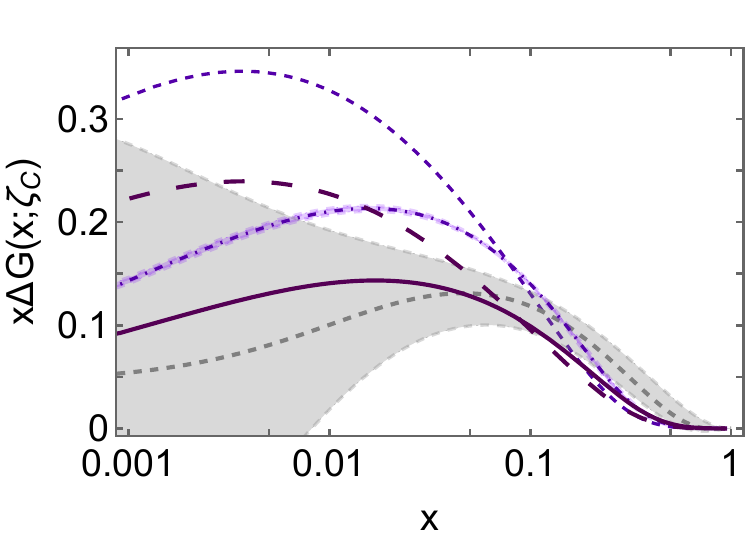}
\vspace*{-1ex}

\leftline{\hspace*{0.5em}{\large{\textsf{B}}}}
\vspace*{-3ex}
\includegraphics[width=0.41\textwidth]{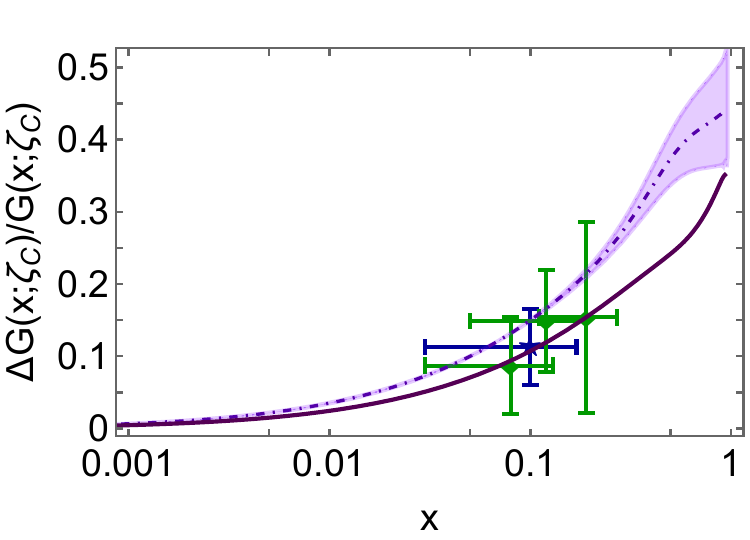}
\caption{\label{FigPolarisedGlue}
\textsf{Panel A}.
Polarised gluon DF, $\Delta G(x;\zeta_{\rm C})$: SCI -- solid purple curve; QCD-kindred \cite{Cheng:2023kmt} -- dot-dashed violet curve within like-coloured band.
Grey dashed curve and broad band: typical result from phenomenological global fit \cite[DSSV14]{deFlorian:2014yva} reported at $\zeta=10\,$GeV.
Evolved to this scale: SCI -- long-dashed purple curve; QCD-kindred central prediction -- dashed violet.
\textsf{Panel B}.
Polarised/unpolarised DF ratio $\Delta G(x;\zeta_{\rm C})/G(x;\zeta_{\rm C})$: SCI -- solid purple curve; QCD-kindred \cite{Cheng:2023kmt} -- dot-dashed violet curve within like-coloured band.
For context, we depict values reported in Ref.\,\cite[COMPASS]{COMPASS:2015pim}.
}
\end{figure}

Beginning with the SCI hadron-scale helicity-dependent DFs in Fig.\,\ref{FigPolarisedzH}, the AO evolution scheme generates the polarised and unpolarised gluon DFs at any scale $\zeta>\zeta_{\cal H}$: the $\zeta=\zeta_{\rm C}$ predictions are drawn in Fig.\,\ref{FigPolarisedGlue}.

Regarding phenomenological DF fits, $\Delta G(x)$ is very poorly constrained.   This is illustrated by the grey band in Fig.\,\ref{FigPolarisedGlue}A, reproduced from Ref.\,\cite[DSSV14]{deFlorian:2014yva} at $\zeta=10\,$GeV.  At this scale, the SCI prediction is indicated by the long-dashed purple curve.  Like the QCD-kindred result, and as also seen with unpolarised DFs, on $x\lesssim 0.05$, internally consistent continuum predictions for \linebreak glue (and sea) DFs are larger in magnitude than those inferred through phenomenological fits \cite{Cui:2020tdf, Lu:2022cjx}.  Notwithstanding this, on the complementary domain one finds
\begin{equation}
\mbox{SCI:} \quad
\int_{0.05}^1 dx \, \Delta G(x;\zeta=10\,{\rm GeV}) = 0.15\,,
\end{equation}
\emph{cf}.\ QCD-kindred $0.199(3)$ \cite{Cheng:2023kmt} and phenomenology \linebreak $0.19(6)$ \cite[DSSV14]{deFlorian:2014yva}.  Notably, under AO evolution, $\Delta G(x;\zeta>\zeta_{\cal H})$ is necessarily positive and the zeroth moment of this DF grows as a positive power of the inverse of the valence quark momentum fraction, with a constant of proportionality fixed by the zeroth moment of the singlet quark polarised DF (denoted $a_{\rm I}$ below) -- see Ref.\,\cite[Eq.\,(29)]{Yin:2023dbw}.

The SCI prediction for the ratio $\Delta G(x;\zeta_{\rm C})/G(x;\zeta_{\rm C})$ is drawn in Fig.\,\ref{FigPolarisedGlue}B: there is good agreement with the results reported in Ref.\,\cite[COMPASS]{COMPASS:2015pim}.  This is highlighted, \emph{e.g}., by noting that the mean value of the SCI result on the domain covered by measurements is $0.13$ \emph{cf}.\ $ 0.113 \pm 0.038 \pm 0.036$ \cite[COMPASS]{COMPASS:2015pim}.  The QCD-kindred value for this average is $0.167(3)$ \cite{Cheng:2023kmt}.  (Qualitatively similar outcomes are reported in Ref.\,\cite{Xu:2023nqv}.)

\section{Spin of the proton}
\label{SecSpin}
Working with the SCI DFs in Fig.\,\ref{FigPolarisedzH}, one obtains the following proton singlet axial charge from Eq.\,\eqref{gACompare}:
\begin{equation}
a_{\rm I} = \langle  \Delta {\mathpzc u}_p \rangle^{\zeta_{\cal H}} + \langle  \Delta {\mathpzc d}_p \rangle^{\zeta_{\cal H}} = 0.49 \, g_A\,, \label{a0SCI}
\end{equation}
a value matching that obtained in the QCD-kindred approach: $a_{\rm I} = 0.52(1)g_A$.  There is a numerical difference between the two final results, however, because the SCI DFs deliver an underestimate of $g_A$.

Proton spin studies measure a conserved current, \emph{viz}.\ a quantity that is proportional to the zeroth moment of the polarised structure function $g_1(x;\zeta)$ \cite{Altarelli:1988nr}.  Mapped into our notation, using the AO evolution \linebreak scheme, we identify the relevant observable as:
{\allowdisplaybreaks
\begin{subequations}
\label{Eqa0E}
\begin{align}
a_{\rm I}^{\rm E}(\zeta) & = a_{\rm I} - n_f \frac{\hat \alpha(\zeta)}{2\pi} \int_0^1 dx\,\Delta G(x;\zeta) \\
& =: a_{\rm I} - n_f \frac{\hat \alpha(\zeta)}{2\pi} \Delta G(\zeta)\,,
\end{align}
\end{subequations}
where $n_f$ is the number of active quark flavours: herein, evolution is defined with $n_f=4$.  Importantly, both terms on the right-hand side of this equation are $\zeta$-independent at leading-order in pQCD; hence, the measured value of the proton helicity may receive a (significant) correction from the gluon helicity despite presence of the running coupling \cite{Altarelli:1988nr}.}

\begin{figure}[t]
\centerline{%
\includegraphics[clip, width=0.44\textwidth]{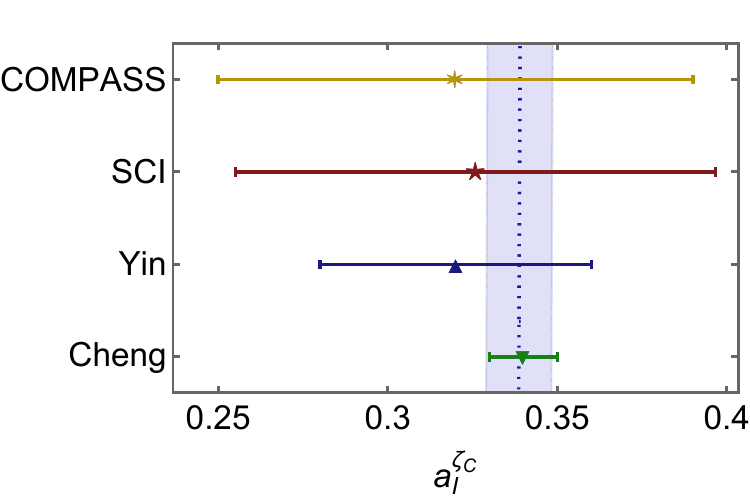}}
\vspace*{-2ex}

\caption{\label{Figa0E}
$a_{\rm I}^{\zeta = \zeta_{\rm C}}$, Eq.\,\eqref{Eqa0E}: non-Abelian anomaly corrected quark helicity contribution to the proton spin.
Data: gold asterisk and error bar -- Ref.\,\cite[COMPASS]{COMPASS:2016jwv}.
Theory:
SCI result and Refs.\,\cite[Cheng]{Cheng:2023kmt}, \cite[Yin]{Yin:2023dbw}.
Vertical dotted blue line within light-coloured band -- error weighted theory average: $0.34(1)$.
(In order to account for the SCI underestimate of $g_A$, we assigned a relative error of $\approx 20$\% to the mean value computed using the SCI and QCD-kindred $g_A$ values, Eqs.\,\eqref{gACompare}.)
}
\end{figure}

Using the SCI prediction for $\Delta G(x;\zeta_{\rm C})$ -- Fig.\,\ref{FigPolarisedGlue} -- to evaluate the right-hand side of Eq.\,\eqref{Eqa0E}, one finds
\begin{equation}
\label{DGSCI}
\Delta G(\zeta_{\rm C}) = 0.99\, g_A\,,
\end{equation} 
which is similar to the QCD-kindred result \cite{Cheng:2023kmt}: \linebreak $1.03(1) g_A$.  Substituting the Eq.\,\eqref{DGSCI} value into Eq.\,\eqref{Eqa0E}:
\begin{equation}
\mbox{SCI:} \quad a_{\rm I}^{\rm E}(\zeta_{\rm C}) = 0.28\,.
\end{equation}
This value compares well with the QCD-kindred result \cite{Cheng:2023kmt}, $0.34(1)$, and that reported in Ref.\,\cite[COMPASS]{COMPASS:2016jwv}: $0.32(7)$.  These recent continuum predictions are compared with the COMPASS result in Fig.\,\ref{Figa0E}.  Evidently, contemporary CSM analyses deliver a viable solution of the proton spin crisis, precipitated by the measurements reported in Ref.\,\cite{EuropeanMuon:1987isl} -- see, also, Ref.\,\cite[Fig.\,4]{Yin:2023dbw}.

It is worth noting that Eq.\,\eqref{Eqa0E} has been challenged \cite{Jaffe:1989jz}.
Whilst we find grounds for disagreement with counterexamples therein, given the identities and results in Refs.\,\cite{Bhagwat:2007ha, Raya:2016yuj, Ding:2018xwy}, we share the view that, owing to the non-Abelian anomaly, it is difficult to supply an unambiguous separation of the proton spin, as measured via the isoscalar axial current, into contributions from quark and gluon partons \cite{Jaffe:1989jz, Bass:2001dg}.
Alternative resolutions of the proton spin puzzle are possible \cite{Aidala:2012mv, Deur:2018roz}.
In this context, we stress that we begin with dressed quasiparticle degrees of freedom, not parton-like gluons and quarks, and therefrom use AO evolution to deliver a result for $a_I^E(\zeta_c)$.
Further examination will reveal whether agreement between our prediction for this quantity and the COMPASS result is accidental or meaningful.

It is relevant here to consider a light-front separation of the proton spin into contributions from quark and gluon spin and orbital angular momenta (OAM) \cite{Jaffe:1989jz}:
\begin{equation}
\frac{1}{2} =: \frac{1}{2} a_{\rm I} + \ell_q(\zeta)+\Delta G(\zeta)+\ell_g(\zeta)\,.
\end{equation}
At the hadron scale: $\ell_q = \ell_u + \ell_d$,
\begin{subequations}
\label{JLS}
\begin{align}
\ell_{u}(\zeta_{\cal H}) + \ell_{d}(\zeta_{\cal H}) & = \frac{1}{2} - \frac{1}{2} 0.494 g_A = 0.27\,, \label{JLSSCI}\\
\Delta G(\zeta_{\cal H})&=0=\ell_g(\zeta_{\cal H})\,.
\end{align}
\end{subequations}
where Eq.\,\eqref{JLSSCI} expresses the SCI result, Eqs.\,\eqref{gACompare}, \eqref{a0SCI}.  Owing to its more realistic value for $g_A$, the QCD-kindred result is $0.175$, \emph{viz}.\ roughly two-thirds of the size of the SCI value.

The individual flavour contributions, $\ell_{u,d}$ in \linebreak Eq.\,\eqref{JLSSCI}, may be extracted from generalised transverse momentum dependent DFs \cite[Sec.\,3.1.11]{Deur:2018roz}.
At the had\-ron scale, where valence degrees-of-free\-dom carry all proton properties, $\ell_{u,d}$ may reliably be estimated using a so-called Wand\-zura-Wilczek approximation \cite{Hatta:2012cs, Bhattacharya:2023hbq}:
\begin{equation}
\label{LWW}
\ell_{\mathpzc f}(x;\zeta_{\cal H}) =
x \int_0^1 dy \frac{1}{y^2}
\left[ y{\mathpzc f}(y;\zeta_{\cal H}) - \Delta{\mathpzc f}(y;\zeta_{\cal H})\right]\,,
\end{equation}
where ${\mathpzc f}$, $\Delta{\mathpzc f}$ are the unpolarised and polarised $f$-flavour DFs.
Using Eq.\,\eqref{LWW}, one obtains the following results for the zeroth moments:
\begin{equation}
\ell_u(\zeta_{\cal H})=0\,, \quad \ell_d(\zeta_{\cal H})=0.27\,.
\end{equation}

Exploiting the character of the hadron scale, $\ell_{u,d}$ may also be determined after calculation of the proton gravitational form factors and their flavour separation.  Following Ref.\,\cite{Xu:2023izo}, this should be possible in the near future.

Using the AO evolution scheme \cite{Yin:2023dbw}, implemented with minor modifications of Eqs.\,(32) in Ref.\,\cite{Deur:2018roz}, one finds
\begin{equation}
\ell_q(\zeta_{\rm C}) = 0.071\,,\,
\Delta G(\zeta_{\rm C}) = 0.91\,,\,
\ell_g(\zeta_{\rm C}) =-0.71\,.
\label{SpinBudget}
\end{equation}
The light-front quark OAM fraction falls steadily with increasing probe scale -- it is not bounded below; and the increasing gluon helicity moment is compensated by a growth in magnitude of the light-front gluon angular momentum fraction.
In these terms, glue accounts for 40\% of the proton spin at $\zeta_{\rm C}$.
Using QCD-kindred DFs, the conclusions are qualitatively similar \cite{Cheng:2023kmt}.
The asymptotic ($\zeta\to\infty$) limits are discussed elsewhere \cite{Ji:1995cu, Chen:2011gn}.

Regarding a flavour decomposition of $\ell_q(\zeta_{\rm C})$, we note that, whilst remaining a useful qualitative guide, the quantitative accuracy of Eq.\,\eqref{LWW} degrades as the glue contribution increases.  Nevertheless, our SCI analysis suggests that Eq.\,\eqref{LWW} is valid at the $\sim 80$\% level for $\zeta=\zeta_{\rm C}$.
Consequently, despite its use of an alternative definition of quark OAM \cite{Ji:2020ena}, based on generalised parton distributions, a semiquantitative comparison with Ref.\,\cite{Alexandrou:2020sml} results may be meaningful at this level because the two definitions are equivalent when Eq.\,\eqref{LWW} is a good approximation \cite{Hatta:2012cs}.
For interest, therefore, we list SCI predictions, obtained using Eq.\,\eqref{LWW}, alongside the Ref.\,\cite{Alexandrou:2020sml} results:
\begin{equation}
\begin{array}{l|llll}
 & \ u & \ d & \ s & \ c \\\hline
\ell_{q+\bar q} & -0.14 & 0.25 & 0.033 & 0.011 \\
L_{q+\bar q}\ \mbox{\cite{Alexandrou:2020sml}} \ & -0.22(3) & 0.26(2) & 0.039(14) & 0.014(10)
\end{array}\,.
\end{equation}

The QCD-kindred result for the total quark OAM is $\ell_q(\zeta_{\rm C}) = -0.027(10)$.  Using Eq.\,\eqref{LWW}, this outcome owes to an increase in the magnitude of the $u$ quark OAM, viz.\ $\ell_{u+\bar u} = -0.29$, with the other results practically unchanged.
Regarding the $u$ quark, at least, there appears to be significant sensitivity to dynamics.

Since $a_I$ itself is the same in both commonly used spin decompositions, then combining Ref.\,\cite{Alexandrou:2020sml} with SCI and QCD-kindred results, one finds that the quark helicity contributes $56\pm 13$\% of the proton spin.  Working only with the continuum results, this becomes $64(3)$\%.  Regarding canonical quark OAM, the fraction lies between $\pm 10$\%.  If one uses accuracy of Eq.\,\eqref{LWW} to justify inclusion of the Ref.\,\cite{Alexandrou:2020sml} result, the quark OAM component is $\sim 4(9)$\%.

\section{Summary and Perspective}
\label{epilogue}
Working with a quark + interacting-diquark approach to baryon structure, we employed a confining, symmetry-preserving formulation of a vector$\,\times\,$vector contact interaction (SCI) to deliver a comprehensive, coherent description of unpolarised and polarised proton structure functions: valence, glue, and four-flavour separated sea. A merit of the SCI is that all analyses are largely algebraic; so, the formulae and results exhibit a high level of transparency.  This enables clear assessments to be made; not just of the SCI outcomes themselves, but also regarding results obtained using more sophisticated frameworks through relevant comparisons.

Following presentation of SCI formulae for the had\-ron-scale, $\zeta_{\cal H}$, unpolarised and polarised valence-quark distribution functions (DFs) [Sec.\,\ref{SecUDF}], numerical results were calculated and plotted [Sec.\,\ref{SeczH}].
By definition, glue and sea DFs vanish at the hadron scale \cite{Yin:2023dbw}.

Meeting expectations engendered by treatments of meson DFs, SCI valence-quark DFs are hard on $x\simeq 1$ and nonzero at $x=0$.
Notwithstanding this, in many respects, each agrees semiquantitatively with the corresponding DF calculated using Schwinger functions with QCD-like characteristics, \emph{i.e}., QCD-kindred results \cite{Chang:2022jri, Lu:2022cjx, Cheng:2023kmt}.
Two points are worth highlighting.
(\emph{a}) Owing to the presence of both isoscalar-scalar and isovector-axial\-vector diquarks in the proton's Poincar\'e-covariant wave function, valence $d$ quark DFs are not simply proportional to those of their $u$ quark partners.  Amongst other differences, their respective maxima are not coincident.
(\emph{b}) For each quark flavour, the polarised:unpolarised ratio of valence-quark DFs is not unity at $x=1$.  Consequently, as in QCD-kindred studies \cite{Cheng:2023kmt}, the notion of helicity retention is disfavoured.

Comparisons with data require evolution to those resolving scales $\zeta > \zeta_{\cal H}$ appropriate to modern experiments.  We accomplish that by using the all-orders \linebreak scheme detailed elsewhere \cite{Yin:2023dbw} [Sec.\,\ref{EvolvedDFs}].  Naturally, such QCD evolution is not truly appropriate in connection with DFs generated by a contact interaction -- see Ref. \cite[Sec.\,5]{Cui:2021mom} .  Nevertheless, there is a purpose to the exercise.  Namely, the results can be used to identify those aspects of proton DFs which are sensitive to the underlying picture of proton structure.  For instance, focusing on the evolved valence-quark DFs, it was found that the isovector difference of valence quark DFs is a poor discriminator between competing descriptions of proton structure [Fig.\,\ref{FigUnPolarisedz3}B].

Glue and sea DFs are nonzero $\forall \zeta > \zeta_{\cal H}$.
In this connection, with Pauli blocking expressed in a modification of the gluon splitting function, SCI DFs provide a straightforward explanation of the asymmetry of antimatter in the proton \cite{SeaQuest:2021zxb}.  In fact, on the domain of existing measurements, SCI and QCD-kindred results are indistinguishable [Fig.\,\ref{ImageSeaUD}B].  This agreement supports the Pauli-blocking picture that we discussed.
Naturally, above a scale threshold defined by the $c$-quark current-mass, the charm quark sea DFs are also nonzero [Fig.\,\ref{ImageGlue}]; and regarding the unpolarised DF, both SCI and QCD-kindred analyses predict a measurable charm quark momentum fraction in the proton [Table~\ref{TabMoments}] without recourse to ``intrinsic charm'' \cite{Brodsky:1980pb}.

Employing the SCI predictions for all unpolarised singlet quark DFs, a comparison with modern data on the neutron:proton structure function ratio was possible [Fig.\,\ref{ImageSeaQuest}].  Whilst SCI and QCD-kindred results are indistinguishable on $x\lesssim 0.25$, significant differences are manifest on the complementary valence-quark domain.  Hence, as long anticipated \cite{Holt:2010vj}, this ratio is a keen discriminator between pictures of proton structure.

Regarding helicity-dependent DFs, there are differences between SCI and QCD-kindred predictions [Sec.\ \ref{SubSecH}]; however, extant phenomenological analyses of data on valence and sea quark distributions deliver results that are typically of insufficient precision to distinguish between them.  This accentuates the importance of analyses of data collected recently \cite[CLAS RGC]{E1206109}, \cite[E12-06-110]{Zheng:2006}.  Similar remarks can be made about glue; nevertheless, it is notable that both SCI and QCD-kindred studies predict that the proton polarised gluon DF, $\Delta G(x;\zeta_{\rm C}=\surd 3\,{\rm GeV})$, is positive and large [Fig.\,\ref{FigPolarisedGlue}].  This outcome impacts heavily on the proton spin question.  In fact, including the SCI prediction in an uncertainty weighted average of contemporary continuum calculations, one may argue that measurements of the proton singlet axial charge should return \linebreak $a_{\rm I}^{\rm E}(\zeta_{\rm C}) = 0.34(1)$, a value in accord with contemporary data [Fig.\,\ref{Figa0E}].  At $\zeta_{\rm C}$, glue accounts for $\approx 40$\% of the proton spin and the quark spin component is $(56 \pm 13)$\% [Eq.\,\eqref{SpinBudget}].  (The quark orbital angular momentum contribution need not be a positive fraction.)

This study completes the first step in a systematic programme that aims to deliver QCD-connected predictions for all proton DFs and their unification with those of the pion and kaon.  The next stage will see the helicity-dependent DFs calculated using QCD-connec\-ted Schwinger functions, thereby testing and, perhaps, improving upon the results in Ref.\,\cite{Cheng:2023kmt}.  Further ahead, one may reasonably anticipate calculations that begin with a Poincar\'e-covariant Faddeev equation treatment of the three-valence-body bound state problem \cite{Eichmann:2009qa, Wang:2018kto}.

\begin{CJK}{UTF8}{song}
\medskip
\noindent\textbf{Acknowledgments} \
%
We are grateful for assistance from D.\ Binosi in reconstructing the phenomenological DF fits in Figs.\,\ref{ImageFits1}, \ref{ImageFits2}; and for constructive comments from Z.-F.\,Cui 
and C.\,Chen. 
Work supported by the
National Natural Science Foundation of China (grant no.\ 12135007).
%
%
%
\end{CJK}

\medskip

\noindent\textbf{Data availability} \  This manuscript has no associated data or the data will not be deposited. [Authors' comment: All information necessary to reproduce the results described herein is contained in the manuscript.]

\appendix

\section{Proton Bound State}
\label{AppendixSM}
%

\subsection{Proton Faddeev equation}
\label{AppFE}
\begin{figure}[t]
\centerline{%
\includegraphics[clip, height=0.14\textwidth, width=0.45\textwidth]{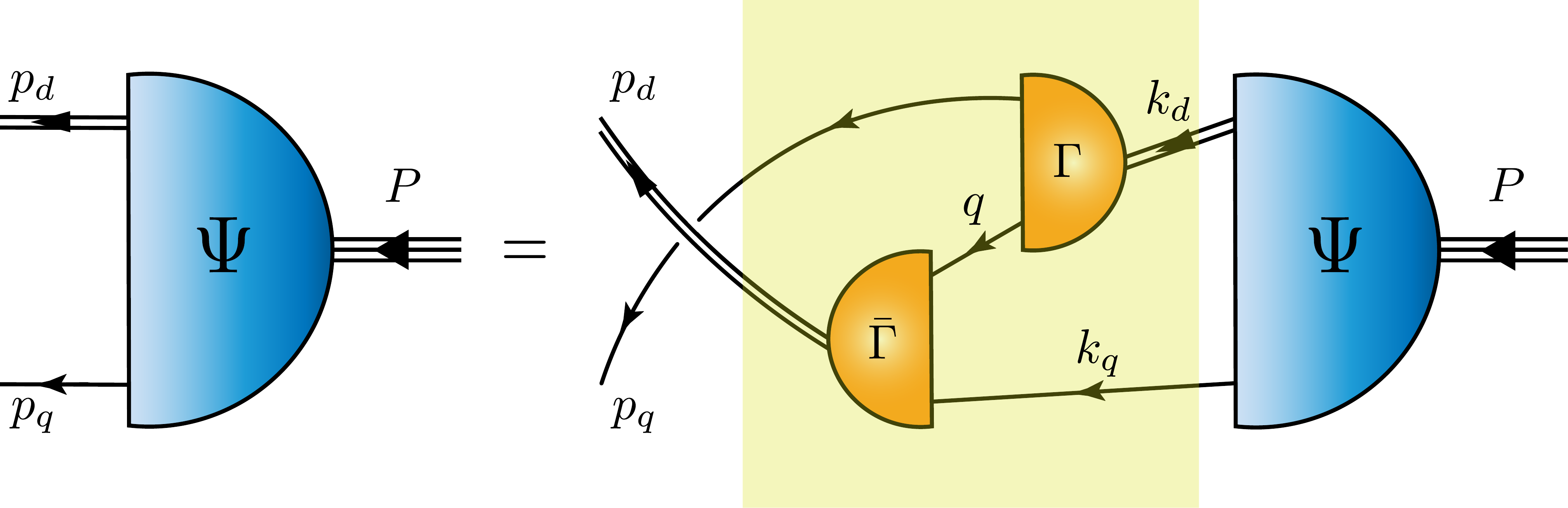}}
\caption{\label{figFaddeev}
Integral equation for the Poincar\'e-covariant matrix-valued function $\Psi$, which is the Faddeev amplitude for a proton with total momentum $P=p_q+p_d=k_q+k_d$ constituted from three valence quarks, two of which are contained in a nonpointlike, interacting diquark correlation. $\Psi$ describes the relative momentum correlation between the dressed-quarks and -diquarks.
Legend. \emph{Shaded rectangle} -- Faddeev kernel;
\emph{single line} -- dressed-quark propagator [\ref{AppendixSCI}];
$\Gamma$ -- diquark correlation amplitude and \emph{double line} -- diquark propagator [\ref{Appendixdiquarks}].
Only isoscalar-scalar diquarks, $[ud]$, and isovector--axialvector diquarks, $\{uu\}$, $\{ud\}$, $\{dd\}$ play a role in nucleons \cite{Barabanov:2020jvn}.}
\end{figure}

Our analysis of proton DFs rests on a solution of the Poincar\'e-covariant Faddeev equation, depicted figuratively in Fig.\,\ref{figFaddeev}.  It expresses the quark--plus--interacting-diquark picture of baryon structure introduced in Refs.\ \cite{Cahill:1988dx, Reinhardt:1989rw, Efimov:1990uz}.  An updated perspective is provided in Refs.\ \cite{Barabanov:2020jvn, Eichmann:2022zxn, Liu:2022ndb, Liu:2022nku}.

According to Fig.\,\ref{figFaddeev}, there are two contributions to binding within the proton \cite{Segovia:2015ufa}.  One forms tight (but not pointlike) quark + quark correlations.  The other is generated by the quark exchange depicted in the shaded area of the image, which ensures that all diquark correlations are fully dynamical.  Thus, no quark is special because each one participates in all diquark correlations to the fullest extent allowed by its quantum numbers. Amongst other things, this continual quark rearrangement guarantees that the proton's dressed-quark wave function complies with Pauli statistics.

\subsection{SCI}
\label{AppendixSCI}
Our implementation of the symmetry-preserving treatment of a vector$\,\times\,$vector contact interaction (SCI) is detailed in Ref.\,\cite[Appendix~A]{Cheng:2022jxe}.  In order to make our presentation self-contained, we repeat some of that material here.

The basic element is the quark + antiquark scattering kernel.  In rainbow-ladder (RL) truncation \cite{Munczek:1994zz, Bender:1996bb}, it can be written:
\begin{subequations}
\label{KDinteraction}
\begin{align}
\mathscr{K}_{\alpha_1\alpha_1',\alpha_2\alpha_2'}  & = {\mathpzc G}_{\mu\nu}(k) [i\gamma_\mu]_{\alpha_1\alpha_1'} [i\gamma_\nu]_{\alpha_2\alpha_2'}\,,\\
 {\mathpzc G}_{\mu\nu}(k)  & = \tilde{\mathpzc G}(k^2) T^k_{\mu\nu}\,,
\end{align}
\end{subequations}
where $p_1-p_1^\prime = k = p_2^\prime -p_2$, with $p_{1,2}$, $p_{1,2}^\prime$ being, respectively, the initial and final momenta of the scatterers, and $k^2T_{\mu\nu}^k = k^2\delta_{\mu\nu} - k_\mu k_\nu$.

The defining element is $\tilde{\mathpzc G}$.  Owing to the emergence of a gluon mass-scale \cite{Gao:2017uox, Cui:2019dwv}, $\tilde{\mathpzc G}$ is nonzero and finite at infrared momenta, \emph{viz}.\
\begin{align}
\tilde{\mathpzc G}(k^2) & \stackrel{k^2 \simeq 0}{=} \frac{4\pi \alpha_{\rm IR}}{m_G^2}\,.
\end{align}
In QCD \cite{Cui:2019dwv, Deur:2023dzc}: $m_G \approx 0.5\,$GeV, $\alpha_{\rm IR} \approx \pi$.
The value of $m_G$ is retained herein and capitalising on the fact that a SCI does not support relative momentum between meson bound-state constituents, we simplify the tensor in Eqs.\,\eqref{KDinteraction}:
\begin{align}
\label{KCI}
\mathscr{K}_{\alpha_1\alpha_1',\alpha_2\alpha_2'}^{\rm CI}  & = \frac{4\pi \alpha_{\rm IR}}{m_G^2}
 [i\gamma_\mu]_{\alpha_1\alpha_1'} [i\gamma_\mu]_{\alpha_2\alpha_2'}\,.
 \end{align}

Confinement is implemented by including an infrared regularising scale, $\Lambda_{\rm ir}$, when solving all integral equations relevant to bound-state problems \cite{Ebert:1996vx}.  This expedient eliminates quark + antiquark production thresholds \cite{Krein:1990sf}.  The standard choice is $\Lambda_{\rm ir} = 0.24\,$GeV\,$=1/[0.82\,{\rm fm}]$ \cite{GutierrezGuerrero:2010md}, \emph{i.e}., a confinement length scale of roughly the same size as the proton radii \cite{Cui:2022fyr}.

All SCI integrals require ultraviolet regularisation.  This eliminates the link between ultraviolet and infrared scales that is characteristic of QCD; so the associated ultraviolet mass-scale, $\Lambda_{\rm uv}$, becomes a physical parameter.  It may be interpreted as an upper bound on the domain whereupon distributions within the associated systems are practically momentum-independent.

\subsubsection{Dressed quarks}
{\allowdisplaybreaks
The SCI gap equation is
\begin{align}
\label{GapEqn}
S^{-1}(p)  & = i\gamma\cdot p +m \nonumber \\
& \quad + \frac{16 \pi}{3} \frac{\alpha_{\rm IR}}{m_G^2}
\int \frac{d^4q}{(2\pi)^4} \gamma_\mu S(q) \gamma_\mu\,,
\end{align}
where $m$ is the light-quark current-mass.  With a Poincar\'e-invariant regularisation, the solution is
\begin{equation}
\label{genS}
S^{-1}(p) = i \gamma\cdot p + M\,.
\end{equation}
The dynamically generated dressed-quark mass, $M$, is obtained by solving
\begin{equation}
M = m + M\frac{4\alpha_{\rm IR}}{3\pi m_G^2}\,\,{\cal C}_0^{\rm iu}(M^2)\,.
\label{gapactual}
\end{equation}
Here
\begin{align}
\nonumber
{\cal C}_0^{\rm iu}(\sigma) &=
\int_0^\infty\! ds \, s \int_{\tau_{\rm uv}^2=1/\Lambda_{\textrm{uv}}^{2}}^{\tau_{\rm ir}^2=1/\Lambda_{\textrm{ir}}^{2}} d\tau\,{\rm e}^{-\tau (s+\sigma)}\\
& =
\sigma \big[\Gamma(-1,\sigma \tau_{\rm uv}^2) - \Gamma(-1,\sigma \tau_{\rm ir}^2)\big],
\label{eq:C0}
\end{align}
where $\Gamma(\alpha,y)$ is the incomplete gamma-function.  The ``iu'' superscript emphasises that the function depends on both the infrared and ultraviolet cutoffs.
Functions of the following type arise in SCI bound-state equations:
\begin{align}
%
%
%
n !\, \overline{\cal C}^{\rm iu}_n(\sigma) & = \Gamma(n-1,\sigma \tau_{\textrm{uv}}^{2}) - \Gamma(n-1,\sigma \tau_{\textrm{ir}}^{2})\,,
\label{eq:Cn}
\end{align}
${\cal C}^{\rm iu}_n(\sigma)=\sigma \overline{\cal C}^{\rm iu}_n(\sigma)$, $n\in {\mathbb Z}^\geq$.}

\begin{table}[t]
\caption{\label{Tab:DressedQuarks}
Coupling, $\alpha_{\rm IR}$, ultraviolet cutoff, $\Lambda_{\rm uv}$, and current-quark mass, $m$, that are part of a good description of flavoured pseudoscalar meson properties, along with the dressed-quark mass, $M$, pion mass, $m_{\pi}$, and leptonic decay constant, $f_{\pi}$, they produce \cite{Xu:2021iwv}; all obtained with $m_G=0.5\,$GeV, $\Lambda_{\rm ir} = 0.24\,$GeV.
Empirically, at a sensible level of precision \cite{Workman:2022ynf}:
$m_\pi =0.14$, $f_\pi=0.092$.
%
(We assume isospin symmetry and list dimensioned quantities in GeV.)}
\begin{center}
\begin{tabular*}
{\hsize}
{
l@{\extracolsep{0ptplus1fil}}|
c@{\extracolsep{0ptplus1fil}}|
c@{\extracolsep{0ptplus1fil}}
c@{\extracolsep{0ptplus1fil}}
c@{\extracolsep{0ptplus1fil}}
|c@{\extracolsep{0ptplus1fil}}
c@{\extracolsep{0ptplus1fil}}
c@{\extracolsep{0ptplus1fil}}}\hline
& $\alpha_{\rm IR}\ $ & $\Lambda_{\rm uv}$ & $m$ &   $M$ &  $m_{\pi}$ & $f_{\pi}$ \\\hline
$\pi\ $  & $1.13\phantom{2}$ & $0.91\ $ & $0.0068\ $ & 0.37$\ $ & 0.14 & 0.10  \\\hline
\end{tabular*}
\end{center}
\end{table}

The SCI parameters were fixed in Ref.\,\cite{Xu:2021iwv}.  Those relevant herein are recorded in Table~\ref{Tab:DressedQuarks}.  All information necessary to specify the dressed-quark propagators that appear when solving Fig.\,\ref{figFaddeev} is now available.

\subsubsection{Diquarks}
\label{Appendixdiquarks}
The next step is to compute the diquark correlation amplitudes.  The relevant Bethe-Salpeter equations are written in
Ref.\,\cite[Sec.\,2.2.2]{Chen:2012qr}
along with the form of their solutions, which can be expressed as follows:
\begin{equation}
\label{DefineBSAs}
^a\Gamma_{fg}^{J^P}(K)
= T_{\bar 3_c}^a \otimes \underline\Gamma_{fg}^{J^P}(K)
= T_{\bar 3_c}^a \otimes t^J_{fg}\otimes\Gamma_{fg}^{J^P}(K)\,,
\end{equation}
where the colour-antitriplet character is seen in $\{T_{\bar 3_c}^a,a=1,2,3\}=\{i\lambda^2, i\lambda^5, i\lambda^7\}$, using Gell-Mann matrices;
the flavour structure is expressed via
{\allowdisplaybreaks
\begin{subequations}
\begin{align}
t_{ud}^0 =
\left[\begin{array}{rcc}
0 & 1 & 0 \\
-1 & 0 & 0 \\
0 & 0 & 0
\end{array}\right], & \quad
t_{uu}^1 =
\left[\begin{array}{ccc}
\surd 2 & 0 & 0 \\
0 & 0 & 0 \\
0 & 0 & 0
\end{array}\right], \\
t_{ud}^1 =
\left[\begin{array}{rcc}
0 & 1 & 0 \\
\phantom{-}1 & 0 & 0 \\
0 & 0 & 0
\end{array}\right], & \quad
 t_{dd}^1 =
\left[\begin{array}{ccc}
0 & 0 & 0 \\
0 & \surd 2 & 0 \\
0 & 0 & 0
\end{array}\right],
\end{align}
\end{subequations}
(we have included $\{dd\}$ only for completeness, because this correlation appears in the neutron);
and the spinor structure is
\begin{subequations}
\label{qqBSAs}
\begin{align}
\Gamma_{ud}^{0^+}(K) & =
\gamma_5\left[ i E_{[ud]} + \frac{\gamma\cdot K}{M} F_{[ud]} \right]C,\\
\Gamma_{fg}^{1^+}(K) & = T_{\mu\nu}^K \gamma_\nu C E_{\{fg\}}\,,
\end{align}
\end{subequations}
$f,g=\{u,d\}$, where $K$ is the correlation's total momentum and $C=\gamma_2\gamma_4$ is the charge conjugation matrix.

The $J^P$ diquark Bethe-Salpeter equations are solved using the dressed-quark propagators described above and the value of $\Lambda_{\rm uv}$ in Table~\ref{Tab:DressedQuarks}.  The calculated diquark masses and canonically normalised amplitudes used herein are listed in Table~\ref{qqBSAsolutions}.

The scalar and axialvector diquark propagators take standard forms:
\begin{subequations}
\label{qqPropagator}
\begin{align}
\Delta^{[fg]}(K) & = \frac{1}{K^2 + m_{[fg]}^2}\,, \\
\Delta^{\{fg\}}_{\mu\nu}(K) & =\delta_{\mu\nu} \frac{1}{K^2 + m_{\{fg\}}^2}\,,
\label{AVdqprop}
\end{align}
\end{subequations}
where the masses are taken from Table~\ref{qqBSAsolutions}.

\begin{table}[t]
\caption{\label{qqBSAsolutions}
Masses and canonically normalised correlation amplitudes obtained by solving the diquark Bethe-Salpeter equations.  Recall that we work in the isospin-symmetry limit; so, $\{uu\}$, $\{ud\}$, $\{dd\}$ correlations are degenerate and have the same correlation amplitudes.
}
\begin{center}
\begin{tabular*}
{\hsize}
{
c@{\extracolsep{0ptplus1fil}}
c@{\extracolsep{0ptplus1fil}}
c@{\extracolsep{0ptplus1fil}}|
c@{\extracolsep{0ptplus1fil}}
c@{\extracolsep{0ptplus1fil}}}\hline
$m_{[ud]=0}/$GeV & $E_{[ud]=0}$ & $F_{[ud]=0}\ $ & $m_{\{uu\}=1}/$GeV & $E_{\{uu\}=1}\ $ \\
$0.78$ & $2.71$ &  $0.31\ $ & $1.06$ & $1.39\ $ \\\hline
\end{tabular*}
\end{center}
\end{table}

\subsubsection{Proton Faddeev amplitude}
\label{AppendixFaddeev}
All elements necessary to construct the proton (nucleon) Faddeev kernel are now in hand.  The task can be completed following Ref.\,\cite[Sec.\,3]{Chen:2012qr}.  We would like to stress that, as therein, a ``static approximation'' is used to represent the quark exchanged between the diquarks, \emph{viz}.
\begin{equation}
\label{static}
S(q) \to \frac{g_8^2}{M}\,,
\end{equation}
$g_8 = 1.18$.  This expedient has the merit of ensuring that the proton Faddeev amplitude is momentum independent.  Eliminating the static approximation leads to few improvements at the cost of many complications to all calculations -- see, \emph{e.g}., Ref.\,\cite{Xu:2015kta}.

Using Eq.\,\eqref{static}, the solution of the resulting Faddeev equation can be written as follows:
\begin{equation}
\label{PsiuP}
\Psi(P) = \psi(P) u(P)\,,
\end{equation}
where the positive energy spinor satisfies
\begin{equation}
\bar u(P)(i\gamma\cdot P+m_p) = 0=(i\gamma\cdot P + m_p)u(P)\,,
\end{equation}
is normalised such that $\bar u(P) u(P) = 2 m_p$, and
\begin{equation}
2 m_p \Lambda_+(P) = \sum_{\sigma=\pm}u(P;\sigma)\bar u(P;\sigma) = m_p-i\gamma\cdot P\,,
\end{equation}
where in this line the spin label is made explicit.  (Details are provided in Ref.\,\cite[Appendix~A]{Chen:2012qr}.)  Using Eq.\ \eqref{PsiuP}, then the SCI solution for $\psi(P)$ is a sum of the following Dirac structures ($\hat P^2=-1$):
\begin{subequations}
\label{SAPD}
\begin{align}
a_{0^+}\psi^{0^+}(P) & = {\mathpzc a}_{0^+}\,\mathbf{I}_{\rm D}\,, \\
a_{1_{\{uu\}}^+} \psi^{1_{\{uu\}}^+}_{\mu}(P)
& = {\mathpzc a}_{1^+}\,i\gamma_{5}\gamma_{\mu}+{\mathpzc a}_{2^+}\gamma_{5}\hat P_{\mu}\,, \\
a_{1_{\{ud\}}^+} \psi^{1_{\{ud\}}^+}_{\mu}(P)
& = {\mathpzc a}_{1^0}\,i\gamma_{5}\gamma_{\mu}+{\mathpzc a}_{2^0}\gamma_{5}\hat P_{\mu}\,.
\end{align}
\end{subequations}
As usual, $\bar \Psi(P) = \Psi(P)^\dagger \gamma_4 = \bar u(P) \gamma_4 \psi(P)^\dagger \gamma_4$.

Faddeev equation dynamics determines the values of the coefficients: $\{{\mathpzc a}_{0^+}, {\mathpzc a}_{1,2}^{0,+}\}$.
%

Solving the Faddeev equations, one obtains the proton mass and amplitude listed in Table~\ref{SolveFaddeev}.  
Regarding the proton mass, we note that the scale is set by $3 \times M$, \emph{viz}.\ the basic source is traceable to the dynamical generation of a large dressed-quark mass.  In Nature, this is a consequence of nonperturbative dynamics in QCD's gauge sector \cite{Binosi:2022djx, Papavassiliou:2022wrb, Ding:2022ows, Ferreira:2023fva}.
Concerning details, we note that the calculated value is deliberately $\approx 0.21\,$GeV above experiment \cite{Workman:2022ynf} because Fig.\,\ref{figFaddeev} describes the \emph{dressed-quark core} of the nucleon.  To build a complete nucleon, resonant contributions should be included in the kernel.  Such ``meson cloud'' effects lower the mass of the nucleon \cite{Hecht:2002ej, Sanchis-Alepuz:2014wea, Garcia-Tecocoatzi:2016rcj, Chen:2017mug}.  Their impact on baryon structure may be estimated using dynamical coupled-channels models \cite{Aznauryan:2012ba, Burkert:2017djo}, but that is beyond the reach of contemporary Faddeev equation analyses.

\begin{table}[t]
\caption{\label{SolveFaddeev}
Proton mass and unit normalised Faddeev amplitude, obtained by solving the Faddeev equations defined by Fig.\,\ref{figFaddeev}.
%
The canonically normalised amplitude, explained in connection with Eq.\,\eqref{CanonicalFA}, is obtained by dividing the amplitude entries in each row by ${\mathpzc n}_c^{p}=0.157$.
(Masses listed in GeV.  Amplitudes are dimensionless. Recall that we work in the isospin-symmetry limit.)}
\begin{center}
\begin{tabular*}
{\hsize}
{
c@{\extracolsep{0ptplus1fil}}
|c@{\extracolsep{0ptplus1fil}}
c@{\extracolsep{0ptplus1fil}}
c@{\extracolsep{0ptplus1fil}}
c@{\extracolsep{0ptplus1fil}}
c@{\extracolsep{0ptplus1fil}}
c@{\extracolsep{0ptplus1fil}}}\hline
  & mass
  & ${\mathpzc a}_{0^+}$ & ${\mathpzc a}_{1^+}$ & ${\mathpzc a}_{2^+}$ & ${\mathpzc a}_{1^0}$ & ${\mathpzc a}_{2^0}$\\\hline
$p\ $ & $1.15$ & $\phantom{-}0.88$ & $-0.38$ & $-0.063$ & $\phantom{-}0.27$ & $\phantom{-}0.044\ $  \\
\hline
\end{tabular*}
\end{center}
\end{table}

The Faddeev amplitude in Table~\ref{SolveFaddeev} is unit normalised.  One must use the canonically normalised amplitude when calculating observables.  That is defined via the proton's Dirac form factor in elastic electromagnetic scattering, $F_1(Q^2=0)$:
\begin{align}
F_1^p(Q^2=0)  & = 2 e_u F_1^{pu}(0) +  e_d F_1^{pd}(0)\,,
\label{CanonicalFA}
\end{align}
where $e_{u,d}$ are the quark electric charges, expressed in units of the positron charge.  It is straightforward to calculate the single constant factor that, when used to rescale the proton unit-normalised Faddeev amplitude, ensures $F_1^{pu}(0)=1=F_1^{pd}(0)$.  So long as one employs a symmetry-preserving treatment of the elastic scattering problem, it is guaranteed that a single factor ensures all flavour-separated electromagnetic form factors are unity at $Q^2=0$.  Explicit examples are provided elsewhere \cite{Wilson:2011aa}.

\section{Proton Distribution Functions}
\label{AppendixDFs}
Using the material in \ref{AppendixSM} and the standard SCI regularisation procedures, the formulae in Sec.\,\ref{SecUDF} evaluate to the expressions listed hereafter.

\subsection{Unpolarised}
{\allowdisplaybreaks
\begin{subequations}
\begin{align}
{\mathpzc u}_{V_{Q_0}}^p & (x; \zeta_{\cal H})  =
 {\mathpzc a}_{0^+}^2 (1-x) \nonumber \\
 & \times \left[
\bar{\cal C}^{\rm iu}_1(\omega_0)  + 2 x (2 M m_p - {\mathpzc t}_0 ) \bar{\cal C}^{\rm iu}_2(\omega_0)/ \omega_0
 \right] , \\
{\mathpzc u}_{V_{Q_1}}^p & (x; \zeta_{\cal H})  = \nonumber \\
& {\mathpzc u}_{11}^p  (x; \zeta_{\cal H}) +
{\mathpzc u}_{12}^p  (x; \zeta_{\cal H}) +
{\mathpzc u}_{22}^p  (x; \zeta_{\cal H}) \,,\\
{\mathpzc u}_{11}^p & (x; \zeta_{\cal H})  =
2 {\mathpzc a}_{1^0}^2 (1-x) \nonumber \\
 & \times \left[
\bar{\cal C}^{\rm iu}_1(\omega_1)  + 2 x (4 M m_p - {\mathpzc t}_1 ) \bar{\cal C}^{\rm iu}_2(\omega_1)/ \omega_1
 \right] , \\
{\mathpzc u}_{12}^p & (x; \zeta_{\cal H})  =
-2 {\mathpzc a}_{1^0} {\mathpzc a}_{2^0} (1-x) \nonumber \\
 & \times \left[
\bar{\cal C}^{\rm iu}_1(\omega_1)  - 2 x (2 M m_p + {\mathpzc t}_1 ) \bar{\cal C}^{\rm iu}_2(\omega_1)/ \omega_1
 \right] , \\
{\mathpzc u}_{22}^p & (x; \zeta_{\cal H})  =
- {\mathpzc a}_{2^0}^2 (1-x) \nonumber \\
 & \times \left[
\bar{\cal C}^{\rm iu}_1(\omega_1)  - 2 x (2 M m_p + {\mathpzc t}_1 ) \bar{\cal C}^{\rm iu}_2(\omega_1)/ \omega_1
 \right] ,
\end{align}
\end{subequations}
where
\begin{subequations}
\begin{align}
\omega_{0,1} & = x m_{0^+,1^+}^2 + (1-x) M^2 - x(1-x) m_p^2 \, ,\\
{\mathpzc t}_{0,1} & = m_{0^+,1^+}^2 - M^2 -m_p^2.
\end{align}
\end{subequations}
Each right-hand side should be multiplied by $1/[16\pi^2]$.
}

The functions ${\mathpzc s}_{0^+,1^+}^p(x;\zeta_{\cal H})$ in Eqs.\,\eqref{convolution0}, \eqref{convolution1} are obtained from the above by using Eq.\,\eqref{Equsexchange}; so, evaluation of ${\mathpzc u}_{V_{D_{0,1}}}^p(x;\zeta_{\cal H})$ requires only that ${\mathpzc u}_V^{0^+,1^+}(x;\zeta_{\cal H})$ be specified.  Following Ref.\,\cite{Lu:2021sgg}, these functions may be obtained from the square of the $0^+$ and $1^+$ diquark distribution amplitudes.  This leads to the following results:
\begin{subequations}
\begin{align}
u_V^{0^+}(x;\zeta_{\cal H}) & = {\mathpzc n}_0
\left[ (M^2 [ E_{0^+} - 2 F_{0^+} ] + \omega_{D_0} F_{0^+}) \bar{\cal C}^{\rm iu}_1(\omega_{D_0}) \right. \nonumber \\
& \quad \left. + F_{0^+} {\cal C}^{\rm iu}_0(\omega_{D_0}) \right]^2 , \label{SCdqDA2} \\
u_V^{1^+}(x;\zeta_{\cal H}) & = {\mathpzc n}_1 E_{1^+}^2
\left[ (2 x (1-x) m_{1^+}^2  + \omega_{D_1} ) \bar{\cal C}^{\rm iu}_1(\omega_{D_1}) \right. \nonumber \\
& \quad \left. + {\cal C}^{\rm iu}_0(\omega_{D_1}) \right]^2 . \label{AVdqDA2}
\end{align}
\end{subequations}
where the diquark correlation amplitudes are listed in Table~\ref{qqBSAsolutions},
${\mathpzc n}_{0,1}$ are constants that ensure
\begin{equation}
\int_0^1 dx\, u_V^{0^+,1^+}(x;\zeta_{\cal H}) = 1\,,
\end{equation}
and
$\omega_{D_{0,1}}= M^2 -x(1-x) m_{0^+,1^+}^2$.

The helicity independent $d$ quark distributions are obtained from the above by using Eqs.\,\eqref{dquarkDF}, \eqref{dpDFs}.

\subsection{Polarised}
{\allowdisplaybreaks
\begin{subequations}
\begin{align}
\Delta& {\mathpzc u}_{V_{Q_0}}^p  (x; \zeta_{\cal H})  =
 {\mathpzc a}_{0^+}^2 (1-x)  \nonumber \\
 & \times \left[
2  (2 x M m_p + \tilde{\mathpzc t}_0 ) \bar{\cal C}^{\rm iu}_2(\omega_0)/ \omega_0
-\bar{\cal C}^{\rm iu}_1(\omega_0)
 \right] , \\
\Delta & {\mathpzc u}_{V_{Q_1}}^p (x; \zeta_{\cal H})  = \nonumber \\
& \Delta{\mathpzc u}_{11}^p  (x; \zeta_{\cal H}) +
\Delta{\mathpzc u}_{12}^p  (x; \zeta_{\cal H}) +
\Delta{\mathpzc u}_{22}^p  (x; \zeta_{\cal H}) \,,\\
\Delta & {\mathpzc u}_{11}^p (x; \zeta_{\cal H})  =
2 {\mathpzc a}_{1^0}^2 (1-x) \nonumber \\
 & \times \left[
\bar{\cal C}^{\rm iu}_1(\omega_1) - 2 \tilde{\mathpzc t}_1  \bar{\cal C}^{\rm iu}_2(\omega_1)/ \omega_1
 \right] , \\
\Delta & {\mathpzc u}_{12}^p (x; \zeta_{\cal H})  =
2 {\mathpzc a}_{1^0} {\mathpzc a}_{2^0} (1-x) \nonumber \\
 & \times \left[
\bar{\cal C}^{\rm iu}_1(\omega_1)  + 2 (2 x M m_p - \tilde {\mathpzc t}_1 ) \bar{\cal C}^{\rm iu}_2(\omega_1)/ \omega_1
 \right] , \\
\Delta & {\mathpzc u}_{22}^p (x; \zeta_{\cal H})  =
 {\mathpzc a}_{2^0}^2 (1-x) \nonumber \\
 & \times \left[
\bar{\cal C}^{\rm iu}_1(\omega_1)  + 2 (2 x M m_p - \tilde {\mathpzc t}_1 ) \bar{\cal C}^{\rm iu}_2(\omega_1)/ \omega_1
 \right] ,
\end{align}
\end{subequations}
where
\begin{equation}
\tilde {\mathpzc t}_{0,1} = x m_{0^+,1^+}^2 +(2-x) M^2 - x(1-2x)m_p^2.
\end{equation}
Each right-hand side should be multiplied by ${\mathpzc A}_0/[16\pi^2]$, where the dressed-quark axial charge is \cite[Eq.\,(A25), (A26)]{Xing:2022sor}:
\begin{equation}
{\mathpzc A}_0 = 1/[1+ (4 M^2 \alpha_{\rm IR}/[3\pi m_G^2]) \bar{\cal C}^{\rm iu}_1(M^2)]
= 0.738\,.
\end{equation}
This level of quenching is an anticipated consequence of phenomena associated with emergent hadron mass \cite[Eq.\,(26)]{Chang:2012cc}.
}

\begin{table}[th]
\caption{\label{Icoeffs}
Interpolation coefficients for hadron scale DFs, to be used in Eq.\,\eqref{InterpolateDFs}.  Each entry should be divided by $10^{4}$.
}
\begin{tabular*}
{\hsize}
{
l@{\extracolsep{0ptplus1fil}}
c@{\extracolsep{0ptplus1fil}}
c@{\extracolsep{0ptplus1fil}}
c@{\extracolsep{0ptplus1fil}}
c@{\extracolsep{0ptplus1fil}}}\hline
${\mathpzc f}$ & ${\mathpzc u}$ & ${\mathpzc d}$ & $\Delta{\mathpzc u}$ & $\Delta{\mathpzc d}$ \\\hline
${\mathpzc n}_{\mathpzc f}$  & $\phantom{-}6858\ $  & $\phantom{-}3427\ $  & $\phantom{-}2429\ $ & $-568.5$  \\
$a_1^{\mathpzc f}$  & $\phantom{-}2059\ $ & $\phantom{-}2676\ $ & $\phantom{-}1595\ $ & $\phantom{-}5084\ $  \\
$a_2^{\mathpzc f}$  & $-1799\ $  & $-1313\ $ & $-1810\ $ & $\phantom{-}217.7$   \\
$a_3^{\mathpzc f}$  & $-925.2$ & $-880.9$ & $-766.8$ & $-1168\ $   \\
$a_4^{\mathpzc f}$  & $\phantom{-}180.6$ & $-25.22\,$ & $\phantom{-}266.1$ & $-573.5$   \\
$a_5^{\mathpzc f}$  & $\phantom{-}239.2$ & $\phantom{0}123.8$ & $\phantom{-}201.0$ & $-40.98\ $   \\
$a_6^{\mathpzc f}$  & $\phantom{-}15.01$ & $\phantom{0}42.05$ & $-45.98\ $ & $\phantom{0}57.58$  \\
$a_7^{\mathpzc f}$  & $-43.92\ $ & $-3.640\ $  & $-51.65\ $ & $\phantom{0}24.09$  \\
$a_8^{\mathpzc f}$  & $-10.91\ $ & $-7.820\ $ & $\phantom{-}4.833\ $ & $\phantom{-}3.180\ $  \\
$a_9^{\mathpzc f}$  & $\phantom{-}6.547$ & $-2.530\ $ &$\phantom{-}14.15$  & $-0.911\ $   \\
$a_{10}^{\mathpzc f}$  & $\phantom{-}3.768$ & $\phantom{-}0.566\ $ & $\phantom{-}1.228\ $ & $\phantom{-}0.034\ $ \\\hline
\end{tabular*}
\end{table}

In calculating the term $3_\Delta$ and $4_\Delta$ contributions in Sec.\,\ref{SecPDF}, we use \cite{Cheng:2022jxe} -- Eqs.\,(A35), (A37b), (A39b), Table~XIV -- adapted to Eq.\,\eqref{AVdqprop} above:
\begin{subequations}
\label{AppAV}
\begin{align}
\Gamma_{5\mu;\alpha\beta}^{AA}(\ell,\ell) & = \phantom{-}0.47\, \varepsilon_{\mu\alpha\beta\delta} 2 \ell_\delta \,, \\
\Gamma_{5\mu;\alpha}^{SA}(\ell,\ell) & = - 0.649\,  [m_{0^+}+m_{1^+}] i \delta_{\mu\alpha}\,.
\end{align}
\end{subequations}
Consequently,
\begin{subequations}  
\label{sterms}
\begin{align}
\Delta& {\mathpzc s}_{1^+}^p(x;\zeta_{\cal H})  =
0.94\, {\mathpzc a}_{1^0}^2 x \left\{
\bar{\cal C}^{\rm iu}_1(\tilde\omega_1)  \right. \nonumber \\
& \left.
+ \left[ (M+[1-x] m_p)^2 - \tilde\omega_1
\right] 2 \bar{\cal C}^{\rm iu}_2(\tilde\omega_1)/\tilde\omega_1
\right\}\,, \\
\Delta& {\mathpzc s}_{0^+1^+}^p(x;\zeta_{\cal H})  = 2.596\, [m_{0^+}+m_{1^+}] {\mathpzc a}_{0^+} {\mathpzc a}_{1^0} (1-x) \nonumber \\
& \times  (M+[1-x]m_p) \int_0^1 dy \, \bar{\cal C}^{\rm iu}_2(\omega_{01})/\omega_{01} \,,
\end{align}
\end{subequations}
where each right-hand side should be multiplied by \linebreak $1/[16\pi^2]$,
$\tilde\omega_1 = \left.\omega_1\right|_{x\to (1-x)}$, and
\begin{align}
\omega_{01} & = x M^2 + (1-y)(1-x)m_{0^+}^2 \nonumber \\
& \quad + y (1-x) m_{1^+}^2 -x (1-x) m_p^2\,.
\end{align}
In Eqs.\,\eqref{sterms}, we have omitted pieces that contribute negligibly to the final numerical results.

The convolution in Eq.\,\eqref{convolutionD1} is computed using Eq.\ \eqref{AVdqDA2}.  The other, Eq.\,\eqref{Eq01TDF}, involves the generalised DF for the $u$ quark exposed in a scalar-axialvector transition.  This has not yet been calculated; so, we report central results obtained using the valence-quark scale-free DF:
\begin{equation}
\label{TransitionDFCentral}
{\mathpzc u}_V^{0^+1^+}(x;\zeta_{\cal H}) = 30 x^2 (1-x)^2\,,
\end{equation}
because, on a sizeable neighbourhood of $x=1/2$, this approximates an average of the DAs in Eqs.\,\eqref{SCdqDA2}, \eqref{AVdqDA2}.
We also used
\begin{subequations}
\label{TransitionDFModels}
\begin{align}
{\mathpzc u}_V^{0^+1^+}(x;\zeta_{\cal H}) & = 6 x (1-x)\,, \\
{\mathpzc u}_V^{0^+1^+}(x;\zeta_{\cal H}) & = 140 x^3 (1-x)^3\,,
\end{align}
\end{subequations}
the first of which lies close to the $0^+$ DF on a large neighbourhood of $x=1/2$ and the second, the $1^+$ DF.

The helicity dependent $d$ quark DFs are obtained from the above by using Eqs.\,\eqref{dquarkPDF}, \eqref{dpPDFs}. \\

\subsection{Interpolation coefficients}
All SCI DFs can reliably be interpreted using the function in Eq.\,\eqref{InterpolateDFs} with the interpolation coefficients listed in Table~\ref{Icoeffs}.


\begin{thebibliography}{143}
\providecommand{\natexlab}[1]{#1}
\providecommand{\url}[1]{\texttt{#1}}
\providecommand{\urlprefix}{URL }
\expandafter\ifx\csname urlstyle\endcsname\relax
  \providecommand{\doi}[1]{doi:\discretionary{}{}{}#1}\else
  \providecommand{\doi}[1]{doi:\discretionary{}{}{}\begingroup
  \urlstyle{rm}\url{#1}\endgroup}\fi
\providecommand{\bibinfo}[2]{#2}

\bibitem[{Ashman et~al.(1988)}]{EuropeanMuon:1987isl}
\bibinfo{author}{J.~Ashman}, et~al., \bibinfo{title}{{A Measurement of the Spin
  Asymmetry and Determination of the Structure Function $g_1$ in Deep Inelastic
  Muon-Proton Scattering}}, \bibinfo{journal}{Phys. Lett. B}
  \bibinfo{volume}{206} (\bibinfo{year}{1988}) \bibinfo{pages}{364}.

\bibitem[{Aidala et~al.(2013)Aidala, Bass, Hasch, and Mallot}]{Aidala:2012mv}
\bibinfo{author}{C.~A. Aidala}, \bibinfo{author}{S.~D. Bass},
  \bibinfo{author}{D.~Hasch}, \bibinfo{author}{G.~K. Mallot},
  \bibinfo{title}{{The Spin Structure of the Nucleon}}, \bibinfo{journal}{Rev.
  Mod. Phys.} \bibinfo{volume}{85} (\bibinfo{year}{2013})
  \bibinfo{pages}{655--691}.

\bibitem[{Deur et~al.(2019)Deur, Brodsky, and De~T\'eramond}]{Deur:2018roz}
\bibinfo{author}{A.~Deur}, \bibinfo{author}{S.~J. Brodsky},
  \bibinfo{author}{G.~F. De~T\'eramond}, \bibinfo{title}{{The Spin Structure of
  the Nucleon}}, \bibinfo{journal}{Rept. Prog. Phys.}
  \bibinfo{volume}{82}~(\bibinfo{number}{076201}).

\bibitem[{Bashir et~al.(2012)Bashir, Chang, Cloet, El-Bennich, Liu, Roberts,
  and Tandy}]{Bashir:2012fs}
\bibinfo{author}{A.~Bashir}, \bibinfo{author}{L.~Chang}, \bibinfo{author}{I.~C.
  Cloet}, \bibinfo{author}{B.~El-Bennich}, \bibinfo{author}{Y.-X. Liu},
  \bibinfo{author}{C.~D. Roberts}, \bibinfo{author}{P.~C. Tandy},
  \bibinfo{title}{{Collective perspective on advances in Dyson-Schwinger
  Equation QCD}}, \bibinfo{journal}{Commun. Theor. Phys.} \bibinfo{volume}{58}
  (\bibinfo{year}{2012}) \bibinfo{pages}{79--134}.

\bibitem[{Roberts(2017)}]{Roberts:2016vyn}
\bibinfo{author}{C.~D. Roberts}, \bibinfo{title}{{Perspective on the origin of
  hadron masses}}, \bibinfo{journal}{Few Body Syst.} \bibinfo{volume}{58}
  (\bibinfo{year}{2017}) \bibinfo{pages}{5}.

\bibitem[{Aguilar et~al.(2019)}]{Aguilar:2019teb}
\bibinfo{author}{A.~C. Aguilar}, et~al., \bibinfo{title}{{Pion and Kaon
  Structure at the Electron-Ion Collider}}, \bibinfo{journal}{Eur. Phys. J. A}
  \bibinfo{volume}{55} (\bibinfo{year}{2019}) \bibinfo{pages}{190}.

\bibitem[{Chen et~al.(2020)Chen, Guo, Roberts, and Wang}]{Chen:2020ijn}
\bibinfo{author}{X.~Chen}, \bibinfo{author}{F.-K. Guo}, \bibinfo{author}{C.~D.
  Roberts}, \bibinfo{author}{R.~Wang}, \bibinfo{title}{{Selected Science
  Opportunities for the EicC}}, \bibinfo{journal}{Few Body Syst.}
  \bibinfo{volume}{61} (\bibinfo{year}{2020}) \bibinfo{pages}{43}.

\bibitem[{Anderle et~al.(2021)}]{Anderle:2021wcy}
\bibinfo{author}{D.~P. Anderle}, et~al., \bibinfo{title}{{Electron-ion collider
  in China}}, \bibinfo{journal}{Front. Phys. (Beijing)}
  \bibinfo{volume}{16}~(\bibinfo{number}{6}) (\bibinfo{year}{2021})
  \bibinfo{pages}{64701}.

\bibitem[{Arrington et~al.(2021)}]{Arrington:2021biu}
\bibinfo{author}{J.~Arrington}, et~al., \bibinfo{title}{{Revealing the
  structure of light pseudoscalar mesons at the electron\textendash{}ion
  collider}}, \bibinfo{journal}{J. Phys. G} \bibinfo{volume}{48}
  (\bibinfo{year}{2021}) \bibinfo{pages}{075106}.

\bibitem[{Abdul~Khalek et~al.(2022)}]{AbdulKhalek:2021gbh}
\bibinfo{author}{R.~Abdul~Khalek}, et~al., \bibinfo{title}{{Science
  Requirements and Detector Concepts for the Electron-Ion Collider: EIC Yellow
  Report}}, \bibinfo{journal}{Nucl. Phys. A} \bibinfo{volume}{1026}
  (\bibinfo{year}{2022}) \bibinfo{pages}{122447}.

\bibitem[{Quintans(2022)}]{Quintans:2022utc}
\bibinfo{author}{C.~Quintans}, \bibinfo{title}{{The New AMBER Experiment at the
  CERN SPS}}, \bibinfo{journal}{Few Body Syst.}
  \bibinfo{volume}{63}~(\bibinfo{number}{4}) (\bibinfo{year}{2022})
  \bibinfo{pages}{72}.

\bibitem[{Carman et~al.(2023)Carman, Gothe, Mokeev, and
  Roberts}]{Carman:2023zke}
\bibinfo{author}{D.~S. Carman}, \bibinfo{author}{R.~W. Gothe},
  \bibinfo{author}{V.~I. Mokeev}, \bibinfo{author}{C.~D. Roberts},
  \bibinfo{title}{{Nucleon Resonance Electroexcitation Amplitudes and Emergent
  Hadron Mass}}, \bibinfo{journal}{Particles}
  \bibinfo{volume}{6}~(\bibinfo{number}{1}) (\bibinfo{year}{2023})
  \bibinfo{pages}{416--439}.

\bibitem[{Workman et~al.(2022)}]{Workman:2022ynf}
\bibinfo{author}{R.~L. Workman}, et~al., \bibinfo{title}{{Review of Particle
  Physics}}, \bibinfo{journal}{PTEP} \bibinfo{volume}{2022}
  (\bibinfo{year}{2022}) \bibinfo{pages}{083C01}.

\bibitem[{Binosi(2022)}]{Binosi:2022djx}
\bibinfo{author}{D.~Binosi}, \bibinfo{title}{{Emergent Hadron Mass in Strong
  Dynamics}}, \bibinfo{journal}{Few Body Syst.}
  \bibinfo{volume}{63}~(\bibinfo{number}{2}) (\bibinfo{year}{2022})
  \bibinfo{pages}{42}.

\bibitem[{Papavassiliou(2022)}]{Papavassiliou:2022wrb}
\bibinfo{author}{J.~Papavassiliou}, \bibinfo{title}{{Emergence of mass in the
  gauge sector of QCD}}, \bibinfo{journal}{Chin. Phys. C}
  \bibinfo{volume}{46}~(\bibinfo{number}{11}) (\bibinfo{year}{2022})
  \bibinfo{pages}{112001}.

\bibitem[{Ding et~al.(2023)Ding, Roberts, and Schmidt}]{Ding:2022ows}
\bibinfo{author}{M.~Ding}, \bibinfo{author}{C.~D. Roberts},
  \bibinfo{author}{S.~M. Schmidt}, \bibinfo{title}{{Emergence of Hadron Mass
  and Structure}}, \bibinfo{journal}{Particles}
  \bibinfo{volume}{6}~(\bibinfo{number}{1}) (\bibinfo{year}{2023})
  \bibinfo{pages}{57--120}.

\bibitem[{Ferreira and Papavassiliou(2023)}]{Ferreira:2023fva}
\bibinfo{author}{M.~N. Ferreira}, \bibinfo{author}{J.~Papavassiliou},
  \bibinfo{title}{{Gauge Sector Dynamics in QCD}}, \bibinfo{journal}{Particles}
  \bibinfo{volume}{6}~(\bibinfo{number}{1}) (\bibinfo{year}{2023})
  \bibinfo{pages}{312--363}.

\bibitem[{Lin et~al.(2018)}]{Lin:2017snn}
\bibinfo{author}{H.-W. Lin}, et~al., \bibinfo{title}{{Parton distributions and
  lattice QCD calculations: a community white paper}}, \bibinfo{journal}{Prog.
  Part. Nucl. Phys.} \bibinfo{volume}{100} (\bibinfo{year}{2018})
  \bibinfo{pages}{107--160}.

\bibitem[{Cui et~al.(2020{\natexlab{a}})Cui, Ding, Gao, Raya, Binosi, Chang,
  Roberts, Rodr\'{\i}guez-Quintero, and Schmidt}]{Cui:2020tdf}
\bibinfo{author}{Z.-F. Cui}, \bibinfo{author}{M.~Ding},
  \bibinfo{author}{F.~Gao}, \bibinfo{author}{K.~Raya},
  \bibinfo{author}{D.~Binosi}, \bibinfo{author}{L.~Chang},
  \bibinfo{author}{C.~D. Roberts},
  \bibinfo{author}{J.~Rodr\'{\i}guez-Quintero}, \bibinfo{author}{S.~M.
  Schmidt}, \bibinfo{title}{{Kaon and pion parton distributions}},
  \bibinfo{journal}{Eur. Phys. J. C} \bibinfo{volume}{80}
  (\bibinfo{year}{2020}{\natexlab{a}}) \bibinfo{pages}{1064}.

\bibitem[{Chang et~al.(2022)Chang, Gao, and Roberts}]{Chang:2022jri}
\bibinfo{author}{L.~Chang}, \bibinfo{author}{F.~Gao}, \bibinfo{author}{C.~D.
  Roberts}, \bibinfo{title}{{Parton distributions of light quarks and
  antiquarks in the proton}}, \bibinfo{journal}{Phys. Lett. B}
  \bibinfo{volume}{829} (\bibinfo{year}{2022}) \bibinfo{pages}{137078}.

\bibitem[{Lu et~al.(2022)Lu, Chang, Raya, Roberts, and
  Rodr\'\i{}guez-Quintero}]{Lu:2022cjx}
\bibinfo{author}{Y.~Lu}, \bibinfo{author}{L.~Chang}, \bibinfo{author}{K.~Raya},
  \bibinfo{author}{C.~D. Roberts},
  \bibinfo{author}{J.~Rodr\'\i{}guez-Quintero}, \bibinfo{title}{{Proton and
  pion distribution functions in counterpoint}}, \bibinfo{journal}{Phys. Lett.
  B} \bibinfo{volume}{830} (\bibinfo{year}{2022}) \bibinfo{pages}{137130}.

\bibitem[{Cheng et~al.(2023)Cheng, Yu, Xing, Chen, Cui, and
  Roberts}]{Cheng:2023kmt}
\bibinfo{author}{P.~Cheng}, \bibinfo{author}{Y.~Yu}, \bibinfo{author}{H.-Y.
  Xing}, \bibinfo{author}{C.~Chen}, \bibinfo{author}{Z.-F. Cui},
  \bibinfo{author}{C.~D. Roberts}, \bibinfo{title}{{Perspective on polarised
  parton distribution functions and proton spin}}, \bibinfo{journal}{Phys.
  Lett. B} \bibinfo{volume}{844} (\bibinfo{year}{2023})
  \bibinfo{pages}{138074}.

\bibitem[{Xing et~al.(2024)Xing, Yao, Li, Binosi, Cui, and
  Roberts}]{Xing:2023pms}
\bibinfo{author}{H.~Y. Xing}, \bibinfo{author}{Z.~Q. Yao},
  \bibinfo{author}{B.~L. Li}, \bibinfo{author}{D.~Binosi},
  \bibinfo{author}{Z.~F. Cui}, \bibinfo{author}{C.~D. Roberts},
  \bibinfo{title}{{Developing predictions for pion fragmentation functions}},
  \bibinfo{journal}{Eur. Phys. J. C} \bibinfo{volume}{84}~(\bibinfo{number}{1})
  (\bibinfo{year}{2024}) \bibinfo{pages}{82}.

\bibitem[{Chang and Roberts(2021)}]{Chang:2021utv}
\bibinfo{author}{L.~Chang}, \bibinfo{author}{C.~D. Roberts},
  \bibinfo{title}{{Regarding the distribution of glue in the pion}},
  \bibinfo{journal}{Chin. Phys. Lett.}
  \bibinfo{volume}{38}~(\bibinfo{number}{8}) (\bibinfo{year}{2021})
  \bibinfo{pages}{081101}.

\bibitem[{Cui et~al.(2022{\natexlab{a}})Cui, Ding, Morgado, Raya, Binosi,
  Chang, Papavassiliou, Roberts, Rodr\'\i{}guez-Quintero, and
  Schmidt}]{Cui:2021mom}
\bibinfo{author}{Z.~F. Cui}, \bibinfo{author}{M.~Ding}, \bibinfo{author}{J.~M.
  Morgado}, \bibinfo{author}{K.~Raya}, \bibinfo{author}{D.~Binosi},
  \bibinfo{author}{L.~Chang}, \bibinfo{author}{J.~Papavassiliou},
  \bibinfo{author}{C.~D. Roberts},
  \bibinfo{author}{J.~Rodr\'\i{}guez-Quintero}, \bibinfo{author}{S.~M.
  Schmidt}, \bibinfo{title}{{Concerning pion parton distributions}},
  \bibinfo{journal}{Eur. Phys. J. A} \bibinfo{volume}{58}~(\bibinfo{number}{1})
  (\bibinfo{year}{2022}{\natexlab{a}}) \bibinfo{pages}{10}.

\bibitem[{Cui et~al.(2022{\natexlab{b}})Cui, Ding, Morgado, Raya, Binosi,
  Chang, De~Soto, Roberts, Rodr\'\i{}guez-Quintero, and Schmidt}]{Cui:2022bxn}
\bibinfo{author}{Z.~F. Cui}, \bibinfo{author}{M.~Ding}, \bibinfo{author}{J.~M.
  Morgado}, \bibinfo{author}{K.~Raya}, \bibinfo{author}{D.~Binosi},
  \bibinfo{author}{L.~Chang}, \bibinfo{author}{F.~De~Soto},
  \bibinfo{author}{C.~D. Roberts},
  \bibinfo{author}{J.~Rodr\'\i{}guez-Quintero}, \bibinfo{author}{S.~M.
  Schmidt}, \bibinfo{title}{{Emergence of pion parton distributions}},
  \bibinfo{journal}{Phys. Rev. D} \bibinfo{volume}{105}~(\bibinfo{number}{9})
  (\bibinfo{year}{2022}{\natexlab{b}}) \bibinfo{pages}{L091502}.

\bibitem[{Lu et~al.(2024)Lu, Xu, Raya, Roberts, and
  Rodr\'\i{}guez-Quintero}]{Lu:2023yna}
\bibinfo{author}{Y.~Lu}, \bibinfo{author}{Y.-Z. Xu}, \bibinfo{author}{K.~Raya},
  \bibinfo{author}{C.~D. Roberts},
  \bibinfo{author}{J.~Rodr\'\i{}guez-Quintero}, \bibinfo{title}{{Pion
  distribution functions from low-order Mellin moments}},
  \bibinfo{journal}{Phys. Lett. B} \bibinfo{volume}{850} (\bibinfo{year}{2024})
  \bibinfo{pages}{138534}.

\bibitem[{Guti{\'e}rrez-Guerrero et~al.(2010)Guti{\'e}rrez-Guerrero, Bashir,
  Cloet, and Roberts}]{GutierrezGuerrero:2010md}
\bibinfo{author}{L.~X. Guti{\'e}rrez-Guerrero}, \bibinfo{author}{A.~Bashir},
  \bibinfo{author}{I.~C. Cloet}, \bibinfo{author}{C.~D. Roberts},
  \bibinfo{title}{{Pion form factor from a contact interaction}},
  \bibinfo{journal}{Phys. Rev. C} \bibinfo{volume}{81} (\bibinfo{year}{2010})
  \bibinfo{pages}{065202}.

\bibitem[{Roberts et~al.(2010)Roberts, Roberts, Bashir, Guti{\'e}rrez-Guerrero,
  and Tandy}]{Roberts:2010rn}
\bibinfo{author}{H.~L.~L. Roberts}, \bibinfo{author}{C.~D. Roberts},
  \bibinfo{author}{A.~Bashir}, \bibinfo{author}{L.~X. Guti{\'e}rrez-Guerrero},
  \bibinfo{author}{P.~C. Tandy}, \bibinfo{title}{{Abelian anomaly and neutral
  pion production}}, \bibinfo{journal}{Phys. Rev. C} \bibinfo{volume}{82}
  (\bibinfo{year}{2010}) \bibinfo{pages}{{\mbox{065202}}}.

\bibitem[{Roberts et~al.(2011)Roberts, Bashir, Guti{\'e}rrez-Guerrero, Roberts,
  and Wilson}]{Roberts:2011wy}
\bibinfo{author}{H.~L.~L. Roberts}, \bibinfo{author}{A.~Bashir},
  \bibinfo{author}{L.~X. Guti{\'e}rrez-Guerrero}, \bibinfo{author}{C.~D.
  Roberts}, \bibinfo{author}{D.~J. Wilson}, \bibinfo{title}{{$\pi$- and
  $\rho$-mesons, and their diquark partners, from a contact interaction}},
  \bibinfo{journal}{Phys. Rev. C} \bibinfo{volume}{83} (\bibinfo{year}{2011})
  \bibinfo{pages}{065206}.

\bibitem[{Wilson et~al.(2012)Wilson, Cloet, Chang, and Roberts}]{Wilson:2011aa}
\bibinfo{author}{D.~J. Wilson}, \bibinfo{author}{I.~C. Cloet},
  \bibinfo{author}{L.~Chang}, \bibinfo{author}{C.~D. Roberts},
  \bibinfo{title}{{Nucleon and Roper electromagnetic elastic and transition
  form factors}}, \bibinfo{journal}{Phys. Rev. C} \bibinfo{volume}{85}
  (\bibinfo{year}{2012}) \bibinfo{pages}{025205}.

\bibitem[{Chen et~al.(2013)Chen, Chang, Roberts, Wan, Schmidt, and
  Wilson}]{Chen:2012txa}
\bibinfo{author}{C.~Chen}, \bibinfo{author}{L.~Chang}, \bibinfo{author}{C.~D.
  Roberts}, \bibinfo{author}{S.-L. Wan}, \bibinfo{author}{S.~M. Schmidt},
  \bibinfo{author}{D.~J. Wilson}, \bibinfo{title}{{Features and flaws of a
  contact interaction treatment of the kaon}}, \bibinfo{journal}{Phys. Rev. C}
  \bibinfo{volume}{87} (\bibinfo{year}{2013}) \bibinfo{pages}{045207}.

\bibitem[{Segovia et~al.(2014)Segovia, Chen, Cloet, Roberts, Schmidt, and
  Wan}]{Segovia:2013uga}
\bibinfo{author}{J.~Segovia}, \bibinfo{author}{C.~Chen}, \bibinfo{author}{I.~C.
  Cloet}, \bibinfo{author}{C.~D. Roberts}, \bibinfo{author}{S.~M. Schmidt},
  \bibinfo{author}{S.-L. Wan}, \bibinfo{title}{{Elastic and transition form
  factors of the \mbox{$\Delta(1232)$}}}, \bibinfo{journal}{Few Body Syst.}
  \bibinfo{volume}{55} (\bibinfo{year}{2014}) \bibinfo{pages}{1--33}.

\bibitem[{Xu et~al.(2015)Xu, Chen, Cloet, Roberts, Segovia, and
  Zong}]{Xu:2015kta}
\bibinfo{author}{S.-S. Xu}, \bibinfo{author}{C.~Chen}, \bibinfo{author}{I.~C.
  Cloet}, \bibinfo{author}{C.~D. Roberts}, \bibinfo{author}{J.~Segovia},
  \bibinfo{author}{H.-S. Zong}, \bibinfo{title}{{Contact-interaction Faddeev
  equation and, \emph{inter alia}, proton tensor charges}},
  \bibinfo{journal}{Phys. Rev. D} \bibinfo{volume}{92} (\bibinfo{year}{2015})
  \bibinfo{pages}{114034}.

\bibitem[{Bedolla et~al.(2016)Bedolla, Raya, Cobos-Mart{\'\i}nez, and
  Bashir}]{Bedolla:2016yxq}
\bibinfo{author}{M.~A. Bedolla}, \bibinfo{author}{K.~Raya},
  \bibinfo{author}{J.~J. Cobos-Mart{\'\i}nez}, \bibinfo{author}{A.~Bashir},
  \bibinfo{title}{{\mbox{$\eta_c$} elastic and transition form factors: Contact
  interaction and algebraic model}}, \bibinfo{journal}{Phys. Rev. D}
  \bibinfo{volume}{93} (\bibinfo{year}{2016}) \bibinfo{pages}{094025}.

\bibitem[{Raya et~al.(2021)Raya, Guti\'errez-Guerrero, Bashir, Chang, Cui, Lu,
  Roberts, and Segovia}]{Raya:2021pyr}
\bibinfo{author}{K.~Raya}, \bibinfo{author}{L.~X. Guti\'errez-Guerrero},
  \bibinfo{author}{A.~Bashir}, \bibinfo{author}{L.~Chang},
  \bibinfo{author}{Z.~F. Cui}, \bibinfo{author}{Y.~Lu}, \bibinfo{author}{C.~D.
  Roberts}, \bibinfo{author}{J.~Segovia}, \bibinfo{title}{{Dynamical diquarks
  in the \mbox{$\gamma^{(\ast)} p\to N(1535)\tfrac{1}{2}^-$} transition}},
  \bibinfo{journal}{Eur. Phys. J. A} \bibinfo{volume}{57}~(\bibinfo{number}{9})
  (\bibinfo{year}{2021}) \bibinfo{pages}{266}.

\bibitem[{Xing et~al.(2022)Xing, Xu, Cui, Roberts, and Xu}]{Xing:2022sor}
\bibinfo{author}{H.-Y. Xing}, \bibinfo{author}{Z.-N. Xu},
  \bibinfo{author}{Z.-F. Cui}, \bibinfo{author}{C.~D. Roberts},
  \bibinfo{author}{C.~Xu}, \bibinfo{title}{{Heavy + heavy and heavy + light
  pseudoscalar to vector semileptonic transitions}}, \bibinfo{journal}{Eur.
  Phys. J. C} \bibinfo{volume}{82}~(\bibinfo{number}{10})
  (\bibinfo{year}{2022}) \bibinfo{pages}{889}.

\bibitem[{Cheng et~al.(2022)Cheng, Serna, Yao, Chen, Cui, and
  Roberts}]{Cheng:2022jxe}
\bibinfo{author}{P.~Cheng}, \bibinfo{author}{F.~E. Serna},
  \bibinfo{author}{Z.-Q. Yao}, \bibinfo{author}{C.~Chen},
  \bibinfo{author}{Z.-F. Cui}, \bibinfo{author}{C.~D. Roberts},
  \bibinfo{title}{{Contact interaction analysis of octet baryon axial-vector
  and pseudoscalar form factors}}, \bibinfo{journal}{Phys. Rev. D}
  \bibinfo{volume}{106}~(\bibinfo{number}{5}) (\bibinfo{year}{2022})
  \bibinfo{pages}{054031}.

\bibitem[{Xing and Chang(2023)}]{Xing:2022jtt}
\bibinfo{author}{Z.~Xing}, \bibinfo{author}{L.~Chang},
  \bibinfo{title}{{Symmetry preserving contact interaction treatment of the
  kaon}}, \bibinfo{journal}{Phys. Rev. D}
  \bibinfo{volume}{107}~(\bibinfo{number}{1}) (\bibinfo{year}{2023})
  \bibinfo{pages}{014019}.

\bibitem[{Lu et~al.(2021)Lu, Binosi, Ding, Roberts, Xing, and Xu}]{Lu:2021sgg}
\bibinfo{author}{Y.~Lu}, \bibinfo{author}{D.~Binosi},
  \bibinfo{author}{M.~Ding}, \bibinfo{author}{C.~D. Roberts},
  \bibinfo{author}{H.-Y. Xing}, \bibinfo{author}{C.~Xu},
  \bibinfo{title}{{Distribution amplitudes of light diquarks}},
  \bibinfo{journal}{Eur. Phys. J A (Lett)}
  \bibinfo{volume}{57}~(\bibinfo{number}{4}) (\bibinfo{year}{2021})
  \bibinfo{pages}{115}.

\bibitem[{Roberts et~al.(2013)Roberts, Holt, and Schmidt}]{Roberts:2013mja}
\bibinfo{author}{C.~D. Roberts}, \bibinfo{author}{R.~J. Holt},
  \bibinfo{author}{S.~M. Schmidt}, \bibinfo{title}{{Nucleon spin structure at
  very high $x$}}, \bibinfo{journal}{Phys. Lett. B} \bibinfo{volume}{727}
  (\bibinfo{year}{2013}) \bibinfo{pages}{249--254}.

\bibitem[{Barabanov et~al.(2021)}]{Barabanov:2020jvn}
\bibinfo{author}{M.~Y. Barabanov}, et~al., \bibinfo{title}{{Diquark
  Correlations in Hadron Physics: Origin, Impact and Evidence}},
  \bibinfo{journal}{Prog. Part. Nucl. Phys.} \bibinfo{volume}{116}
  (\bibinfo{year}{2021}) \bibinfo{pages}{103835}.

\bibitem[{Yin et~al.(2023)Yin, Xu, Cui, Roberts, and
  Rodr\'\i{}guez-Quintero}]{Yin:2023dbw}
\bibinfo{author}{P.-L. Yin}, \bibinfo{author}{Y.-Z. Xu}, \bibinfo{author}{Z.-F.
  Cui}, \bibinfo{author}{C.~D. Roberts},
  \bibinfo{author}{J.~Rodr\'\i{}guez-Quintero}, \bibinfo{title}{{All-Orders
  Evolution of Parton Distributions: Principle, Practice, and Predictions}},
  \bibinfo{journal}{Chin. Phys. Lett. \emph{Express}}
  \bibinfo{volume}{40}~(\bibinfo{number}{9}) (\bibinfo{year}{2023})
  \bibinfo{pages}{091201}.

\bibitem[{Chen et~al.(2012)Chen, Chang, Roberts, Wan, and Wilson}]{Chen:2012qr}
\bibinfo{author}{C.~Chen}, \bibinfo{author}{L.~Chang}, \bibinfo{author}{C.~D.
  Roberts}, \bibinfo{author}{S.-L. Wan}, \bibinfo{author}{D.~J. Wilson},
  \bibinfo{title}{{Spectrum of hadrons with strangeness}},
  \bibinfo{journal}{Few Body Syst.} \bibinfo{volume}{53} (\bibinfo{year}{2012})
  \bibinfo{pages}{293--326}.

\bibitem[{Chang et~al.(2013)Chang, Roberts, and Schmidt}]{Chang:2012cc}
\bibinfo{author}{L.~Chang}, \bibinfo{author}{C.~D. Roberts},
  \bibinfo{author}{S.~M. Schmidt}, \bibinfo{title}{{Dressed-quarks and the
  nucleon's axial charge}}, \bibinfo{journal}{Phys. Rev. C}
  \bibinfo{volume}{87} (\bibinfo{year}{2013}) \bibinfo{pages}{015203}.

\bibitem[{Brodsky et~al.(1995)Brodsky, Burkardt, and Schmidt}]{Brodsky:1994kg}
\bibinfo{author}{S.~J. Brodsky}, \bibinfo{author}{M.~Burkardt},
  \bibinfo{author}{I.~Schmidt}, \bibinfo{title}{{Perturbative QCD constraints
  on the shape of polarized quark and gluon distributions}},
  \bibinfo{journal}{Nucl. Phys. B} \bibinfo{volume}{441} (\bibinfo{year}{1995})
  \bibinfo{pages}{197--214}.

\bibitem[{Yuan(2004)}]{Yuan:2003fs}
\bibinfo{author}{F.~Yuan}, \bibinfo{title}{{Generalized parton distributions at
  $x \to 1$}}, \bibinfo{journal}{Phys. Rev. D} \bibinfo{volume}{69}
  (\bibinfo{year}{2004}) \bibinfo{pages}{051501}.

\bibitem[{Abrams et~al.(2022)}]{Abrams:2021xum}
\bibinfo{author}{D.~Abrams}, et~al., \bibinfo{title}{{Measurement of the
  Nucleon $F^n_2/F^p_2$ Structure Function Ratio by the Jefferson Lab MARATHON
  Tritium/Helium-3 Deep Inelastic Scattering Experiment}},
  \bibinfo{journal}{Phys. Rev. Lett.}
  \bibinfo{volume}{128}~(\bibinfo{number}{13}) (\bibinfo{year}{2022})
  \bibinfo{pages}{132003}.

\bibitem[{Holt and Roberts(2010)}]{Holt:2010vj}
\bibinfo{author}{R.~J. Holt}, \bibinfo{author}{C.~D. Roberts},
  \bibinfo{title}{{Distribution Functions of the Nucleon and Pion in the
  Valence Region}}, \bibinfo{journal}{Rev. Mod. Phys.} \bibinfo{volume}{82}
  (\bibinfo{year}{2010}) \bibinfo{pages}{2991--3044}.

\bibitem[{Cui et~al.(2021)Cui, Ding, Gao, Raya, Binosi, Chang, Roberts,
  Rodr\'{\i}guez-Quintero, and Schmidt}]{Cui:2020dlm}
\bibinfo{author}{Z.-F. Cui}, \bibinfo{author}{M.~Ding},
  \bibinfo{author}{F.~Gao}, \bibinfo{author}{K.~Raya},
  \bibinfo{author}{D.~Binosi}, \bibinfo{author}{L.~Chang},
  \bibinfo{author}{C.~D. Roberts},
  \bibinfo{author}{J.~Rodr\'{\i}guez-Quintero}, \bibinfo{author}{S.~M.
  Schmidt}, \bibinfo{title}{{Higgs modulation of emergent mass as revealed in
  kaon and pion parton distributions}}, \bibinfo{journal}{Eur. Phys. J. A
  (Lett.)} \bibinfo{volume}{57}~(\bibinfo{number}{1}) (\bibinfo{year}{2021})
  \bibinfo{pages}{5}.

\bibitem[{Close and Thomas(1988)}]{Close:1988br}
\bibinfo{author}{F.~E. Close}, \bibinfo{author}{A.~W. Thomas},
  \bibinfo{title}{{The Spin and Flavor Dependence of Parton Distribution
  Functions}}, \bibinfo{journal}{Phys. Lett. B} \bibinfo{volume}{212}
  (\bibinfo{year}{1988}) \bibinfo{pages}{227}.

\bibitem[{Hughes and Voss(1999)}]{Hughes:1999wr}
\bibinfo{author}{E.~W. Hughes}, \bibinfo{author}{R.~Voss},
  \bibinfo{title}{{Spin structure functions}}, \bibinfo{journal}{Ann. Rev.
  Nucl. Part. Sci.} \bibinfo{volume}{49} (\bibinfo{year}{1999})
  \bibinfo{pages}{303--339}.

\bibitem[{Farrar and Jackson(1975)}]{Farrar:1975yb}
\bibinfo{author}{G.~R. Farrar}, \bibinfo{author}{D.~R. Jackson},
  \bibinfo{title}{{Pion and Nucleon Structure Functions Near $x=1$}},
  \bibinfo{journal}{Phys. Rev. Lett.} \bibinfo{volume}{35}
  (\bibinfo{year}{1975}) \bibinfo{pages}{1416}.

\bibitem[{Airapetian et~al.(2005)}]{HERMES:2004zsh}
\bibinfo{author}{A.~Airapetian}, et~al., \bibinfo{title}{{Quark helicity
  distributions in the nucleon for up, down, and strange quarks from
  semi-inclusive deep-inelastic scattering}}, \bibinfo{journal}{Phys. Rev. D}
  \bibinfo{volume}{71} (\bibinfo{year}{2005}) \bibinfo{pages}{012003}.

\bibitem[{Alekseev et~al.(2010)}]{COMPASS:2010hwr}
\bibinfo{author}{M.~G. Alekseev}, et~al., \bibinfo{title}{{Quark helicity
  distributions from longitudinal spin asymmetries in muon-proton and
  muon-deuteron scattering}}, \bibinfo{journal}{Phys. Lett. B}
  \bibinfo{volume}{693} (\bibinfo{year}{2010}) \bibinfo{pages}{227--235}.

\bibitem[{Dharmawardane et~al.(2006)}]{CLAS:2006ozz}
\bibinfo{author}{K.~V. Dharmawardane}, et~al., \bibinfo{title}{{Measurement of
  the $x$- and $Q^2$-dependence of the asymmetry $A_1$ on the nucleon}},
  \bibinfo{journal}{Phys. Lett. B} \bibinfo{volume}{641} (\bibinfo{year}{2006})
  \bibinfo{pages}{11--17}.

\bibitem[{Prok et~al.(2009)}]{CLAS:2008xos}
\bibinfo{author}{Y.~Prok}, et~al., \bibinfo{title}{{Moments of the Spin
  Structure Functions $g_1^p$ and $g_1^d$ for $0.05 < Q^2 < 3$-GeV$^2$}},
  \bibinfo{journal}{Phys. Lett. B} \bibinfo{volume}{672} (\bibinfo{year}{2009})
  \bibinfo{pages}{12--16}.

\bibitem[{Guler et~al.(2015)}]{CLAS:2015otq}
\bibinfo{author}{N.~Guler}, et~al., \bibinfo{title}{{Precise determination of
  the deuteron spin structure at low to moderate $Q^2$ with CLAS and extraction
  of the neutron contribution}}, \bibinfo{journal}{Phys. Rev. C}
  \bibinfo{volume}{92}~(\bibinfo{number}{5}) (\bibinfo{year}{2015})
  \bibinfo{pages}{055201}.

\bibitem[{Fersch et~al.(2017)}]{CLAS:2017qga}
\bibinfo{author}{R.~Fersch}, et~al., \bibinfo{title}{{Determination of the
  Proton Spin Structure Functions for $0.05 < Q^{2} < 5\, {\rm GeV}^{2}$ using
  CLAS}}, \bibinfo{journal}{Phys. Rev. C}
  \bibinfo{volume}{96}~(\bibinfo{number}{6}) (\bibinfo{year}{2017})
  \bibinfo{pages}{065208}.

\bibitem[{Parno et~al.(2015)}]{JeffersonLabHallA:2014mam}
\bibinfo{author}{D.~S. Parno}, et~al., \bibinfo{title}{{Precision Measurements
  of $A_1^n$ in the Deep Inelastic Regime}}, \bibinfo{journal}{Phys. Lett. B}
  \bibinfo{volume}{744} (\bibinfo{year}{2015}) \bibinfo{pages}{309--314}.

\bibitem[{Zheng et~al.(2004{\natexlab{a}})}]{JeffersonLabHallA:2003joy}
\bibinfo{author}{X.~Zheng}, et~al., \bibinfo{title}{{Precision measurement of
  the neutron spin asymmetry $A_1^N$ and spin flavor decomposition in the
  valence quark region}}, \bibinfo{journal}{Phys. Rev. Lett.}
  \bibinfo{volume}{92} (\bibinfo{year}{2004}{\natexlab{a}})
  \bibinfo{pages}{012004}.

\bibitem[{Zheng et~al.(2004{\natexlab{b}})}]{JeffersonLabHallA:2004tea}
\bibinfo{author}{X.~Zheng}, et~al., \bibinfo{title}{{Precision measurement of
  the neutron spin asymmetries and spin-dependent structure functions in the
  valence quark region}}, \bibinfo{journal}{Phys. Rev. C} \bibinfo{volume}{70}
  (\bibinfo{year}{2004}{\natexlab{b}}) \bibinfo{pages}{065207}.

\bibitem[{Kuhn et~al.( 109)}]{E1206109}
\bibinfo{author}{S.~Kuhn}, et~al., \bibinfo{note}{{\emph{The Longitudinal Spin
  Structure of the Nucleon}}}, \bibinfo{year}{CLAS Collaboration (E12-06-109)}.

\bibitem[{Zheng et~al.(tion)}]{Zheng:2006}
\bibinfo{author}{X.~Zheng}, et~al., \bibinfo{note}{{\emph{Measurement of
  Neutron Spin Asymmetry A$_1^n$ in the Valence Quark Region using an 11 GeV
  Beam and a Polarized $^3$He Target in Hall C}}}, \bibinfo{year}{E12-06-110
  Collaboration}.

\bibitem[{Bali et~al.(2016)}]{Bali:2015ykx}
\bibinfo{author}{G.~S. Bali}, et~al., \bibinfo{title}{{Light-cone distribution
  amplitudes of the baryon octet}}, \bibinfo{journal}{JHEP}
  \bibinfo{volume}{02} (\bibinfo{year}{2016}) \bibinfo{pages}{070}.

\bibitem[{Mezrag et~al.(2018)Mezrag, Segovia, Chang, and
  Roberts}]{Mezrag:2017znp}
\bibinfo{author}{C.~Mezrag}, \bibinfo{author}{J.~Segovia},
  \bibinfo{author}{L.~Chang}, \bibinfo{author}{C.~D. Roberts},
  \bibinfo{title}{{Parton distribution amplitudes: Revealing correlations
  within the proton and Roper}}, \bibinfo{journal}{Phys. Lett. B}
  \bibinfo{volume}{783} (\bibinfo{year}{2018}) \bibinfo{pages}{263--267}.

\bibitem[{Lepage and Brodsky(1980)}]{Lepage:1980fj}
\bibinfo{author}{G.~P. Lepage}, \bibinfo{author}{S.~J. Brodsky},
  \bibinfo{title}{{Exclusive Processes in Perturbative Quantum
  Chromodynamics}}, \bibinfo{journal}{Phys. Rev. D} \bibinfo{volume}{22}
  (\bibinfo{year}{1980}) \bibinfo{pages}{2157--2198}.

\bibitem[{Cui et~al.(2022{\natexlab{c}})Cui, Gao, Binosi, Chang, Roberts, and
  Schmidt}]{Cui:2021gzg}
\bibinfo{author}{Z.-F. Cui}, \bibinfo{author}{F.~Gao},
  \bibinfo{author}{D.~Binosi}, \bibinfo{author}{L.~Chang},
  \bibinfo{author}{C.~D. Roberts}, \bibinfo{author}{S.~M. Schmidt},
  \bibinfo{title}{{Valence quark ratio in the proton}}, \bibinfo{journal}{Chin.
  Phys. Lett. \emph{Express}} \bibinfo{volume}{39}~(\bibinfo{number}{04})
  (\bibinfo{year}{2022}{\natexlab{c}}) \bibinfo{pages}{041401}.

\bibitem[{Grunberg(1980)}]{Grunberg:1980ja}
\bibinfo{author}{G.~Grunberg}, \bibinfo{title}{{Renormalization Group Improved
  Perturbative QCD}}, \bibinfo{journal}{Phys. Lett. B} \bibinfo{volume}{95}
  (\bibinfo{year}{1980}) \bibinfo{pages}{70}, \bibinfo{note}{[Erratum: Phys.
  Lett. B 110, 501 (1982)]}.

\bibitem[{Grunberg(1984)}]{Grunberg:1982fw}
\bibinfo{author}{G.~Grunberg}, \bibinfo{title}{{Renormalization Scheme
  Independent QCD and QED: The Method of Effective Charges}},
  \bibinfo{journal}{Phys. Rev. D} \bibinfo{volume}{29} (\bibinfo{year}{1984})
  \bibinfo{pages}{2315}.

\bibitem[{Deur et~al.(2024)Deur, Brodsky, and Roberts}]{Deur:2023dzc}
\bibinfo{author}{A.~Deur}, \bibinfo{author}{S.~J. Brodsky},
  \bibinfo{author}{C.~D. Roberts}, \bibinfo{title}{{QCD Running Couplings and
  Effective Charges}}, \bibinfo{journal}{Prog. Part. Nucl. Phys.}
  \bibinfo{volume}{134} (\bibinfo{year}{2024}) \bibinfo{pages}{104081}.

\bibitem[{Binosi et~al.(2017)Binosi, Mezrag, Papavassiliou, Roberts, and
  Rodr{\'i}guez-Quintero}]{Binosi:2016nme}
\bibinfo{author}{D.~Binosi}, \bibinfo{author}{C.~Mezrag},
  \bibinfo{author}{J.~Papavassiliou}, \bibinfo{author}{C.~D. Roberts},
  \bibinfo{author}{J.~Rodr{\'i}guez-Quintero},
  \bibinfo{title}{{Process-independent strong running coupling}},
  \bibinfo{journal}{Phys. Rev. D} \bibinfo{volume}{96} (\bibinfo{year}{2017})
  \bibinfo{pages}{054026}.

\bibitem[{Cui et~al.(2020{\natexlab{b}})Cui, Zhang, Binosi, de~Soto, Mezrag,
  Papavassiliou, Roberts, Rodr{\'{\i}}guez-Quintero, Segovia, and
  Zafeiropoulos}]{Cui:2019dwv}
\bibinfo{author}{Z.-F. Cui}, \bibinfo{author}{J.-L. Zhang},
  \bibinfo{author}{D.~Binosi}, \bibinfo{author}{F.~de~Soto},
  \bibinfo{author}{C.~Mezrag}, \bibinfo{author}{J.~Papavassiliou},
  \bibinfo{author}{C.~D. Roberts},
  \bibinfo{author}{J.~Rodr{\'{\i}}guez-Quintero}, \bibinfo{author}{J.~Segovia},
  \bibinfo{author}{S.~Zafeiropoulos}, \bibinfo{title}{{Effective charge from
  lattice QCD}}, \bibinfo{journal}{Chin. Phys. C} \bibinfo{volume}{44}
  (\bibinfo{year}{2020}{\natexlab{b}}) \bibinfo{pages}{083102}.

\bibitem[{Brodsky et~al.(2024)Brodsky, Deur, and Roberts}]{Brodsky:2024zev}
\bibinfo{author}{S.~J. Brodsky}, \bibinfo{author}{A.~Deur},
  \bibinfo{author}{C.~D. Roberts}, \bibinfo{title}{{The Secret to the Strongest
  Force in the Universe}}, \bibinfo{journal}{Sci. Am.} \bibinfo{volume}{5
  (May)} (\bibinfo{year}{2024}) \bibinfo{pages}{32--39}.

\bibitem[{Lin et~al.(2020)Lin, Chen, and Zhang}]{Lin:2020fsj}
\bibinfo{author}{H.-W. Lin}, \bibinfo{author}{J.-W. Chen},
  \bibinfo{author}{R.~Zhang}, \bibinfo{title}{{Lattice$\!$ Nucleon$\!$
  Isovector$\!$ Unpolarized$\!$ Parton$\!$ Distribution$\!$ in$\!$ the$\!$
  Physical-Continuum Limit -- arXiv:2011.14971 [hep-lat]}} .

\bibitem[{Alexandrou et~al.(2021)Alexandrou, Constantinou, Hadjiyiannakou,
  Jansen, and Manigrasso}]{Alexandrou:2021oih}
\bibinfo{author}{C.~Alexandrou}, \bibinfo{author}{M.~Constantinou},
  \bibinfo{author}{K.~Hadjiyiannakou}, \bibinfo{author}{K.~Jansen},
  \bibinfo{author}{F.~Manigrasso}, \bibinfo{title}{{Flavor decomposition of the
  nucleon unpolarized, helicity, and transversity parton distribution functions
  from lattice QCD simulations}}, \bibinfo{journal}{Phys. Rev. D}
  \bibinfo{volume}{104}~(\bibinfo{number}{5}) (\bibinfo{year}{2021})
  \bibinfo{pages}{054503}.

\bibitem[{Alexandrou et~al.(2013)Alexandrou, Constantinou, Drach,
  Hatziyiannakou, Jansen, Kallidonis, Koutsou, Leontiou, and
  Vaquero}]{Alexandrou:2013cda}
\bibinfo{author}{C.~Alexandrou}, \bibinfo{author}{M.~Constantinou},
  \bibinfo{author}{V.~Drach}, \bibinfo{author}{K.~Hatziyiannakou},
  \bibinfo{author}{K.~Jansen}, \bibinfo{author}{C.~Kallidonis},
  \bibinfo{author}{G.~Koutsou}, \bibinfo{author}{T.~Leontiou},
  \bibinfo{author}{A.~Vaquero}, \bibinfo{title}{{Nucleon Structure using
  lattice QCD}}, \bibinfo{journal}{Nuovo Cim. C}
  \bibinfo{volume}{036}~(\bibinfo{number}{05}) (\bibinfo{year}{2013})
  \bibinfo{pages}{111--120}.

\bibitem[{Dove et~al.(2021)}]{SeaQuest:2021zxb}
\bibinfo{author}{J.~Dove}, et~al., \bibinfo{title}{{The asymmetry of antimatter
  in the proton}}, \bibinfo{journal}{Nature}
  \bibinfo{volume}{590}~(\bibinfo{number}{7847}) (\bibinfo{year}{2021})
  \bibinfo{pages}{561--565}.

\bibitem[{Amaudruz et~al.(1991)}]{NewMuon:1991hlj}
\bibinfo{author}{P.~Amaudruz}, et~al., \bibinfo{title}{{The Gottfried sum from
  the ratio $F_2^n/F_2^p$}}, \bibinfo{journal}{Phys. Rev. Lett.}
  \bibinfo{volume}{66} (\bibinfo{year}{1991}) \bibinfo{pages}{2712--2715}.

\bibitem[{Arneodo et~al.(1994)}]{NewMuon:1993oys}
\bibinfo{author}{M.~Arneodo}, et~al., \bibinfo{title}{{A Reevaluation of the
  Gottfried sum}}, \bibinfo{journal}{Phys. Rev. D} \bibinfo{volume}{50}
  (\bibinfo{year}{1994}) \bibinfo{pages}{R1--R3}.

\bibitem[{Baldit et~al.(1994)}]{NA51:1994xrz}
\bibinfo{author}{A.~Baldit}, et~al., \bibinfo{title}{{Study of the isospin
  symmetry breaking in the light quark sea of the nucleon from the Drell-Yan
  process}}, \bibinfo{journal}{Phys. Lett. B} \bibinfo{volume}{332}
  (\bibinfo{year}{1994}) \bibinfo{pages}{244--250}.

\bibitem[{Towell et~al.(2001)}]{NuSea:2001idv}
\bibinfo{author}{R.~S. Towell}, et~al., \bibinfo{title}{{Improved measurement
  of the $\bar d/\bar u$ asymmetry in the nucleon sea}},
  \bibinfo{journal}{Phys. Rev. D} \bibinfo{volume}{64} (\bibinfo{year}{2001})
  \bibinfo{pages}{052002}.

\bibitem[{Gottfried(1967)}]{Gottfried:1967kk}
\bibinfo{author}{K.~Gottfried}, \bibinfo{title}{{Sum rule for high-energy
  electron - proton scattering}}, \bibinfo{journal}{Phys. Rev. Lett.}
  \bibinfo{volume}{18} (\bibinfo{year}{1967}) \bibinfo{pages}{1174}.

\bibitem[{Brock et~al.(1995)}]{Brock:1993sz}
\bibinfo{author}{R.~Brock}, et~al., \bibinfo{title}{{Handbook of perturbative
  QCD: Version 1.0}}, \bibinfo{journal}{Rev. Mod. Phys.} \bibinfo{volume}{67}
  (\bibinfo{year}{1995}) \bibinfo{pages}{157--248}.

\bibitem[{Hou et~al.(2021)}]{Hou:2019efy}
\bibinfo{author}{T.-J. Hou}, et~al., \bibinfo{title}{{New CTEQ global analysis
  of quantum chromodynamics with high-precision data from the LHC}},
  \bibinfo{journal}{Phys. Rev. D} \bibinfo{volume}{103}~(\bibinfo{number}{1})
  (\bibinfo{year}{2021}) \bibinfo{pages}{014013}.

\bibitem[{Field and Feynman(1977)}]{Field:1976ve}
\bibinfo{author}{R.~D. Field}, \bibinfo{author}{R.~P. Feynman},
  \bibinfo{title}{{Quark Elastic Scattering as a Source of High Transverse
  Momentum Mesons}}, \bibinfo{journal}{Phys. Rev. D} \bibinfo{volume}{15}
  (\bibinfo{year}{1977}) \bibinfo{pages}{2590--2616}.

\bibitem[{Tkachenko et~al.(2014)}]{CLAS:2014jvt}
\bibinfo{author}{S.~Tkachenko}, et~al., \bibinfo{title}{{Measurement of the
  structure function of the nearly free neutron using spectator tagging in
  inelastic $^2$H(e, e'p)X scattering with CLAS}}, \bibinfo{journal}{Phys. Rev.
  C} \bibinfo{volume}{89} (\bibinfo{year}{2014}) \bibinfo{pages}{045206},
  \bibinfo{note}{[Addendum: Phys. Rev. C 90 (2014) 059901]}.

\bibitem[{Brodsky et~al.(1980)Brodsky, Hoyer, Peterson, and
  Sakai}]{Brodsky:1980pb}
\bibinfo{author}{S.~J. Brodsky}, \bibinfo{author}{P.~Hoyer},
  \bibinfo{author}{C.~Peterson}, \bibinfo{author}{N.~Sakai},
  \bibinfo{title}{{The Intrinsic Charm of the Proton}}, \bibinfo{journal}{Phys.
  Lett. B} \bibinfo{volume}{93} (\bibinfo{year}{1980})
  \bibinfo{pages}{451--455}.

\bibitem[{Ball et~al.(2022{\natexlab{a}})Ball, Candido, Cruz-Martinez, Forte,
  Giani, Hekhorn, Kudashkin, Magni, and Rojo}]{Ball:2022qks}
\bibinfo{author}{R.~D. Ball}, \bibinfo{author}{A.~Candido},
  \bibinfo{author}{J.~Cruz-Martinez}, \bibinfo{author}{S.~Forte},
  \bibinfo{author}{T.~Giani}, \bibinfo{author}{F.~Hekhorn},
  \bibinfo{author}{K.~Kudashkin}, \bibinfo{author}{G.~Magni},
  \bibinfo{author}{J.~Rojo}, \bibinfo{title}{{Evidence for intrinsic charm
  quarks in the proton}}, \bibinfo{journal}{Nature}
  \bibinfo{volume}{608}~(\bibinfo{number}{7923})
  (\bibinfo{year}{2022}{\natexlab{a}}) \bibinfo{pages}{483--487}.

\bibitem[{Ball et~al.(2022{\natexlab{b}})}]{NNPDF:2021njg}
\bibinfo{author}{R.~D. Ball}, et~al., \bibinfo{title}{{The path to proton
  structure at 1\% accuracy}}, \bibinfo{journal}{Eur. Phys. J. C}
  \bibinfo{volume}{82}~(\bibinfo{number}{5})
  (\bibinfo{year}{2022}{\natexlab{b}}) \bibinfo{pages}{428}.

\bibitem[{Prok et~al.(2014)}]{CLAS:2014qtg}
\bibinfo{author}{Y.~Prok}, et~al., \bibinfo{title}{{Precision measurements of
  $g_1$ of the proton and the deuteron with 6 GeV electrons}},
  \bibinfo{journal}{Phys. Rev. C} \bibinfo{volume}{90}~(\bibinfo{number}{2})
  (\bibinfo{year}{2014}) \bibinfo{pages}{025212}.

\bibitem[{Anthony et~al.(1999)}]{E155:1999pwm}
\bibinfo{author}{P.~L. Anthony}, et~al., \bibinfo{title}{{Measurement of the
  deuteron spin structure function $g_1^d(x)$ for 1-(GeV/c)$^2< Q^2<
  40$-(GeV/c)$^2$}}, \bibinfo{journal}{Phys. Lett. B} \bibinfo{volume}{463}
  (\bibinfo{year}{1999}) \bibinfo{pages}{339--345}.

\bibitem[{Anthony et~al.(2000)}]{E155:2000qdr}
\bibinfo{author}{P.~L. Anthony}, et~al., \bibinfo{title}{{Measurements of the
  $Q^2$ dependence of the proton and neutron spin structure functions $g_1^p$
  and $g_1^n$}}, \bibinfo{journal}{Phys. Lett. B} \bibinfo{volume}{493}
  (\bibinfo{year}{2000}) \bibinfo{pages}{19--28}.

\bibitem[{Airapetian et~al.(1998)}]{HERMES:1998cbu}
\bibinfo{author}{A.~Airapetian}, et~al., \bibinfo{title}{{Measurement of the
  proton spin structure function $g_1^p$ with a pure hydrogen target}},
  \bibinfo{journal}{Phys. Lett. B} \bibinfo{volume}{442} (\bibinfo{year}{1998})
  \bibinfo{pages}{484--492}.

\bibitem[{Abe et~al.(1995)}]{E143:1995clm}
\bibinfo{author}{K.~Abe}, et~al., \bibinfo{title}{{Measurements of the $Q^2$
  dependence of the proton and deuteron spin structure functions $g_1^p$ and
  $g_1^d$}}, \bibinfo{journal}{Phys. Lett. B} \bibinfo{volume}{364}
  (\bibinfo{year}{1995}) \bibinfo{pages}{61--68}.

\bibitem[{Abe et~al.(1997{\natexlab{a}})}]{E143:1996vck}
\bibinfo{author}{K.~Abe}, et~al., \bibinfo{title}{{Measurements of the proton
  and deuteron spin structure function $g_1$ in the resonance region}},
  \bibinfo{journal}{Phys. Rev. Lett.} \bibinfo{volume}{78}
  (\bibinfo{year}{1997}{\natexlab{a}}) \bibinfo{pages}{815--819}.

\bibitem[{Abe et~al.(1998)}]{E143:1998hbs}
\bibinfo{author}{K.~Abe}, et~al., \bibinfo{title}{{Measurements of the proton
  and deuteron spin structure functions $g_1$ and $g_2$}},
  \bibinfo{journal}{Phys. Rev. D} \bibinfo{volume}{58} (\bibinfo{year}{1998})
  \bibinfo{pages}{112003}.

\bibitem[{Adams et~al.(1994)}]{SpinMuonSMC:1994met}
\bibinfo{author}{D.~Adams}, et~al., \bibinfo{title}{{Measurement of the spin
  dependent structure function $g_1(x)$ of the proton}},
  \bibinfo{journal}{Phys. Lett. B} \bibinfo{volume}{329} (\bibinfo{year}{1994})
  \bibinfo{pages}{399--406}, \bibinfo{note}{[Erratum: Phys. Lett. B 339,
  332--333 (1994)]}.

\bibitem[{Adams et~al.(1997)}]{SpinMuonSMC:1997mkb}
\bibinfo{author}{D.~Adams}, et~al., \bibinfo{title}{{Spin structure of the
  proton from polarized inclusive deep inelastic muon - proton scattering}},
  \bibinfo{journal}{Phys. Rev. D} \bibinfo{volume}{56} (\bibinfo{year}{1997})
  \bibinfo{pages}{5330--5358}.

\bibitem[{Ackerstaff et~al.(1997)}]{HERMES:1997hjr}
\bibinfo{author}{K.~Ackerstaff}, et~al., \bibinfo{title}{{Measurement of the
  neutron spin structure function $g_1^n$ with a polarized $\,^3$He internal
  target}}, \bibinfo{journal}{Phys. Lett. B} \bibinfo{volume}{404}
  (\bibinfo{year}{1997}) \bibinfo{pages}{383--389}.

\bibitem[{Abe et~al.(1997{\natexlab{b}})}]{E154:1997xfa}
\bibinfo{author}{K.~Abe}, et~al., \bibinfo{title}{{Precision determination of
  the neutron spin structure function $g_1^n$}}, \bibinfo{journal}{Phys. Rev.
  Lett.} \bibinfo{volume}{79} (\bibinfo{year}{1997}{\natexlab{b}})
  \bibinfo{pages}{26--30}.

\bibitem[{Abe et~al.(1997{\natexlab{c}})}]{E154:1997ysl}
\bibinfo{author}{K.~Abe}, et~al., \bibinfo{title}{{Next-to-leading order QCD
  analysis of polarized deep inelastic scattering data}},
  \bibinfo{journal}{Phys. Lett. B} \bibinfo{volume}{405}
  (\bibinfo{year}{1997}{\natexlab{c}}) \bibinfo{pages}{180--190}.

\bibitem[{Anthony et~al.(1996)}]{E142:1996thl}
\bibinfo{author}{P.~L. Anthony}, et~al., \bibinfo{title}{{Deep inelastic
  scattering of polarized electrons by polarized $\,^3$He and the study of the
  neutron spin structure}}, \bibinfo{journal}{Phys. Rev. D}
  \bibinfo{volume}{54} (\bibinfo{year}{1996}) \bibinfo{pages}{6620--6650}.

\bibitem[{Ellis et~al.(1991)Ellis, Stirling, and Webber}]{Ellis:1991qj}
\bibinfo{author}{R.~K. Ellis}, \bibinfo{author}{W.~J. Stirling},
  \bibinfo{author}{B.~R. Webber}, \bibinfo{title}{{\mbox{$\;$}QCD and collider
  physics}}, \bibinfo{publisher}{Cambridge University Press, Cambridge, UK},
  \bibinfo{year}{1991}.

\bibitem[{Xu et~al.(2023)Xu, Mondal, Zhao, Li, and Vary}]{Xu:2023nqv}
\bibinfo{author}{S.~Xu}, \bibinfo{author}{C.~Mondal},
  \bibinfo{author}{X.~Zhao}, \bibinfo{author}{Y.~Li}, \bibinfo{author}{J.~P.
  Vary}, \bibinfo{title}{{Quark and gluon spin and orbital angular momentum in
  the proton}}, \bibinfo{journal}{Phys. Rev. D}
  \bibinfo{volume}{108}~(\bibinfo{number}{9}) (\bibinfo{year}{2023})
  \bibinfo{pages}{094002}.

\bibitem[{de~Florian et~al.(2014)de~Florian, Sassot, Stratmann, and
  Vogelsang}]{deFlorian:2014yva}
\bibinfo{author}{D.~de~Florian}, \bibinfo{author}{R.~Sassot},
  \bibinfo{author}{M.~Stratmann}, \bibinfo{author}{W.~Vogelsang},
  \bibinfo{title}{{Evidence for polarization of gluons in the proton}},
  \bibinfo{journal}{Phys. Rev. Lett.}
  \bibinfo{volume}{113}~(\bibinfo{number}{1}) (\bibinfo{year}{2014})
  \bibinfo{pages}{012001}.

\bibitem[{Adolph et~al.(2017{\natexlab{a}})}]{COMPASS:2015pim}
\bibinfo{author}{C.~Adolph}, et~al., \bibinfo{title}{{Leading-order
  determination of the gluon polarisation from semi-inclusive deep inelastic
  scattering data}}, \bibinfo{journal}{Eur. Phys. J. C}
  \bibinfo{volume}{77}~(\bibinfo{number}{4})
  (\bibinfo{year}{2017}{\natexlab{a}}) \bibinfo{pages}{209}.

\bibitem[{Altarelli and Ross(1988)}]{Altarelli:1988nr}
\bibinfo{author}{G.~Altarelli}, \bibinfo{author}{G.~G. Ross},
  \bibinfo{title}{{The Anomalous Gluon Contribution to Polarized
  Leptoproduction}}, \bibinfo{journal}{Phys. Lett. B} \bibinfo{volume}{212}
  (\bibinfo{year}{1988}) \bibinfo{pages}{391--396}.

\bibitem[{Adolph et~al.(2017{\natexlab{b}})}]{COMPASS:2016jwv}
\bibinfo{author}{C.~Adolph}, et~al., \bibinfo{title}{{Final COMPASS results on
  the deuteron spin-dependent structure function $g_1^{\rm d}$ and the Bjorken
  sum rule}}, \bibinfo{journal}{Phys. Lett. B} \bibinfo{volume}{769}
  (\bibinfo{year}{2017}{\natexlab{b}}) \bibinfo{pages}{34--41}.

\bibitem[{Jaffe and Manohar(1990)}]{Jaffe:1989jz}
\bibinfo{author}{R.~L. Jaffe}, \bibinfo{author}{A.~Manohar},
  \bibinfo{title}{{The G(1) Problem: Fact and Fantasy on the Spin of the
  Proton}}, \bibinfo{journal}{Nucl. Phys. B} \bibinfo{volume}{337}
  (\bibinfo{year}{1990}) \bibinfo{pages}{509--546}.

\bibitem[{Bhagwat et~al.(2007)Bhagwat, Chang, Liu, Roberts, and
  Tandy}]{Bhagwat:2007ha}
\bibinfo{author}{M.~S. Bhagwat}, \bibinfo{author}{L.~Chang},
  \bibinfo{author}{Y.-X. Liu}, \bibinfo{author}{C.~D. Roberts},
  \bibinfo{author}{P.~C. Tandy}, \bibinfo{title}{{Flavour symmetry breaking and
  meson masses}}, \bibinfo{journal}{Phys. Rev. C} \bibinfo{volume}{76}
  (\bibinfo{year}{2007}) \bibinfo{pages}{045203}.

\bibitem[{Raya et~al.(2017)Raya, Ding, Bashir, Chang, and
  Roberts}]{Raya:2016yuj}
\bibinfo{author}{K.~Raya}, \bibinfo{author}{M.~Ding},
  \bibinfo{author}{A.~Bashir}, \bibinfo{author}{L.~Chang},
  \bibinfo{author}{C.~D. Roberts}, \bibinfo{title}{{Partonic structure of
  neutral pseudoscalars via two photon transition form factors}},
  \bibinfo{journal}{Phys. Rev. D} \bibinfo{volume}{95} (\bibinfo{year}{2017})
  \bibinfo{pages}{074014}.

\bibitem[{Ding et~al.(2019)Ding, Raya, Bashir, Binosi, Chang, Chen, and
  Roberts}]{Ding:2018xwy}
\bibinfo{author}{M.~Ding}, \bibinfo{author}{K.~Raya},
  \bibinfo{author}{A.~Bashir}, \bibinfo{author}{D.~Binosi},
  \bibinfo{author}{L.~Chang}, \bibinfo{author}{M.~Chen}, \bibinfo{author}{C.~D.
  Roberts}, \bibinfo{title}{{$\gamma^\ast \gamma \to \eta, \eta^\prime$
  transition form factors}}, \bibinfo{journal}{Phys. Rev. D}
  \bibinfo{volume}{99} (\bibinfo{year}{2019}) \bibinfo{pages}{014014}.

\bibitem[{Bass(2002)}]{Bass:2001dg}
\bibinfo{author}{S.~D. Bass}, \bibinfo{title}{{Spin structure in nonforward
  partons}}, \bibinfo{journal}{Phys. Rev. D} \bibinfo{volume}{65}
  (\bibinfo{year}{2002}) \bibinfo{pages}{074025}.

\bibitem[{Hatta and Yoshida(2012)}]{Hatta:2012cs}
\bibinfo{author}{Y.~Hatta}, \bibinfo{author}{S.~Yoshida},
  \bibinfo{title}{{Twist analysis of the nucleon spin in QCD}},
  \bibinfo{journal}{JHEP} \bibinfo{volume}{10} (\bibinfo{year}{2012})
  \bibinfo{pages}{080}.

\bibitem[{Bhattacharya et~al.(2023)Bhattacharya, Zheng, and
  Zhou}]{Bhattacharya:2023hbq}
\bibinfo{author}{S.~Bhattacharya}, \bibinfo{author}{D.~Zheng},
  \bibinfo{author}{J.~Zhou}, \bibinfo{title}{{Probing quark orbital angular
  momentum at EIC and EicC -- arXiv:2312.01309 [hep-ph]}} .

\bibitem[{Xu et~al.(2024)Xu, Ding, Raya, Roberts, Rodr\'\i{}guez-Quintero, and
  Schmidt}]{Xu:2023izo}
\bibinfo{author}{Y.-Z. Xu}, \bibinfo{author}{M.~Ding},
  \bibinfo{author}{K.~Raya}, \bibinfo{author}{C.~D. Roberts},
  \bibinfo{author}{J.~Rodr\'\i{}guez-Quintero}, \bibinfo{author}{S.~M.
  Schmidt}, \bibinfo{title}{{Pion and kaon electromagnetic and gravitational
  form factors}}, \bibinfo{journal}{Eur. Phys. J. C}
  \bibinfo{volume}{84}~(\bibinfo{number}{2}) (\bibinfo{year}{2024})
  \bibinfo{pages}{191}.

\bibitem[{Ji et~al.(1996)Ji, Tang, and Hoodbhoy}]{Ji:1995cu}
\bibinfo{author}{X.-D. Ji}, \bibinfo{author}{J.~Tang},
  \bibinfo{author}{P.~Hoodbhoy}, \bibinfo{title}{{The spin structure of the
  nucleon in the asymptotic limit}}, \bibinfo{journal}{Phys. Rev. Lett.}
  \bibinfo{volume}{76} (\bibinfo{year}{1996}) \bibinfo{pages}{740--743}.

\bibitem[{Chen et~al.(2011)Chen, Sun, Wang, and Goldman}]{Chen:2011gn}
\bibinfo{author}{X.-S. Chen}, \bibinfo{author}{W.-M. Sun},
  \bibinfo{author}{F.~Wang}, \bibinfo{author}{T.~Goldman},
  \bibinfo{title}{{Proper identification of the gluon spin}},
  \bibinfo{journal}{Phys. Lett. B} \bibinfo{volume}{700} (\bibinfo{year}{2011})
  \bibinfo{pages}{21--24}.

\bibitem[{Ji et~al.(2021)Ji, Yuan, and Zhao}]{Ji:2020ena}
\bibinfo{author}{X.~Ji}, \bibinfo{author}{F.~Yuan}, \bibinfo{author}{Y.~Zhao},
  \bibinfo{title}{{What we know and what we don\textquoteright{}t know about
  the proton spin after 30 years}}, \bibinfo{journal}{Nature Rev. Phys.}
  \bibinfo{volume}{3}~(\bibinfo{number}{1}) (\bibinfo{year}{2021})
  \bibinfo{pages}{27--38}.

\bibitem[{Alexandrou et~al.(2020)Alexandrou, Bacchio, Constantinou, Finkenrath,
  Hadjiyiannakou, Jansen, Koutsou, Panagopoulos, and
  Spanoudes}]{Alexandrou:2020sml}
\bibinfo{author}{C.~Alexandrou}, \bibinfo{author}{S.~Bacchio},
  \bibinfo{author}{M.~Constantinou}, \bibinfo{author}{J.~Finkenrath},
  \bibinfo{author}{K.~Hadjiyiannakou}, \bibinfo{author}{K.~Jansen},
  \bibinfo{author}{G.~Koutsou}, \bibinfo{author}{H.~Panagopoulos},
  \bibinfo{author}{G.~Spanoudes}, \bibinfo{title}{{Complete flavor
  decomposition of the spin and momentum fraction of the proton using lattice
  QCD simulations at physical pion mass}}, \bibinfo{journal}{Phys. Rev. D}
  \bibinfo{volume}{101}~(\bibinfo{number}{9}) (\bibinfo{year}{2020})
  \bibinfo{pages}{094513}.

\bibitem[{Eichmann et~al.(2010)Eichmann, Alkofer, Krassnigg, and
  Nicmorus}]{Eichmann:2009qa}
\bibinfo{author}{G.~Eichmann}, \bibinfo{author}{R.~Alkofer},
  \bibinfo{author}{A.~Krassnigg}, \bibinfo{author}{D.~Nicmorus},
  \bibinfo{title}{{Nucleon mass from a covariant three-quark Faddeev
  equation}}, \bibinfo{journal}{Phys. Rev. Lett.} \bibinfo{volume}{104}
  (\bibinfo{year}{2010}) \bibinfo{pages}{201601}.

\bibitem[{Wang et~al.(2018)Wang, Qin, Roberts, and Schmidt}]{Wang:2018kto}
\bibinfo{author}{Q.-W. Wang}, \bibinfo{author}{S.-X. Qin},
  \bibinfo{author}{C.~D. Roberts}, \bibinfo{author}{S.~M. Schmidt},
  \bibinfo{title}{{Proton tensor charges from a Poincar{\'e}-covariant Faddeev
  equation}}, \bibinfo{journal}{Phys. Rev. D} \bibinfo{volume}{98}
  (\bibinfo{year}{2018}) \bibinfo{pages}{054019}.

\bibitem[{Cahill et~al.(1989)Cahill, Roberts, and Praschifka}]{Cahill:1988dx}
\bibinfo{author}{R.~T. Cahill}, \bibinfo{author}{C.~D. Roberts},
  \bibinfo{author}{J.~Praschifka}, \bibinfo{title}{{Baryon structure and QCD}},
  \bibinfo{journal}{Austral. J. Phys.} \bibinfo{volume}{42}
  (\bibinfo{year}{1989}) \bibinfo{pages}{129--145}.

\bibitem[{Reinhardt(1990)}]{Reinhardt:1989rw}
\bibinfo{author}{H.~Reinhardt}, \bibinfo{title}{{Hadronization of Quark Flavor
  Dynamics}}, \bibinfo{journal}{Phys. Lett. B} \bibinfo{volume}{244}
  (\bibinfo{year}{1990}) \bibinfo{pages}{316--326}.

\bibitem[{Efimov et~al.(1990)Efimov, Ivanov, and Lyubovitskij}]{Efimov:1990uz}
\bibinfo{author}{G.~V. Efimov}, \bibinfo{author}{M.~A. Ivanov},
  \bibinfo{author}{V.~E. Lyubovitskij}, \bibinfo{title}{{Quark - diquark
  approximation of the three quark structure of baryons in the quark
  confinement model}}, \bibinfo{journal}{Z. Phys. C} \bibinfo{volume}{47}
  (\bibinfo{year}{1990}) \bibinfo{pages}{583--594}.

\bibitem[{Eichmann(2022)}]{Eichmann:2022zxn}
\bibinfo{author}{G.~Eichmann}, \bibinfo{title}{{Theory Introduction to Baryon
  Spectroscopy}}, \bibinfo{journal}{Few Body Syst.}
  \bibinfo{volume}{63}~(\bibinfo{number}{3}) (\bibinfo{year}{2022})
  \bibinfo{pages}{57}.

\bibitem[{Liu et~al.(2022)Liu, Chen, Lu, Roberts, and Segovia}]{Liu:2022ndb}
\bibinfo{author}{L.~Liu}, \bibinfo{author}{C.~Chen}, \bibinfo{author}{Y.~Lu},
  \bibinfo{author}{C.~D. Roberts}, \bibinfo{author}{J.~Segovia},
  \bibinfo{title}{{Composition of low-lying $J=\tfrac{3}{2}^\pm$
  \ensuremath{\Delta}-baryons}}, \bibinfo{journal}{Phys. Rev. D}
  \bibinfo{volume}{105}~(\bibinfo{number}{11}) (\bibinfo{year}{2022})
  \bibinfo{pages}{114047}.

\bibitem[{Liu et~al.(2023)Liu, Chen, and Roberts}]{Liu:2022nku}
\bibinfo{author}{L.~Liu}, \bibinfo{author}{C.~Chen}, \bibinfo{author}{C.~D.
  Roberts}, \bibinfo{title}{{Wave functions of
  $(I,J^P)=(\tfrac{1}{2},\tfrac{3}{2}^\mp)$ baryons}}, \bibinfo{journal}{Phys.
  Rev. D} \bibinfo{volume}{107}~(\bibinfo{number}{1}) (\bibinfo{year}{2023})
  \bibinfo{pages}{014002}.

\bibitem[{Segovia et~al.(2015)Segovia, Roberts, and Schmidt}]{Segovia:2015ufa}
\bibinfo{author}{J.~Segovia}, \bibinfo{author}{C.~D. Roberts},
  \bibinfo{author}{S.~M. Schmidt}, \bibinfo{title}{{Understanding the nucleon
  as a Borromean bound-state}}, \bibinfo{journal}{Phys. Lett. B}
  \bibinfo{volume}{750} (\bibinfo{year}{2015}) \bibinfo{pages}{100--106}.

\bibitem[{Munczek(1995)}]{Munczek:1994zz}
\bibinfo{author}{H.~J. Munczek}, \bibinfo{title}{{Dynamical chiral symmetry
  breaking, Goldstone's theorem and the consistency of the Schwinger-Dyson and
  Bethe-Salpeter Equations}}, \bibinfo{journal}{Phys. Rev. D}
  \bibinfo{volume}{52} (\bibinfo{year}{1995}) \bibinfo{pages}{4736--4740}.

\bibitem[{Bender et~al.(1996)Bender, Roberts, and von Smekal}]{Bender:1996bb}
\bibinfo{author}{A.~Bender}, \bibinfo{author}{C.~D. Roberts},
  \bibinfo{author}{L.~von Smekal}, \bibinfo{title}{{Goldstone Theorem and
  Diquark Confinement Beyond Rainbow- Ladder Approximation}},
  \bibinfo{journal}{Phys. Lett. B} \bibinfo{volume}{380} (\bibinfo{year}{1996})
  \bibinfo{pages}{7--12}.

\bibitem[{Gao et~al.(2018)Gao, Qin, Roberts, and
  Rodr{\'{\i}}guez-Quintero}]{Gao:2017uox}
\bibinfo{author}{F.~Gao}, \bibinfo{author}{S.-X. Qin}, \bibinfo{author}{C.~D.
  Roberts}, \bibinfo{author}{J.~Rodr{\'{\i}}guez-Quintero},
  \bibinfo{title}{{Locating the Gribov horizon}}, \bibinfo{journal}{Phys. Rev.
  D} \bibinfo{volume}{97} (\bibinfo{year}{2018}) \bibinfo{pages}{034010}.

\bibitem[{Ebert et~al.(1996)Ebert, Feldmann, and Reinhardt}]{Ebert:1996vx}
\bibinfo{author}{D.~Ebert}, \bibinfo{author}{T.~Feldmann},
  \bibinfo{author}{H.~Reinhardt}, \bibinfo{title}{{Extended NJL model for light
  and heavy mesons without $q \bar q$ thresholds}}, \bibinfo{journal}{Phys.
  Lett. B} \bibinfo{volume}{388} (\bibinfo{year}{1996})
  \bibinfo{pages}{154--160}.

\bibitem[{Roberts et~al.(1992)Roberts, Williams, and Krein}]{Krein:1990sf}
\bibinfo{author}{C.~D. Roberts}, \bibinfo{author}{A.~G. Williams},
  \bibinfo{author}{G.~Krein}, \bibinfo{title}{{On the implications of
  confinement}}, \bibinfo{journal}{Int. J. Mod. Phys. A} \bibinfo{volume}{7}
  (\bibinfo{year}{1992}) \bibinfo{pages}{5607--5624}.

\bibitem[{Cui et~al.(2022{\natexlab{d}})Cui, Binosi, Roberts, and
  Schmidt}]{Cui:2022fyr}
\bibinfo{author}{Z.-F. Cui}, \bibinfo{author}{D.~Binosi},
  \bibinfo{author}{C.~D. Roberts}, \bibinfo{author}{S.~M. Schmidt},
  \bibinfo{title}{{Hadron and light nucleus radii from electron scattering}},
  \bibinfo{journal}{Chin. Phys. C} \bibinfo{volume}{46}~(\bibinfo{number}{12})
  (\bibinfo{year}{2022}{\natexlab{d}}) \bibinfo{pages}{122001}.

\bibitem[{Xu et~al.(2021)Xu, Cui, Roberts, and Xu}]{Xu:2021iwv}
\bibinfo{author}{Z.-N. Xu}, \bibinfo{author}{Z.-F. Cui}, \bibinfo{author}{C.~D.
  Roberts}, \bibinfo{author}{C.~Xu}, \bibinfo{title}{{Heavy + light
  pseudoscalar meson semileptonic transitions}}, \bibinfo{journal}{Eur. Phys.
  J. C} \bibinfo{volume}{81}~(\bibinfo{number}{12}) (\bibinfo{year}{2021})
  \bibinfo{pages}{1105}.

\bibitem[{Hecht et~al.(2002)Hecht, Oettel, Roberts, Schmidt, Tandy, and
  Thomas}]{Hecht:2002ej}
\bibinfo{author}{M.~B. Hecht}, \bibinfo{author}{M.~Oettel},
  \bibinfo{author}{C.~D. Roberts}, \bibinfo{author}{S.~M. Schmidt},
  \bibinfo{author}{P.~C. Tandy}, \bibinfo{author}{A.~W. Thomas},
  \bibinfo{title}{{Nucleon mass and pion loops}}, \bibinfo{journal}{Phys. Rev.
  C} \bibinfo{volume}{65} (\bibinfo{year}{2002}) \bibinfo{pages}{055204}.

\bibitem[{Sanchis-Alepuz et~al.(2014)Sanchis-Alepuz, Fischer, and
  Kubrak}]{Sanchis-Alepuz:2014wea}
\bibinfo{author}{H.~Sanchis-Alepuz}, \bibinfo{author}{C.~S. Fischer},
  \bibinfo{author}{S.~Kubrak}, \bibinfo{title}{{Pion cloud effects on baryon
  masses}}, \bibinfo{journal}{Phys. Lett. B} \bibinfo{volume}{733}
  (\bibinfo{year}{2014}) \bibinfo{pages}{151--157}.

\bibitem[{Garc\'\i{}a-Tecocoatzi et~al.(2017)Garc\'\i{}a-Tecocoatzi, Bijker,
  Ferretti, and Santopinto}]{Garcia-Tecocoatzi:2016rcj}
\bibinfo{author}{H.~Garc\'\i{}a-Tecocoatzi}, \bibinfo{author}{R.~Bijker},
  \bibinfo{author}{J.~Ferretti}, \bibinfo{author}{E.~Santopinto},
  \bibinfo{title}{{Self-energies of octet and decuplet baryons due to the
  coupling to the baryon-meson continuum}}, \bibinfo{journal}{Eur. Phys. J. A}
  \bibinfo{volume}{53}~(\bibinfo{number}{6}) (\bibinfo{year}{2017})
  \bibinfo{pages}{115}.

\bibitem[{Chen et~al.(2018)Chen, Ping, Roberts, and Segovia}]{Chen:2017mug}
\bibinfo{author}{X.~Chen}, \bibinfo{author}{J.~Ping}, \bibinfo{author}{C.~D.
  Roberts}, \bibinfo{author}{J.~Segovia}, \bibinfo{title}{{Light-meson masses
  in an unquenched quark model}}, \bibinfo{journal}{Phys. Rev. D}
  \bibinfo{volume}{97} (\bibinfo{year}{2018}) \bibinfo{pages}{094016}.

\bibitem[{Aznauryan et~al.(2013)}]{Aznauryan:2012ba}
\bibinfo{author}{I.~G. Aznauryan}, et~al., \bibinfo{title}{{Studies of Nucleon
  Resonance Structure in Exclusive Meson Electroproduction}},
  \bibinfo{journal}{Int. J. Mod. Phys. E} \bibinfo{volume}{22}
  (\bibinfo{year}{2013}) \bibinfo{pages}{1330015}.

\bibitem[{Burkert and Roberts(2019)}]{Burkert:2017djo}
\bibinfo{author}{V.~D. Burkert}, \bibinfo{author}{C.~D. Roberts},
  \bibinfo{title}{{Roper resonance: Toward a solution to the fifty-year
  puzzle}}, \bibinfo{journal}{Rev. Mod. Phys.} \bibinfo{volume}{91}
  (\bibinfo{year}{2019}) \bibinfo{pages}{011003}.

\end{thebibliography}

\end{document}